\documentclass[structabstract]{aa}  
\usepackage{graphicx}
\usepackage{enumerate}
\usepackage{natbib}
 \usepackage{lscape}

\bibpunct{(}{)}{;}{a}{}{,} % to follow the A&A style
\usepackage{txfonts}
\usepackage{enumitem}
\usepackage{float}
\usepackage{array}
\usepackage{txfonts}
\usepackage{epstopdf}
\begin{document}
    \title{
    A gas density drop in the inner 6 AU of the transition disk around the Herbig Ae star HD~139614
    }
    \subtitle{Further evidence for a giant planet inside the disk?}
   \author{A. Carmona
          \inst{1, 2, 3}\fnmsep \thanks{Part of this research has been done by A. Carmona under the frame of ESO's scientist visitor program during November 2013 
          and at Universit\'e Grenoble Alpes, Institut de Plan\'etologie et d'Astrophysique de Grenoble (IPAG), F-38000 Grenoble, France. Current address:  Institut de Recherche en Astrophysique et Plan\'etologie (IRAP), 14 avenue E. Belin, Toulouse, F-31400, France.}
           \fnmsep\thanks{Based on CRIRES observations collected at the VLTI and VLT (European Southern Observatory, Paranal, Chile) with program 091.C-0671(B).}
          \and
          {W.F. Thi}
          \inst{4}
          \and
          {I. Kamp}
          \inst{5}   
          \and
          {C. Baruteau}
          \inst{1}
          \and       
           {A. Matter}
          \inst{6}
             \and
           {M. van den Ancker}
          \inst{7}
	\and
          {C. Pinte}
          \inst{8,9}
          \and
          {A. K\'osp\'al}
          \inst{2} 
           \and
          {M. Audard}
          \inst{10}  
          \and
          {A. Liebhart}
          \inst{11}
          \and
          {A. Sicilia-Aguilar}
          \inst{12}
           \and
          {P. Pinilla}
          \inst{13}
           \and
          {Zs. Reg\'aly}
          \inst{2}
           \and
          {M. G\"udel}
          \inst{11}   
          \and
          {Th. Henning}
          \inst{14}
	 \and
          {L.A. Cieza}
          \inst{15}
          \and
          {C. Baldovin-Saavedra}
          \inst{11}
          \and          
          {G. Meeus}
          \inst{3}
          \and
          {C. Eiroa}
          \inst{3,16}
                    }
    \institute{
    Universit\'e de Toulouse, UPS-OMP, IRAP, 14 avenue E. Belin, Toulouse, F-31400, France
    \email{andres.carmona@irap.omp.eu}   
    \and
      Konkoly Observatory, Research Centre for Astronomy and Earth Sciences, Hungarian Academy of Sciences, P.O. Box 67, H-1525 Budapest, Hungary
    \and
    Departamento de F\'isica Te\'orica, Universidad Aut\'onoma de Madrid, Campus Cantoblanco, 28049, Madrid, Spain
    \and
   Max Planck Institute for Extraterrestrial Physics, Giessenbachstrasse 1, D-85748 Garching bei M\"unchen, Germany
    \and
    Kapteyn Astronomical Institute, Postbus 800, 9700 AV, Groningen, The Netherlands
    \and 
    Laboratoire Lagrange, Universit\'e C\^ote d'Azur, Observatoire de la C\^ote d'Azur, CNRS, Boulevard de l'Observatoire, CS 34229, 06304 Nice Cedex 4, France.
    \and
    European Southern Observatory, Karl-Schwarzschild-Str. 2, D-85748 Garching bei M\"unchen, Germany
    \and
    UMI-FCA, CNRS/INSU, France (UMI 3386), and Dept. de Astronom\'ia, Universidad de Chile, Santiago, Chile
    \and
    Univ. Grenoble Alpes, IPAG, 38000, Grenoble, France; CNRS, IPAG, 38000, Grenoble, France
     \and
    Department of Astronomy, University of Geneva, Ch. d'Ecogia 16, 1290 Versoix, Switzerland
    \and
    University of Vienna, Department of Astronomy, T\"urkenschanzstrasse 17, 1180, Vienna, Austria
    \and
    SUPA, School of Physics and Astronomy, University of St
    Andrews, North Haugh, St Andrews KY16 9SS, UK
        \and
     Leiden Observatory, Leiden University, P.O. Box 9513, 2300RA Leiden, The Netherlands
     \and   
      Max-Planck-Institut f\"ur Astronomie, K\"onigstuhl 17, D-69117 Heidelberg, Germany
       \and
        N\'ucleo de Astronom\'ia,  Facultad de Ingenier\'ia, Universidad Diego Portales,  Av. Ejercito 441, Santiago, Chile
        \and
    Unidad Asociada Astro-UAM/CSIC
    }
                            
   \date{}

% \abstract{}{}{}{}{} 
% 5 {} token are mandatory
 
  \abstract
  {Quantifying the gas surface density inside the dust cavities and gaps of transition disks is important to establish their origin.  
   }
  % aims heading (mandatory)
   {We seek to constrain the surface density of warm gas in the inner disk of HD~139614, an accreting 9 Myr Herbig Ae star with a (pre-)transition disk 
   exhibiting a dust gap from 2.3$\pm$0.1 to 5.3$\pm$0.3 AU.
   }
  % methods heading (mandatory)
   {We observed HD~139614 with ESO/VLT CRIRES and obtained high-resolution (R$\sim$90\,000) spectra of 
   CO ro-vibrational emission at 4.7 $\mu$m. 
   We derived constraints on the disk's structure by modeling the CO isotopolog line-profiles, the spectroastrometric signal, 
   and the rotational diagrams using grids of flat Keplerian disk models.
   }
 {
We detected $\upsilon=1\rightarrow0~^{12}$CO,  2$\rightarrow$1~$^{12}$CO, 1$\rightarrow$0~$^{13}$CO, 1$\rightarrow$0~C$^{18}$O, 
and 1$\rightarrow$0~C$^{17}$O ro-vibrational lines. 
Lines are consistent with disk emission and thermal excitation.
$^{12}$CO $\upsilon=1\rightarrow0$  lines have an average width of 14 km s$^{-1}$,  $T_{\rm gas}$ of 450~K and an emitting region from 1 to 15 AU.  
$^{13}$CO and C$^{18}$O lines are on average 70 and 100 K colder,  1 and 4 km s$^{-1}$ narrower than $^{12}$CO 
$\upsilon=1\rightarrow0$, and are dominated by emission at R$\geq$ 6 AU.
The $^{12}$CO $\upsilon=1\rightarrow0$ composite line-profile indicates that if there is a gap devoid of gas it must have a width narrower than 2 AU.
We find that a drop in the gas surface density  ($\delta_{\rm gas}$) at $R<5-6$ AU is required to be able 
to  simultaneously reproduce the line-profiles and rotational diagrams of the three CO isotopologs.
Models without a gas density drop generate $^{13}$CO and C$^{18}$O  emission lines that are too broad and warm.
The value of $\delta_{\rm gas}$ can range from $10^{-2}$ to $10^{-4}$ depending on the gas-to-dust ratio of the outer disk.
We find that the gas surface density profile at $1<R<$ 6 AU is flat or increases with radius.
We derive a gas column density at $1<R<6$ AU of $N_H=3\times10^{19} - 10^{21}$ cm$^{-2}$ ($7\times10^{-5} - 2.4\times10^{-3}$ g cm$^{-2}$) assuming $N_{\rm CO} =  10^{-4} N_H$. 
We find a 5$\sigma$ upper limit on the CO column density $N_{\rm CO}$ at R$\leq$1 AU of $5\times10^{15}$ cm$^{-2}$ ($N_H\leq5\times10^{19}$ cm$^{-2}$).
   }
   {
   The dust gap in the disk of HD~139614 has molecular gas.
   The distribution and amount of gas at R$\leq6$ AU in HD~139614 is very different from that of a primordial disk.
The gas surface density in the disk at $R\leq1$ AU and at $1<R<6$ AU is significantly lower than the surface density that would be expected 
from the accretion rate of HD~139614 ($10^{-8}$ M$_\odot$ yr$^{-1}$) assuming a standard viscous $\alpha$-disk model.
The gas density drop, the non-negative density gradient in the gas inside 6 AU, and the absence of a wide ($>$ 2 AU) gas gap, 
suggest the presence of an embedded $<2$ M$_{\rm J}$ planet at around 4 AU.

     }

\titlerunning{CO ro-vibrational emission in the transition disk HD~139614.}
   \maketitle
%
%________________________________________________________________

\section{Introduction}
{Transition disks are protoplanetary disks that exhibit a deficit of continuum 
emission at near- and/or mid-IR wavelengths in their spectral energy distribution 
\citep[for a recent review, see][]{Espaillat2014}.
This deficit of emission is commonly interpreted as evidence of a  dust 
gap, a dust cavity, or a dust hole inside the disk\footnote{We call a dust hole, 
when no dust emission is detected inside a determined radius in the disk
at {\it all} wavelengths.
We call a dust cavity, a region where there is a drop in the dust density. 
Inside the dust cavity radius dust is still present (i.e. continuum emission
is detected inside the cavity radius, for instance at IR wavelengths).
We call a dust gap when continuum emission is detected at radii smaller and larger than the location of the gap.
A dust cavity can have a dust gap inside it.}.
Sub-mm interferometry observations have confirmed the existence of dust cavities by spatially resolving the thermal 
emission from cold large ($\sim$mm) grains at tens of AU in transition disks \citep[e.g.,][]{Pietu2006,Brown2009,Andrews2011,Cieza2012,Casassus2013,Perez2014}}. 
Observations of scattered light in the near-IR using adaptive optics have further confirmed the dust cavities
in micron-sized dust grains. 
These high-spatial resolution observations show that the cavity size in small grains can be smaller than that in large grains 
\citep[e.g.,][]{Muto2012, Garufi2013, Follette2013, Pinilla2015J16}.
Furthermore, near-IR scattered light imaging and sub-mm interferometry observations 
have revealed that a large fraction of transition disks has asymmetries  in the dust distribution
(e.g.  spirals, blobs, and horseshoe shapes), although, the presence and shape of asymmetries appear to be different 
depending on the wavelength of the observations and  thus the dust sizes traced
\citep[e.g.,][]{Muto2012,vanderMarel2013,Isella2013,Perez2014,Benisty2015,Follete2015}.

\begin{table*}
\begin{center}
\caption{Stellar properties}
\begin{tabular}{cccccccccccc}
\hline
\hline
 Star & Sp. Type & T${\rm eff}$  &  d    & Mass           & Radius  & RV    &  W2  & Age  & i$_{\rm disk}$ & $L_{\rm X}$  & $\dot{M}$ \\
         &                &  [K]               & [pc] &  $[M_{\odot}]$ & $[R_{\odot}]$ & [km s$^{-1}$] & [mag]   &  [Myr] & [$^\circ$] &  [erg s$^{-1}$] & [M$_{\odot}$ yr$^{-1}$]\\[1mm]  
\hline
 {HD~139614}      &  A7Ve $^a$ & 7600$\pm$300 $^{b}$  & {131$\pm$5} $^{c}$ &  1.76$^{+0.15}_{-0.08}$ $^b$  & 2.06$\pm$0.42 $^b$  &0.3$\pm$2.3 $^{b}$  & 5.1 $^d$ &
 8.8$^{ +4.5}_{-1.9}$ $^b$ & 20 $^e$ & 1.2$\times10^{29}$ $^f$ & $10^{-8}$ $^g$ \\[1mm]	
 & &  7850 $^h$ &  & 1.7$\pm$0.3 $^h$  & 1.6~$^h$& & & $>$7 $^a$ & \\[1mm]	
\hline
\end{tabular}
\tablefoot{
$^a$~\citet[][]{vanBoekel2005};
$^b$~\citet[][]{Folsom2012,Alecian2013};
$^c$~{\citet[][]{Gaia2016}};
$^d$~4.6 $\mu$m, WISE satellite release 2012 \citep[][]{Cutri2012};
$^e$ \citet[][]{Matter2016};
$^f$ G\"udel et al. (in prep.) see Sect.~\ref{origin}; 
$^g$ \citet[][]{GarciaLopez2006};
$^h$ \citet[][]{vanBoekel2005} stellar properties used in \citet[][]{Matter2016}.
\\[3mm]}
\label{stellar-properties}
\caption{Log of the science and calibrator observations} 
{
\begin{tabular}{lccccccccc}
\hline
\hline
Star 			& UT Date Obs.	  & $t_{\rm exp}$ & Airmass & Seeing & RV$_{\rm bary}$$^a$  & PSF$_{\rm FWHM}$$^{b}$ & S/N$^{~b,c}$ & \multicolumn{2}{c}{sensitivity 3$\sigma$$~^{b,d}$} \\ 
			& [y-m-d]  	&	     		 [s]		       	&   & 	['']  	& 	[km s$^{-1}$]		& [mas] 			      & 	& \multicolumn{2}{c}{[10$^{-15}$ erg s$^{-1}$ cm$^{-2}$]}\\[1mm]
			& & & & & & & & 3.3 km s$^{-1}$& 20 km s$^{-1}$ \\
\hline
HD~139614         & 2013-06-15 & 2400  & $1.07 - 1.13$ & $ 0.93 - 1.23$ & $9.74\pm0.02$ & 178$\pm$10 & 160 $-$ 100 & 0.2 $-$ 0.3 & 1.2 $-$ 2.0\\
CAL HIP 76829	 & 2013-06-15 & 320    & $1.16 - 1.18$ & $ 0.87 - 1.06$ &  $9.36\pm0.01$& 172$\pm$10 & 310 $-$ 200\\[1mm]
\hline
\end{tabular}
\tablefoot{\small $^a$ {Radial velocity due to the rotation of the Earth, the motion of the Earth about the Earth-Moon barycenter, and the motion of the Earth around the Sun}; $^b$ measured in one nod position; 
$^c$ for the science spectra the S/N is measured in the telluric-corrected spectrum, note that the S/N decreases from chip 1 to chip 4;
$^d$ integrated flux sensitivity limits are given for a spectrally unresolved line of width 3.3 km s$^{-1}$ and a line of width 20 km s$^{-1}$.}\\[3mm]
}
\label{table_observations}
\end{center}
\end{table*}

The origin of the dust cavities and gaps in transition disks is a matter of intense debate in the literature: 
scenarios such as grain growth (e.g., \citealt{DullemondDominik2005}; but see \citealt{Birnstiel2012}),
size-dependent dust radial drift \citep[e.g.,][]{PinteLaibe2014},
dust dynamics at the boundary of the dead-zone \citep[][]{Regaly2012},
photoevaporation \citep[e.g.,][]{Clarke2001,AlexanderArmitage2007,Owen2012},  
{giant planet(s) \cite[e.g.,][]{MarshMahoney1992,Lubow1999,Rice2003,Quillen2004,Varniere2006,Zhu2011}}, 
dynamical interactions in multiple systems \citep[e.g.,][]{ArtymowiczLubow1996,IrelandKraus2008, Fang2014},
and magneto-hydrodynamical phenomena 
\citep[][]{ChiangMurrayClay2007} have all been proposed. 

Accretion signatures in many transition disks \citep[e.g.,][]{Fang2009,SiciliaAguilar2013,Manara2014}
and emission of warm \cite[e.g,][]{Bary2003,Pontoppidan2008, Pontoppidan2011, Salyk2009, Salyk2011} and cold \citep[][]{Casassus2013,Bruderer2014,Perez2015HD142527,Canovas2015,vanderMarel2015,vanderMarel2016}
molecular gas indicate that the dust cavities in accreting transition disks contain gas.
Radiative transfer modeling of CO ro-vibrational emission \citep[][]{Carmona2014} and CO pure rotational emission \citep[][]{Bruderer2013,Perez2015HD142527,vanderMarel2015,vanderMarel2016}
further suggests a gas surface density drop ($\delta_{\rm gas}$) inside the dust cavity, with $\delta_{\rm gas}$ values varying from 0.1 up to 10$^{-5}$ (see Table~\ref{table_gasdrop}).
Some of the transition disks are not accreting and thus do not seem to have gas \citep[][]{SiciliaAguilar2010}.
There is also a substantial difference in the global structure and/or disk
mass between accreting and non-accreting transition disks, with the non-accreting disks
being significantly more evolved (lower masses, flatter disks) as seen with Herschel \citep[][]{SiciliaAguilar2015}.

The different spatial locations of dust grains of different sizes,
the gas inside the sub-mm dust cavities, together with 
the different surface density profiles of gas and dust strongly favor
the planet(s) scenario. 
However, we probably witness several coexisting mechanisms, 
because planet formation might affect the dynamics of the dust in the disk \citep[e.g., ][]{Rice2003, Zhu2011, Pinilla2012, Pinilla2015} 
or favor the onset of photoevaporation, when the accretion rate has decreased \citep[e.g.,][]{Rosotti2013,Dittkrist2014}.
A large portion of studies of transition disks have focused on investigating disks that are bright in the sub-mm 
and that have large dust cavities of tens of AU \citep[e.g.,][]{Andrews2011,vanderMarel2015b}.
Because a single Jovian planet interacting with the disk is expected to open a gap only a few AU wide \citep[e.g.,][]{Kley1999,Crida2007},
multiple (unseen) giant planets have been postulated as a possible explanation 
for the observed large dust cavities 
\citep[][]{Zhu2011, Dodson-RobinsonSalyk2011}. 

{In a recent near-  and mid-IR interferometry campaign,
\citet[][]{Matter2014,Matter2016} have revealed that the 9 Myr old \cite[][]{Alecian2013} accreting \citep[$10^{-8}$ M$_{\odot}$/yr, ][]{GarciaLopez2006} Herbig A7Ve star HD~139614 has a transition
disk with a narrow dust gap extending from 2.3$\pm$0.1 to 5.3$\pm$0.3 AU \footnote{The dust gap limits derived in \citet[][]{Matter2016} are 2.5$\pm$0.1 to 5.7$\pm$0.3 AU. They were 
calculated using a distance of 140 pc. The values in the text are the values corrected by the new Gaia distance. Both values are consistent within the uncertainties.}.
and a dust density drop $\delta_{\rm dust}$ at R$<$6 AU of 10$^{-4}$
(see Table~\ref{stellar-properties} for a summary of the stellar properties).
HD~139614 is one of the first objects with a spatially resolved dust gap with a width of only a few AU,
thus it might be the case of a transition disk where the dust gap has been opened by a single giant planet.

HD~139614 is located within the Sco OB2-3 association \citep[][]{Acke2005}  {at a distance of 131$\pm$5 pc} \citep[][]{Gaia2016}.
HD~139614 has peculiar chemical abundances in its photosphere \citep[][]{Folsom2012}, 
with depletions of heavier refractory elements, while C, N, and O are approximately solar.  
HD~139614 belongs to the group I Herbig Ae stars according to the Spectral Energy Distribution (SED) classification scheme of 
\citet[][]{Meeus2001}, which suggests that its outer disk is flared.
\citet[][]{Matter2016} derived a dust disk mass of 10$^{-4}$ M$_\odot$ based on a fit to the SED.
The {\it Spitzer} mid-IR spectra of HD~139614 exhibit a weak amorphous silicate feature at 10~$\mu$m 
\citep[][]{Juhasz2010} and Polycyclic Aromatic Hydrocarbons (PAH) emission \cite[][]{Acke2010}.
The disk's mid-IR continuum  has been spatially resolved at 18 $\mu$m ({\it FWHM} of 17 $\pm$4 AU)
but it is not resolved at 12 $\mu$m \citep[][]{Marinas2011}.
\citet[][]{Kospal2012} reported that the ISOPHOT-S, {\it Spitzer} and TIMMI-2/ESO 3.6m  
mid-infrared spectra taken at different epochs agree within the measurement uncertainties, thus suggesting that there is no strong mid-IR variability in the source. 
Emission from cold CO gas in the outer disk of HD~139614 has been reported in JCMT single-dish observations
by \citet[][]{Dent2005} and \citet[][]{PanicHogerheijde2009}.
Emission of [\ion{O}{I}] at 63 $\mu$m from the disk has been detected by Herschel  \citep[][]{Meeus2012,Fedele2013}.
The  [\ion{O}{I}] 63 $\mu$m line flux of HD~139614 is among the weakest of the whole Herbig Ae sample observed by Herschel.
No emission of [\ion{O}{I}] at 145 $\mu$m, [\ion{C}{II}] at 157 $\mu$m,  CO, H$_2$O, OH or CH$^{+}$ in the 50 $-$200$~\mu$m region 
was detected by Herschel \citep[][]{Meeus2012,Meeus2013, Fedele2013}.

}

In this paper we present the results of high-resolution spectroscopy 
observations of CO ro-vibrational emission at 4.7 $\mu$m towards HD~139614 obtained
with the ESO/VLT CRIRES instrument \citep[][]{Kaufl2004}.
Our aim is to use CO isotopolog spectra to constrain the warm gas content in the inner disk 
of HD~139614 and address the following questions:
What is the gas distribution in the inner disk of HD~139614?
Does the HD~139614 disk have a gas-hole, a gas-density drop or a gap in the gas?
How does the gas distribution compare with the dust distribution?
What is the most likely explanation for the observed gas and dust distributions in HD~139614?

The paper is organized as follows.
We start by describing the observations and data reduction in Sect. 2.
In Sect. 3, we present the observational results.
In Sect. 4, we derive the CO-emitting region, the average temperature and column density of the emitting gas,  
and the gas surface density and temperature distribution.
In Sect. 5, we discuss our results in the context of the proposed scenarios for the origin of transition disks
and compare HD~139614 with other transition disks.
Sect. 6 summarizes our work and provides our conclusions.

\section{Observations and data reduction}
HD~139614 was observed with the high-resolution near-IR spectrograph CRIRES at the ESO Very Large Telescope 
at  Cerro Paranal Chile in June 2013. 
{CRIRES has a pixel scale of 0.086 arcsec/pixel in the spatial direction {(11 AU at 131 pc)}
and {2.246$\times10^{-6}$ $\mu$m} in the wavelength direction (0.14 km s$^{-1}$ at 4.7 $\mu$m)}.
Observations were performed with a 0.2" slit oriented north$-$south 
using adaptive optics and the target as a natural guide star.
{Observations were performed in the CRIRES ELEV mode, which maintains the slit at the same north-south position angle during the whole observing sequence}.
A standard ABBA nodding sequence was executed using
a nodding throw of 12" along the slit  and two ABBA nodding cycles.
{Observations used a wavelength setting centered on 4.780~$\mu$m, 
covering a wavelength range from 4.713 $\mu$m to 4.818  $\mu$m}. 
The telluric standard star HIP~76829 was observed immediately following the
science observations. 
We provide a summary of the observations in Table~\ref{table_observations}.

We reduced the data with the CRIRES pipeline version 2.3.1\footnote{https://www.eso.org/sci/software/pipelines}
and a custom set of IDL routines for improved 1D spectrum-merging from the two nodding positions, 
accurate telluric correction and wavelength calibration.
Nodding sequences were corrected for non-linear effects, flat-fielded, and combined 
using the CRIRES pipeline.
A combined 2D spectrum for the nod A and nod B positions was generated individually.
Each combined 2D spectrum was corrected for combination residuals
(due to small fluctuations in the sky brightness between nods) 
by subtracting a background spectrum at each position.
This residual background spectrum was obtained by computing at each wavelength the median
of two background windows each 20 pixels wide at either sides of the PSF.
Before subtraction, 
the residual background spectrum was smoothed in the wavelength direction with a three-pixel box.

A 1D spectrum was extracted from each combined 2D spectrum of nod A and nod B
using the optimal extraction method implemented within the CRIRES pipeline. 
Bad pixels and cosmic rays in the 1D spectrum of each nod 
were removed manually using the information of the 1D spectrum of the other nod.
The 1D spectra of both nods were merged taking their average.
Before merging, 
the 1D spectrum of nod B was shifted a fraction of a pixel such that the cross-correlation between the 1D spectrum of nod A and nod B was maximized.
This was done to correct for small sub-pixel differences in wavelength that are due to the tilt of the spectra in the spatial direction.
The merged 1D spectrum was wavelength calibrated using the telluric absorption lines
by cross-correlation with a HITRAN atmospheric spectrum of Paranal.
{The accuracy in the wavelength calibration is 0.15 $-$ 0.2 km s$^{-1}$}.

A 1D telluric standard star spectrum was obtained {from the telluric standard observation} 
following the same procedure as was used for the 1D science spectrum. 
The science 1D spectrum was then corrected for telluric absorption by dividing it by the 1D spectrum of the standard
star. Two adjustments in the 1D standard star spectrum were performed before the telluric correction.
First, the 1D standard star spectrum was shifted in the wavelength direction a fraction of a pixel, such that the cross-correlation  
with the science spectrum was maximized.
Second, the differences on the depth of the telluric lines of the 1D standard star with respect to the 1D science spectrum spectrum were corrected.
For this we found the parameter $f$ in 
\begin{equation}
I_{\rm STD~corrected} = I_{0}~e^{-f\,\tau}
\end{equation}
\label{tau_correction}
that gave the smallest $\chi^2$ statistic between the normalized
science spectrum and normalized standard star spectrum.
The factor $f$ controls the depth of the atmospheric absorption.
{The goal is to find the value of $f$ that makes the depth of telluric lines of the standard star the same as in the science spectrum}. 
The $\chi^2$ between the normalized science spectrum and normalized standard star spectrum (thus the value of the factor $f$) 
was calculated for each chip independently.
A region with several unsaturated sky absorption lines and without CO ro-vibrational emission was selected in each chip for this.
The optical depth $\tau$ was estimated using 
\begin{equation}
\tau=-{\rm ln}~(I_{\rm STD~observed}/I_{0})
\end{equation}
\begin{equation}
I_{0}=\bar{I}_{\rm STD~observed}+3\sigma.
\end{equation}
Here $\bar{I}_{\rm STD~observed}$ is median of the standard star flux at the wavelengths within 95\% and 100\% of the atmospheric transmission and
$\sigma$ is the noise in the standard star spectrum in  the same wavelength range.

\begin{figure}
\begin{center}
\includegraphics[width=0.5\textwidth]{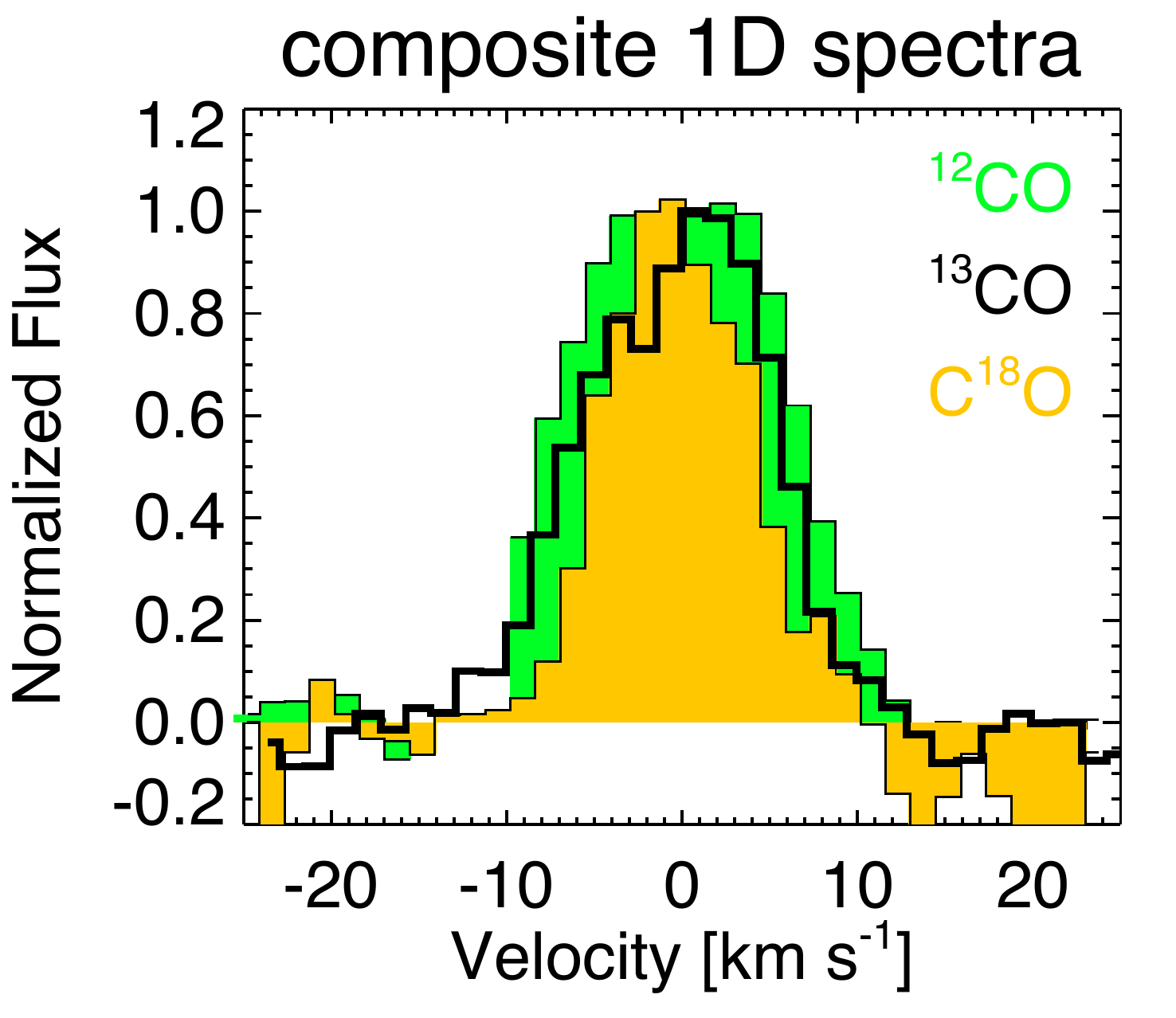} 
\includegraphics[width=0.5\textwidth]{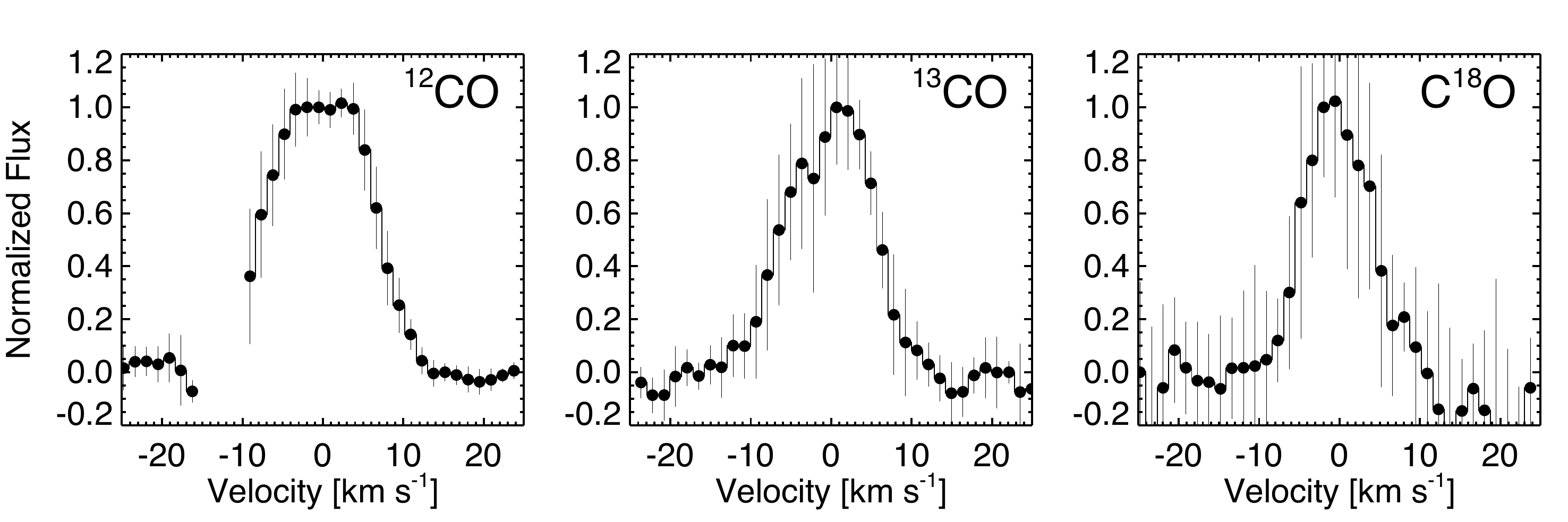}
\caption{Composite normalized spectrum of the $\upsilon$=$1\rightarrow0$ $^{12}$CO, $^{13}$CO, and  C$^{18}$O  lines.
Error bars are 1$\sigma$ in each spectrum.
}
\label{composite_spectrum}
\end{center}
\end{figure}
The telluric corrected 1D spectrum was flux calibrated by first normalizing it with a polynomial fit to the continuum
and then multiplying the normalized spectrum by the expected flux of the WISE W2 (4.6 $\mu$m) magnitude of HD~139614 \citep[5.1 mag, WISE release 2012,][]{Cutri2012}.
To convert the magnitude into flux we used the 4.7 $\mu$m photometry and the zero points of \citet[][]{Johnson1966}\footnote{The WISE and Johnson 1966 
zero-points differ by 10\%, which is a value lower than the uncertainties due to slit losses and systematic errors.}.
Errors in the final flux-calibrated spectra are dominated by slit losses and systematic errors in the telluric correction and are around 20\%.
Finally, the flux-calibrated 1D spectrum was corrected for the radial velocity (RV) of the star \citep[0.3$\pm$2.3 km s$^{-1}$,][]{Alecian2013}
and the radial velocity due to the rotation of Earth, {the motion of Earth around the Sun}, and the motion of Earth about the Earth-Moon barycenter, using the velocities given by the IRAF task {\it rvcorrect} (RV$_{\rm bary}$ = $-1\times V_{helio}$).
Integrated line fluxes, line-profile centers and {\it FWHM} were measured in the telluric-corrected 
1D spectrum using a Gaussian fit to the line-profiles.
The errors on these quantities are the errors on the Gaussian fit.

To produce a merged 2D spectrum,
we employed the following procedure.
The 2D nod A and nod B spectra were corrected for the tilt of the PSF along the wavelength axis using a second-degree polynomial.
The 2D nod B spectrum was shifted by a fraction of a pixel in the wavelength direction with a value 
equal to the shift found for the 1D spectrum extracted for nod B .
A 2D section of $\pm$20 pixels from the PSF center was extracted from the nod A and nod B 2D spectra,
and both sections were averaged to obtain a merged 2D spectrum. 
The merged 2D spectrum was corrected for telluric absorption by diving it by the 1D spectrum of  the standard star.

The photocenter (i.e., spectro-astrometric signature) was calculated from the merged 2D spectrum by employing
 the formalism described by \citet[][]{Pontoppidan2011}.
The {\it PSF-FWHM} as a function of the wavelength was calculated by fitting a Gaussian in the spatial direction of the merged 2D spectrum.
We calculated the composite 1D line-profiles, photocenter and {\it PSF$-$FWHM}  for each isotopolog 
by averaging the data of individual detected lines.
This was done to increase the signal of the CO line with respect to the continuum.
The averaging procedure was performed using the velocity as wavelength scale. 
The theoretical wavelength center of each transition was used as $v=0$ km s$^{-1}$ velocity reference. 
We selected only emission lines that were not blended with other transitions.
For each velocity channel we selected the data in regions with atmospheric transmission 
higher than 20\%, and calculated the average flux when at least three data points were available.
Channels with fewer than three data points were defined as NaN to exclude data from 
regions of poor atmospheric transmission.
The error in each channel was defined as the standard deviation of the values in each channel.
For further analysis, 
the composite data were recentered such that the center of the 1D spectrum was at $v=0$ km s$^{-1}$,
and the 1D spectrum was continuum subtracted and normalized by the peak flux (median of the flux within $\pm$2 km s$^{-1}$).

\begin{figure*}
 \includegraphics[width=\textwidth]{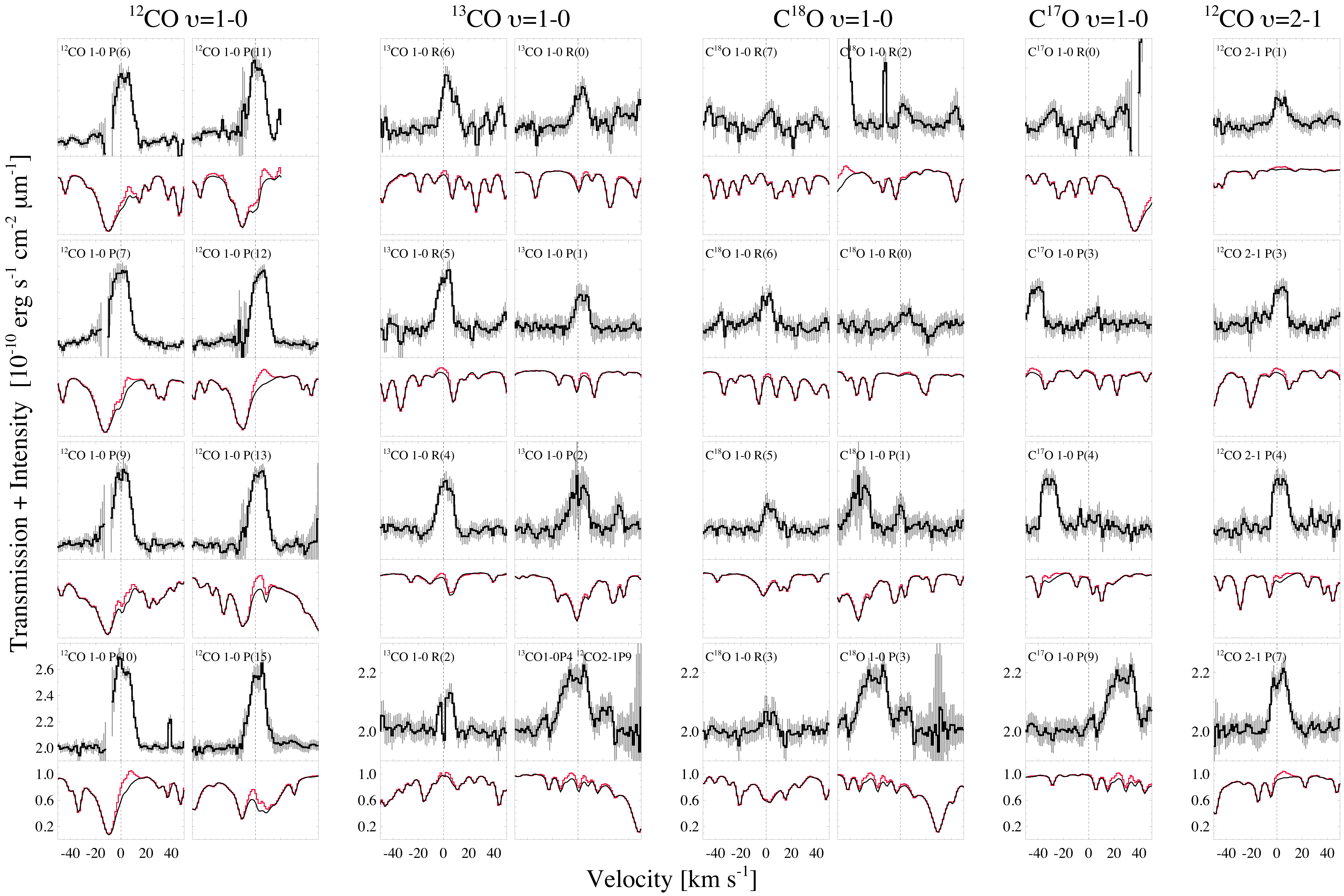} 
 \caption{Examples of the $\upsilon=1\rightarrow0$ $^{12}$CO, $^{13}$CO, C$^{18}$O, C$^{17}$O and $\upsilon=2\rightarrow1$ $^{12}$CO lines observed. The lower panels
 display the normalized spectrum of the target (in red) and the spectrum of the telluric standard (in black). 
 The spectra are presented corrected by the radial velocity of HD~139614 and the barycentric velocity.
 The references for $v=0$ km s$^{-1}$ are the theoretical wavelengths of each of the transitions. 
 Note that the flux scale is larger for the $\upsilon=1\rightarrow0$ $^{12}$CO lines.
 Error bars are 3$\sigma$. Several $\upsilon=3\rightarrow2$ $^{12}$CO lines were covered in the spectra but none were detected. 
 See Table~\ref{table_fluxes} for a summary of the centers, fluxes, flux upper limits and {\it FWHM} of the lines.}
 \label{spectrum_summary}    
 \end{figure*}

 \section{Observational results}
 
\label{Results}
We have detected $\upsilon=1\rightarrow0\,^{12}$CO, $^{13}$CO, C$^{18}$O, C$^{17}$O,
and  $\upsilon=2\rightarrow1\,^{12}$CO emission lines.
The $\upsilon=3\rightarrow2~^{12}$CO emission lines are not detected in the spectrum.
We display  
a summary of the CO lines detected together with the atmospheric transmission in  Fig.~\ref{spectrum_summary}.
In Table~\ref{table_fluxes} in the Appendix, we summarize the observed lines,  their centers,
integrated fluxes, {\it FWHM} 
and the average line ratios with respect to 1$\rightarrow$0\,$^{12}$CO emission.
To keep the notation short in the remaining of the paper, 
we mean by $^{12}$CO, $^{13}$CO, C$^{18}$O emission $\upsilon=1\rightarrow0$  
$^{12}$CO, $^{13}$CO, C$^{18}$O emission unless otherwise specified.

We reached a 3 $\sigma$ sensitivity of 2$\times10^{-16}$ erg s$^{-1}$ cm$^{-2}$ for a line width of 3.3 km s$^{-1}$,
and 1.2$\times10^{-15}$ erg s$^{-1}$ cm$^{-2}$ for a line width of 20 km s$^{-1}$ (equivalent widths of 0.01 and 0.075 \AA~respectively).
We achieved a spectral resolution of $\sim$3.3~km s$^{-1}$  (R$\sim\,10^{5}$) as measured in an unresolved unsaturated sky-absorption line.
The centers of the CO emission lines in the barycentric asnd radial-velocity-corrected spectra are located on average at 
$v=2\pm1$ km s$^{-1}$
(see Table~\ref{table_fluxes}). As this value is close to zero and is lower than the uncertainty of $\pm$2.3 km s$^{-1}$ in the radial velocity \citep[][]{Alecian2013},
we conclude that the CO emission is at the stellar velocity and therefore most likely originates in the disk.
{The $^{12}$CO composite line-profile has a flat top and does not display evidence of asymmetries\footnote{
We note that the 1$\sigma$ error in the flux in the $^{12}$CO composite line-profile is slightly larger at negative velocities.
This is because the left side of the line is located in a region with a lower atmospheric transmission. 
The small differences in the flux between negative and positive velocities in the line are mostly due 
to differences in the atmospheric transmission.}.
The composite $^{13}$CO  and C$^{18}$O lines are single peaked. 
Some asymmetric sub-structures are present in both lines but they are consistent with noise.

{C$^{18}$O emission is, at the 2 $\sigma$ level,
4 km s$^{-1}$ narrower than $^{12}$CO emission.
$^{13}$CO and $2\rightarrow1~^{12}$CO  lines are
1  km s$^{-1}$  narrower than the $^{12}$CO line, at the 1 $\sigma$ level.
To further test whether the $^{12}$CO, $^{13}$CO and C$^{18}$O 
line-profiles are different, 
we ran a two-sample 
Kolmogorov-Smirnov (K-S) test\footnote{KSTWO function in IDL.} on the composite 1D spectra in the $\pm$15 km s$^{-1}$ interval,
after normalization by the line peaks. 
The K-S significance between the $^{12}$CO and C$^{18}$O  line-profiles is 8\%,
between the $^{12}$CO and $^{13}$CO line-profile is 30\%, and 
between the C$^{18}$O and $^{13}$CO line-profile is 97\%.
The K-S test indicates that the $^{12}$CO and C$^{18}$O profiles are different (the C$^{18}$O is narrower),
and that statistically the $^{13}$CO profile resembles the C$^{18}$O profile more than the $^{12}$CO profile.
}

{The 1$\sigma$ average error obtained in the stacked photocenter is 0.06 pixels,
which is equivalent to 5 mas or 0.7 AU at $d=$131 pc (see Fig. \ref{flat_model}). 
We note that different channels have different error bars and the $1\sigma$ error quoted is an average value.
No displacement of the photocenter centroid is detected at the position of the $^{12}$CO lines (the interpretation of this constraint requires modeling and is discussed in the next section).}

The single-nod {\it PSF-FWHM} continuum of HD~139614 (178$\pm$10 mas) and the telluric standard (172$\pm$10) are consistent within the errors, which means that there is no evidence of extended continuum emission at 4.7 $\mu$m.
We measured a stacked continuum {\it PSF-FWHM} of 2.40$\pm$0.05 pixels (1$\sigma$), 
equivalent to  206$\pm$4 mas or {27$\pm$0.6 AU at $d=$ 131 pc (29 AU at 140 pc)} (see Fig. \ref{flat_model}).
The difference of 30 mas ($\sim$1/3 pixel) between the stacked continuum {\it PSF-FWHM} and the single-nod {\it PSF-FWHM}
corresponds to systematic errors introduced during the merging of the 2D nod A and nod B spectra, and to small differences between the {\it PSF-FWHM} at the location of the continuum of the different CO transitions.
The composite {\it PSF-FWHM} at the location of the line appears constant as a function of the wavelength. 
This directly indicates that there is no $^{12}$CO emission extending to spatial scales larger than $\sim$ 30 AU. 
More stringent limits are deduced in the next section.

\begin{table*}
\caption{Types of models used to interpret the observations.}
\begin{center}
\begin{tabular}{|l|l|l|}
\hline
{\bf Model} & {\bf Data modeled}&  {\bf Constraint} \\ 
\hline
1. flat disk with a power-law intensity & composite $^{12}$CO line-profile, photocenter and & $\bullet$ emitting region \\
& {\it PSF$-$FWHM}  & $\bullet$ limits on the gap width\\
\hline
2. 1D LTE slab with single $N_{H}$ and single $T_{\rm gas}$ & $^{12}$CO, $^{13}$CO, and C$^{18}$O rotational diagrams & $\bullet$ average $N_{H}$ and $T_{\rm gas}$ for each isotopolog \\ 
& $^{12}$CO and $\upsilon=2\rightarrow1$ $^{12}$CO rotational diagrams   & $\bullet$ CO excitation mechanism\\
\hline  
3. flat disk with a power-law column density, & $^{12}$CO, $^{13}$CO, C$^{18}$O rotational diagrams & $\bullet$ column density distribution \\
temperature distribution, and LTE excitation     & and $^{12}$CO P(9), $^{13}$CO R(4), and C$^{18}$O R(6) & $\bullet$ temperature distribution\\
&   line-profiles  &  $\bullet$ depth of the gas density drop \\
& &  $\bullet$ limits on the gas gap width and depth \\
\hline
\end{tabular}
\end{center}
\label{table_analysis}
\end{table*}%

\section{Analysis}

We derived constraints on the disk structure from our CRIRES data using models with an increasing complexity.
First,
we deduce the extent of the CO-emitting region from the composite $^{12}$CO spectrum and spectro-astrometric 
signature, using a flat Keplerian disk with a parametric power-law intensity.
Then, we constrain the average column density of the gas and the temperature of each isotopolog 
from the rotational diagrams, using an 1D local thermodynamic equilibrium (LTE) slab model with single temperature and single column density.
Finally, we derive the column density and temperature distribution of the gas as a function of the radius 
from the simultaneous fit of the line-profiles and  rotational diagrams of the three CO isotopologs. 
For this, we use a large grid of 1D flat Keplerian LTE disk models with a power-law temperature and column density distribution.
A summary of our analysis strategy is given in Table~\ref{table_analysis}.

\subsection{Extent of the CO ro-vibrational emitting region}
\label{COemitting_region}
The simplest way to model a line-profile and spectro-astrometric signature and deduce the emitting area
is to assume a flat Keplerian disk with a power-law intensity as a function of the radius:
\begin{equation}
I(R) = I_0 (R/R_{\rm in})^\alpha,
\end{equation}
extending from the an inner radius $R_{\rm in}$ to an outer radius 
$R_{\rm out}$, where $I_0$ is the intensity at $R_{\rm in}$ which is assumed initially to be 1.
The exponent $\alpha$ is obtained for each pair of $R_{\rm in}$ and $R_{\rm out}$ 
such that $I(R_{\rm out}) = 0.01\times I_0$.
In this model, all the physics of the excitation of the line is in the exponent $\alpha$.
The 1\% limit on the intensity was chosen because the line-profile does not change significantly when integrating to a lower percentage.

We modeled the composite $^{12}$CO line-profile, photocenter,  and {\it PSF$-$FWHM}  
with this simple flat Keplerian disk with parametrized intensity.
We provide the details of the model in Sect.~\ref{flat_disk_model} of the Appendix.
The model includes the effect of the disk inclination, 
the effects of the slit width, 
the spectral broadening due to the CRIRES resolution, 
and the spatial resolution during the observations.
In the models, we used a central stellar mass of 1.7 M$_\odot$, 
an inclination $i=20^\circ$, 
a PA = 292$^\circ$
\citep[][]{Matter2016}, and a north$-$south slit orientation.
{Models assume a distance of 140 pc as 
calculations were performed before the recent 
Gaia distance measurement of 131$\pm$5 pc}.

\begin{figure}
\begin{center}
{\large ~~~~~~~~~~~composite $\upsilon=1\rightarrow0~^{12}$CO data} \\
\includegraphics[width=0.4\textwidth]{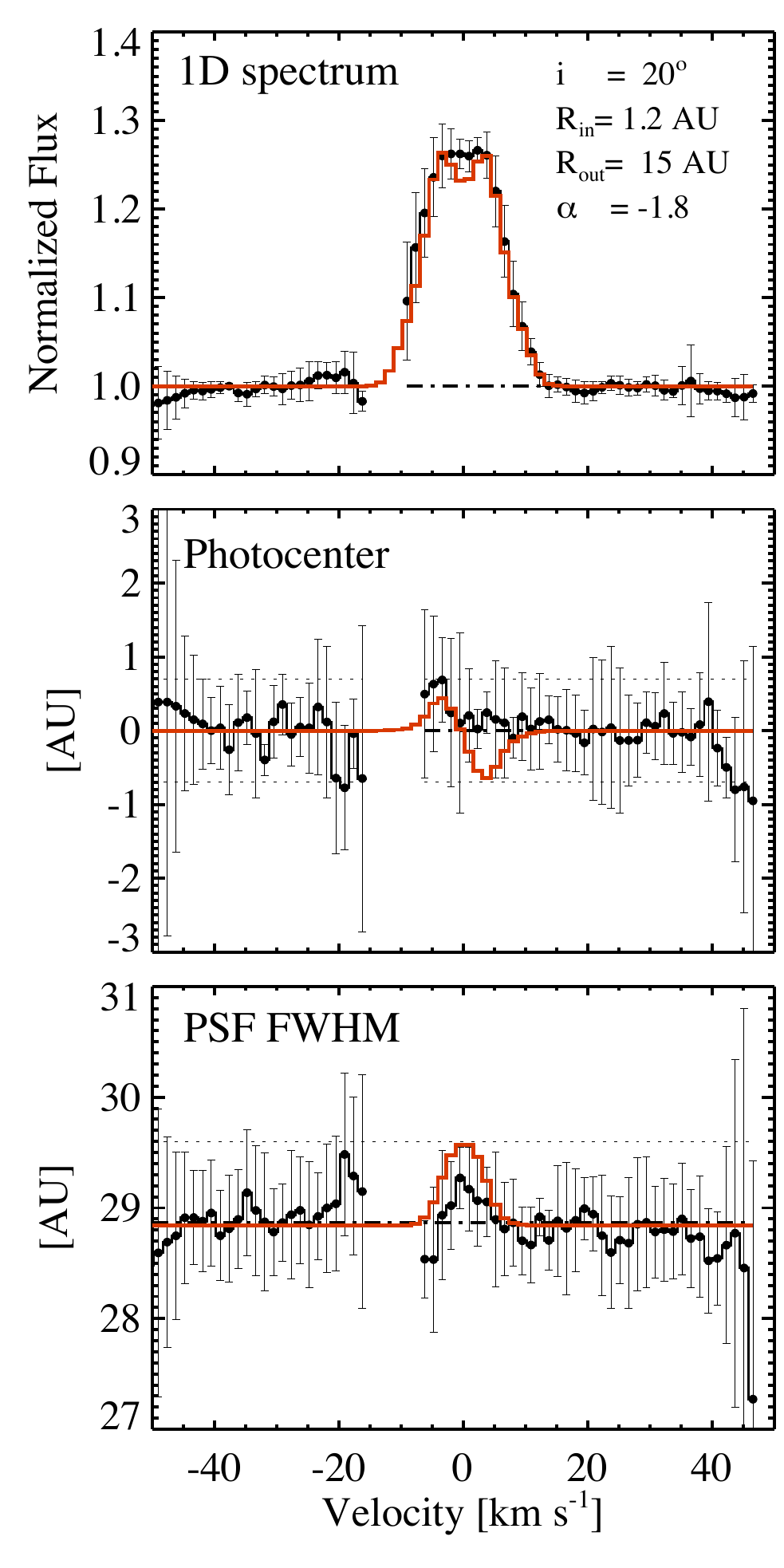} 
\caption{Observed composite of the  normalized $\upsilon=1\rightarrow0~^{12}$CO line-profile, photocenter, and {\it PSF$-$ FWHM}.
In red we show the same quantities produced by a flat disk model with a power-law intensity with $R_{\rm in}$=1.2 AU
and $R_{\rm out}$=15 AU (black cross in Fig.~\ref{chi2_plot}). 
Error bars on the composite line-profile are 1$\sigma$.
Horizontal dotted lines are the
average 1$\sigma$ errors in the +20 to +40 km s$^{-1}$ region.
{Note that this plot assumes a distance of 140 pc for HD~139614. 
Using the recently announced Gaia distance of 131 $\pm$ 5 pc,
the mean of {\it PSF-FWHM} is 27 AU.}
}
\label{flat_model}
\end{center}
\end{figure}

\begin{figure}
\begin{center}
\includegraphics[width=0.45\textwidth]{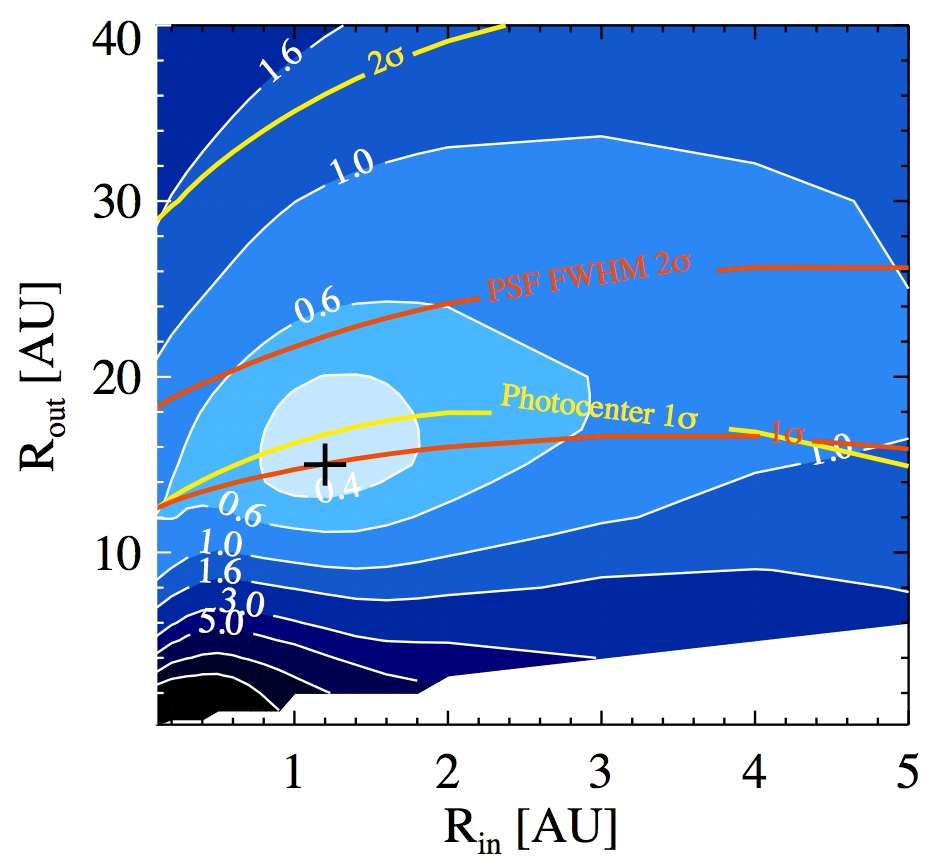} 
\caption{$\chi^2_{\rm red}$ contour plot for the grid of flat disk Keplerian models with a power-law intensity.
The black cross displays the model with the lowest $\chi^2_{\rm red}$ (0.35).
The yellow and red curves show, for each $R_{\rm in}$, the value of $R_{\rm out}$ that would generate a $1\sigma$  
spectro-astrometric signal or a $1\sigma$ {\it PSF-FHWM} broadening, respectively.}
\label{chi2_plot}
\end{center}
\end{figure}

We calculated a grid of disk models varying $R_{\rm in}$ between 0.1 and 70 AU
and $R_{\rm out}$ between 0.2 and 100 AU.
In Fig.~\ref{chi2_plot}
we present the contour plots of the $\chi^2_{\rm red}$ reduced statistic (assuming three free parameters:
$R_{\rm in}$, $R_{\rm out}$, and $I_0$)
of the model of the composite $^{12}$CO line-profile.
Disk models with $0.9<R_{\rm in}<1.8$ AU, and $13<R_{\rm out}<20$ AU gave the best fit to 
the $^{12}$CO composite line-profile. 
The model that displays the smallest $\chi^2_{\rm red}$ has $R_{\rm in}$=1.2 AU and $R_{\rm out}$= 15 AU
and an $\alpha$ exponent of the intensity $-1.8$.

{Although disk models with $R_{\rm out}$ as large as 20 AU provide a good fit to the $^{12}$CO composite line-profile,
star + disk models with a CO-emitting region with $R_{\rm out}> 18$ AU generate a photocenter displacement 
at the position of the CO line that is larger than the 1$\sigma$ limit of the observations (see the yellow curve in Fig.~\ref{chi2_plot}, 
which displays for each $R_{\rm in}$ the value of $R_{\rm out}$ that would generate a displacement  of the photocenter by 1$\sigma$ ). 
In a similar way,  star + disk models with a CO-emitting region with $R_{\rm out}> 15$ AU generate a {\it PSF-FWHM} broadening
at the position of the line that is 1 $\sigma$ larger than the observations (see orange line in Fig.~\ref{chi2_plot},
which displays for each $R_{\rm in}$ the $R_{\rm out}$ that generates a {\it PSF-FWHM} larger than 1$\sigma$ at the line position).
The non-detections of the photocenter displacement and the {\it PSF-FWHM} broadening constrain $R_{\rm out}$ to less than 15 AU.}
The model that best fits the $^{12}$CO line-profile is compatible with the non-detection
of the astrometric signature and PSF broadening at the position of the line (see Fig.~\ref{flat_model}).

{Our simple flat-disk model accurately describes the overall line-profile, the  line width, and the line wings (emission at 5$<$v$<$15 km s$^{-1}$)}.
However, the model appears to slightly underpredict the emission at velocities near zero.
This suggests that a weak emission component at large radii might be present.
Nevertheless, if present, this component does not generate a detectable spectrometric signature.
The zero-velocity component could be an additional emission component from the outer disk at $R\geq6$ AU 
that is not captured in our simple flat-disk model,
or a disk wind emission component as seen in CO ro-vibrational in other protoplanetary disks \citep[e.g.,][]{Pontoppidan2011,
HeinBertelsen2016}.
Although presence of a wind cannot be ruled-out completely,  the symmetry of the line 
(i.e. the lack of an emission shoulder in the blue), the lack of a spectroastrometric
signature, and the very fact that the line-profile is well described by a disk model
suggest that the observed CO emission is consistent with disk emission.

\begin{figure}[t]
\begin{center}
\begin{tabular}{c}
\includegraphics[width=0.35\textwidth]{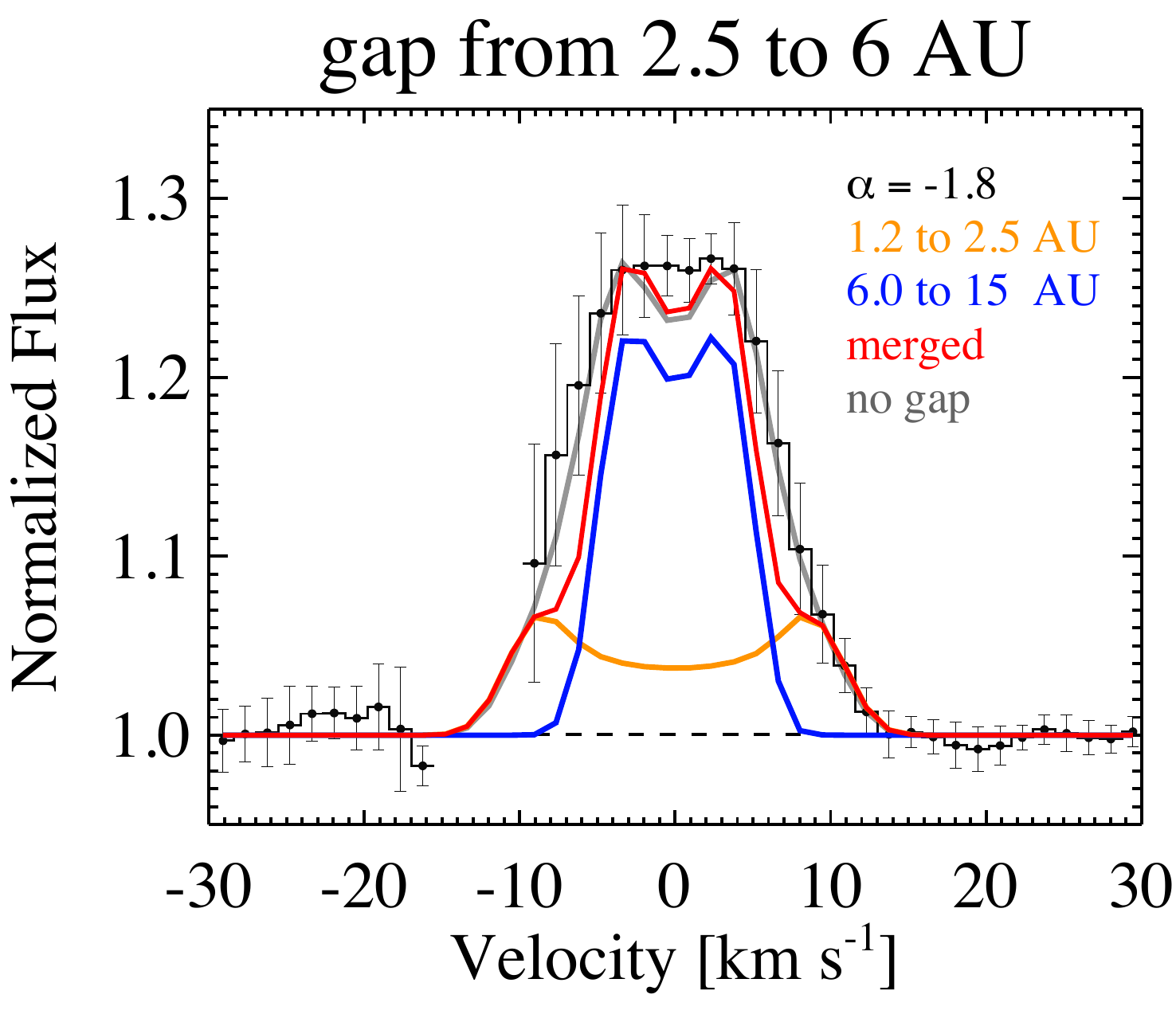} \\
{\bf\large~~~~~~a)} \\[3mm]
\includegraphics[width=0.35\textwidth]{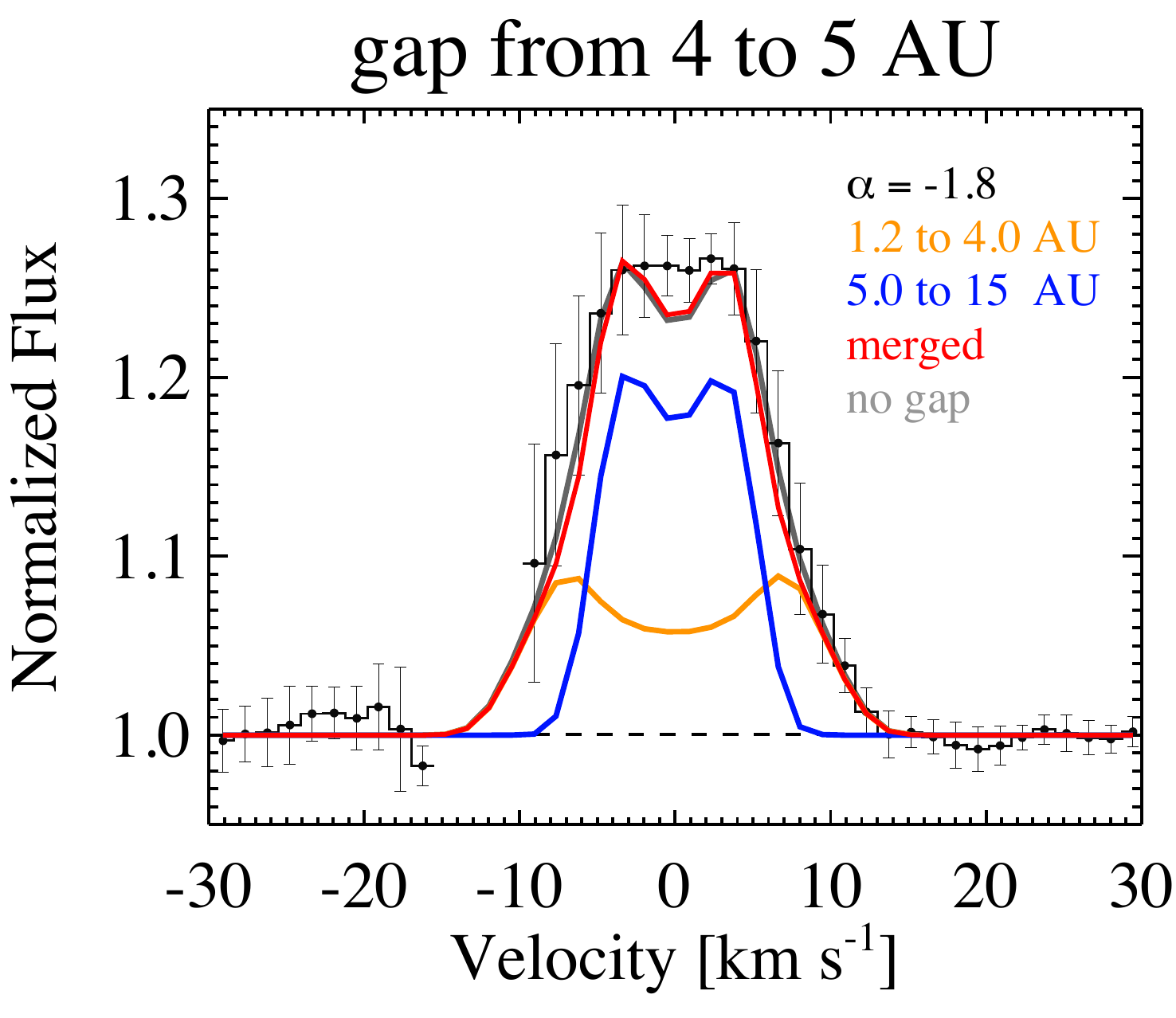} \\
{\bf\large~~~~~~b)} \\[3mm]
\end{tabular}
\caption{Line-profiles predicted for models with a gap in the intensity distribution:
{\bf a)} line-profiles expected for a intensity distribution with a gap devoid of gas from 2.5 AU to 6 AU,
in orange the contribution from 1.2 to 2.5 AU, in blue the contribution from 6 to 15 AU,
and in red the total line-profile;
{\bf b)} similar plot but for a gap devoid of gas of width 1 AU extending from 5 to 6 AU.
Error bars on the spectrum are 1$\sigma$.
}
\label{gap_plot1}
\end{center}
\end{figure}

\subsubsection{The inner radius of the CO emission}
The models that best reproduce the $^{12}$CO composite line-profile have an inner radius around 1 AU.
However, some models with smaller $R_{\rm in}$ are also compatible with the data.
In Fig.~\ref{inner_radius}a,b in the Appendix 
we display the line-profiles expected for disks
with  $R_{\rm in}$ ranging from 0.1 to 1.2 AU.
In panel (a) we show the results of the models with $R_{\rm out}$ fixed to 15 AU ($\alpha$ adjusted such that $I(R_{\rm out}) = 0.01\times I(R_{\rm in}$)).
In panel (b) the line-profiles with $\alpha$ fixed to $-$1.8 ($R_{\rm out}$ set such that $I(R_{\rm out}) = 0.01\times I(R_{\rm in}$)).
Depending on the value of $\alpha$, models with $R_{\rm in}$ as low as 0.3 AU can be compatible 
with the observed $^{12}$CO composite line-profile.

In all our power-law intensity models, 
we have assumed a sharp inner edge, thus an abrupt increase in the intensity from zero to $I_0$ at R$_{\rm in}$.
If instead we assume a soft inner edge, thus a smooth increase of the intensity from R$_{\rm in}$ up to the radius 
of the maximum intensity R$_{\rm I_{\rm max}}$ = 1.2 AU,
then R$_{\rm in}$ can be as small as 0.01 AU and the line
profile would still be compatible with the data (see Fig.~\ref{inner_radius}c in the Appendix).
The R$_{\rm in}$ constraint from a power-law intensity model with a sharp inner edge corresponds to the 
radius of the maximum intensity.
CO gas can still be present farther in if the inner edge is soft.
In Sect. \ref{NH_upper_limits_1au} we provide upper limits to the gas column density at $R<$ 1 AU based
on the $^{12}$CO line-profile shape.

\begin{figure}[t]
\begin{center}
\includegraphics[width=0.5\textwidth]{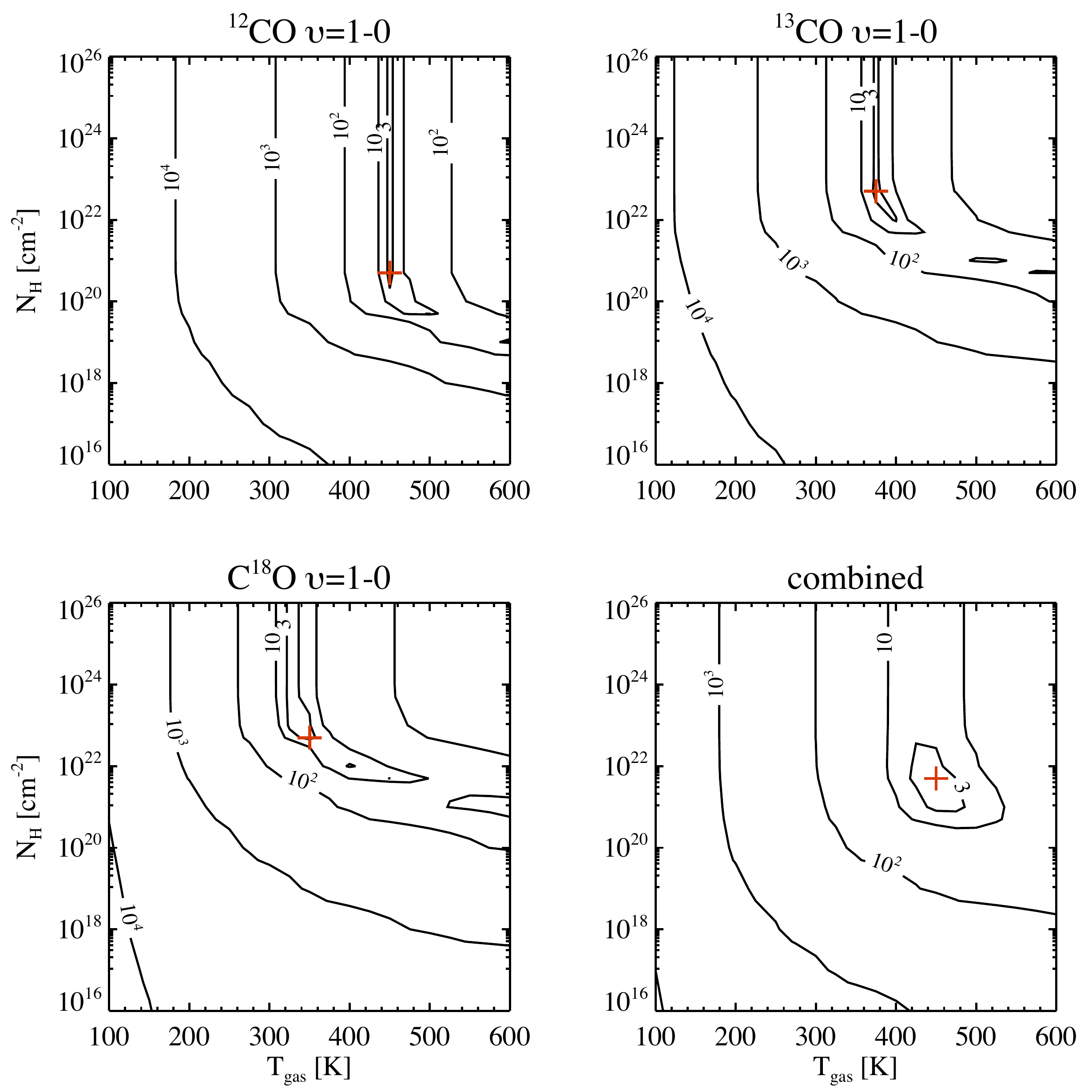}
\caption{ 
$\chi^2_{\rm red}$ contour plot of the modeled rotational diagrams for $^{12}$CO, $^{13}$CO, and C$^{18}$O using a single
temperature and surface density slab model.
The  combined $\chi^2_{\rm red}$ contour plot is obtained by  simultaneously using all the data of the three isotopologs.
The cross indicates the location of the  $\chi^2_{\rm red}$ minimum in each panel. 
The numbers inside the contour indicate $n$-times the value of the minimum  $\chi^2_{\rm red}$. 
}
\label{contour_plots_slab}
\end{center}
\end{figure}

\begin{figure}
\begin{center}
\includegraphics[width=0.45\textwidth]{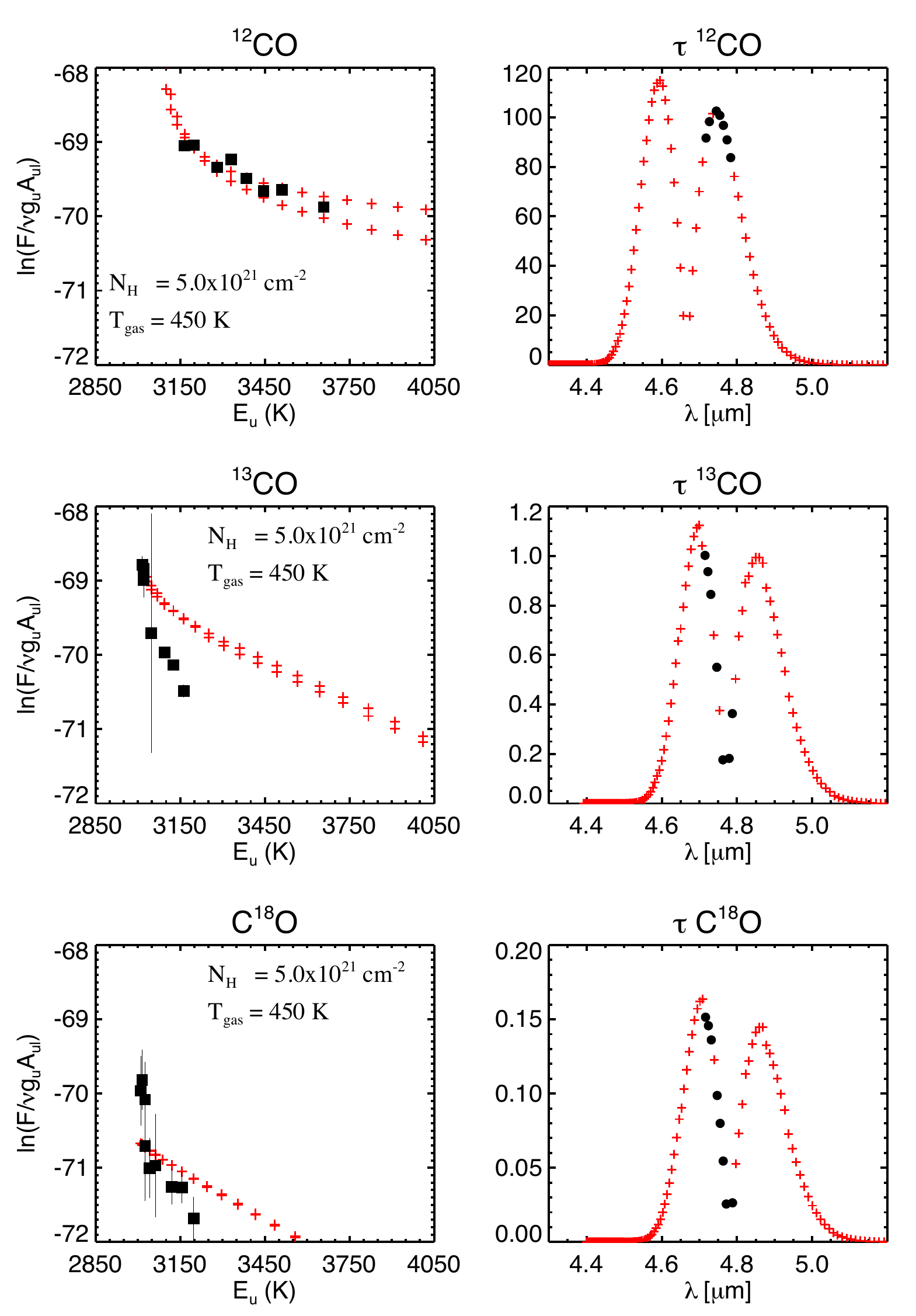}
\caption{Rotational diagrams and optical depths of the slab model ($N_{H} =5  \times10^{21}$ cm$^{-2}$, $T_{\rm gas} = 450$ K) 
with the lowest $\chi^2_{\rm red}$ combining all the  $^{12}$CO, $^{13}$CO, and C$^{18}$O data (cross in Fig.~\ref{contour_plots_slab}). 
In the left panels observations are shown in black and models in red. 
In the right panels the optical depths of the observed transitions are shown in black, other transitions are plotted in red.
A single-temperature and single-density slab does not correctly describe the rotational diagrams of the
three CO isotopologs simultaneously. Error bars on the rotational diagram are 1$\sigma$.
}
\label{combined_rotational}
\end{center}
\end{figure}

\subsubsection{ A continuous or a gapped gas distribution? }
\label{gap_section}
The $^{12}$CO line-profile data is well described by a continuous and smooth intensity profile 
from 1.2 AU up to 15 AU.
As mentioned in the introduction, 
\citet[][]{Matter2014, Matter2016}  resolved a gap in the dust from 2.5 AU to 6 AU based on near- 
and mid-IR VLT interferometric observations.
This raises the question whether the $^{12}$CO  line-profile could  be described by an
intensity distribution with a gap.

The $^{12}$CO line-profile clearly indicates that there is emission at $R<6$ AU,
otherwise the line-profile would have been much narrower (see the blue line in panel a of Fig.~\ref{gap_plot1}).
{As a consequence, an inner gas hole of 6 AU radius is ruled out.}
Furthermore, the line-profile rules out a CO-emitting region 
confined to a narrow ring between 1.2 and 2.5 AU, otherwise the line would have been much broader 
(see the yellow curve in panel a of Fig.~\ref{gap_plot1}).

We have tested the scenario in which the intensity distribution of the best solution of the power-law intensity model 
has a gap (i.e., no emission) between 2.5 AU and 6 AU. 
In this case (Fig.~\ref{gap_plot1}a), 
the velocity channels between 3 and 8 km s$^{-1}$ are not well reproduced.
Such a large gap of  3.5 AU is not compatible with the observed $^{12}$CO line.
If the gap in the gas is smaller than 2 AU, 
then the line-profile could be consistent with the observations
(Fig.~\ref{gap_plot1}b). Given the CRIRES resolution, a small gap of 1$-$2 AU in the intensity would not be detectable.
{In Sect.~\ref{Gap_section} we derive constraints on the CO column density inside a potential gap}.

\subsection{Average temperature and column density}
\label{slab_section}
The detection of ro-vibrational emission of the CO isotopologs C$^{18}$O and C$^{17}$O 
indicates that the emitting medium must be dense and warm.
To constrain the average temperature and column density of the gas probed 
in the CRIRES spectra,
we modeled the $^{12}$CO, $^{13}$CO and C$^{18}$O line fluxes
using a simple semi-infinite slab model in LTE with a single gas temperature and column density.
{We did not model the C$^{17}$O observations because only two line fluxes are available and no reliable
estimate of the temperature can be derived}.
We wrote a CO LTE slab model using 
the frequencies, energy levels, and Einstein coefficients from \citet[][]{Chandra1996}.

We generated a grid of LTE slab models by varying the hydrogen-nuclei column density $N_{H}$ from 10$^{18}$ to 10$^{25}$ cm$^{-2}$ with steps of 0.25 dex, and the gas temperature $T_\mathrm{g}$ from 100 to 1000 K with steps of 25 K.
We assumed a turbulent line broadening of 0.1 km s$^{-1}$.
We used 
a $^{12}$CO abundance of $10^{-4}$ {($N_{^{12}{\rm CO}}=1.0\times10^{-4} N_H$)},
a $^{12}$CO/$^{13}$CO ratio of 100
and a $^{12}$CO/C$^{18}$O  ratio of 690 following \citet[][]{Smith2009}\footnote{The 
$^{12}$CO/$^{13}$CO and $^{12}$CO/C$^{18}$O ratios of \citet[][]{Smith2009} were deduced 
from high-resolution spectroscopy of CO ro-vibrational lines detected in absorption 
toward VV CrA, a binary T Tauri star in the Corona Australis molecular cloud.
The $^{12}$C/$^{13}$C from Smith et al. is nearly twice the expected interstellar medium (ISM) ratio,
and the $^{12}$CO/C$^{18}$O ratio is $\times$1.4 the ISM ratio. 
We used the  $^{12}$CO/$^{13}$CO and $^{12}$CO/C$^{18}$O ratios of Smith et al. 
instead of the ISM ratios because we consider that they are more representative of the isotopologs ratios 
that would be expected in the inner disk of HD~139614.}.

The output of a slab model is in units of line flux per steradian and the optical depth of each transition. 
To compare slab calculations with the observed line fluxes, 
an {average} solid angle of the emitting region needs to be prescribed.
A model with a single temperature and a single column density model is equivalent to assuming 
a disk model with constant intensity with radius (i.e., $\alpha=0$).
We tested Keplerian disk models with a constant intensity and found that the
line-profile, photocenter and {\it PSF$-$FWHM} could be reproduced by a flat disk
of $R_{\rm in}$=0, $R_{\rm out}$=6.5 AU.
We therefore used an average emitting region radius of 6.5 AU and a distance of 140 pc\footnote{Models were calculated before the recent Gaia distance measurement of 131$\pm$5 pc,
the difference in distance of 5$-$9 pc does not change the conclusions reached.}  
to determine the solid angle for the 1D slab model\footnote{We note that the R$_{\rm out}$ of 15 AU found in Sect.~\ref{COemitting_region}
cannot be used here because this emitting region was found with a decreasing power-law intensity, which is equivalent 
to assuming a radial decreasing temperature and column density distribution. The 1D slab model has a single temperature and column density.}.

For each $N_H$ and $T_{\rm gas}$ slab model, a rotational diagram for each CO isotopolog was calculated.
We used the $X$ and $Y$ coordinates $Y=~$ln$(F_{ul}/\nu g_u A_{ul})$ and $X = E_u$.
Here $F_{ul}$ is the line flux of the transition between the upper level $u$ and the lower level $l$, $g_u$ is the degeneracy of the upper level ($2J +1$),
$A_{ul}$ is the Einstein coefficient of transition and $E_u$ the upper energy level of the transition.
For the rotational diagram of each isotopolog we calculated the statistical quantity
$\chi^2_{\rm red}=\frac{1}{N-4}\sum\limits_{i}^{}(Y_{{\rm slab}~i} - Y_{{\rm obs}~i})^2/\sigma^2_{Y_{{\rm obs}~i}}$.
Here  $\sigma_{Y_{{\rm obs}~i}}$ is the difference between $Y$ calculated using the observed line flux 
and $Y$ calculated using the observed line flux plus $3\sigma$.
The $N-4$ corresponds to three degrees of freedom ($N_H$, $T_{\rm gas}$, $\Omega$), 
and $N$ the number of data points (8 for  $^{12}$CO, 7 for $^{13}$CO and 9 for C$^{18}$O).
In Fig.~\ref{contour_plots_slab}
we display the $\chi^2_{\rm red}$ contour plots for each CO isotopolog
and one $\chi^2_{\rm red}$ combining the data of the three isotopologs.

We find that the best fit to the rotational diagram is different for each CO isotopolog. 
$^{12}$CO is best described with $T_{\rm gas} \sim 450$ K  and $N_H$ $>10^{20} $ cm$^{-2}$.
$^{13}$CO  is best reproduced by lower temperatures ($T_{\rm gas} \sim 380$ K) and $N_H$ column densities of at least $5 \times 10^{22} $ cm$^{-2}$.
The C$^{18}$O is best fit by even lower temperatures $\sim 350$ K and $N_H$ column densities higher than $5 \times 10^{22} $ cm$^{-2}$.
The errors on T$_{\rm gas}$ are 10$-$20 K.
The colder temperatures for $^{13}$CO and C$^{18}$O emission indicate that they 
are produced at larger radii than the $^{12}$CO emission or at lower vertical scale heights.

The best-fit combined model for the rotational diagrams of the three isotopologs emission is 
a model with $N_H = 5 \times10^{21}$ cm$^{-2}$ and $T_{\rm gas} = 450$ K.
In this model the $^{12}$CO emission is optically thick and the $^{13}$CO and C$^{18}$O optically thin (Fig.~\ref{combined_rotational}).
The combined solution only satisfactorily describes the $^{12}$CO rotational diagram, however.
The slopes of the $^{13}$CO and C$^{18}$O rotational diagrams are not well reproduced.
The curvature on the rotational diagrams suggests that there  $^{13}$CO and C$^{18}$O emissions are 
optically thick or that there is a gradient in temperature.
This is also suggested by the $\chi^2_{\rm red}$ contours, 
because solutions are degenerate 
with respect to $N_H$, giving only lower limits for the column density.

\begin{figure}[t]
\begin{center}
\includegraphics[width=0.42\textwidth]{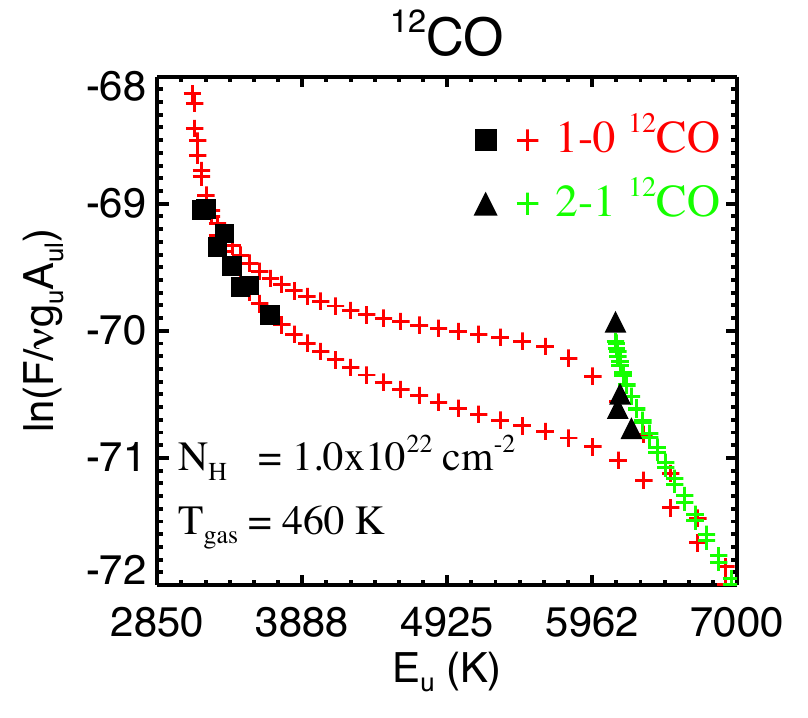}
\caption{Rotational diagrams of the $\upsilon=1\rightarrow0~^{12}$CO (squares) and $\upsilon=2\rightarrow1~^{12}$CO (triangles) observed emission.
Overplotted is a slab model with T$_{\rm rot}$=T$_{\rm vib}$=460 K and N$_H$ = $10^{22}$ cm$^{-2}$,
in red for $\upsilon=1\rightarrow0~^{12}$CO emission and in green for $\upsilon=2\rightarrow1~^{12}$CO emission.}
\label{2-1rotational_diagram}
\end{center}
\end{figure}

\begin{figure*}
\begin{center}
\includegraphics[width=0.8\textwidth]{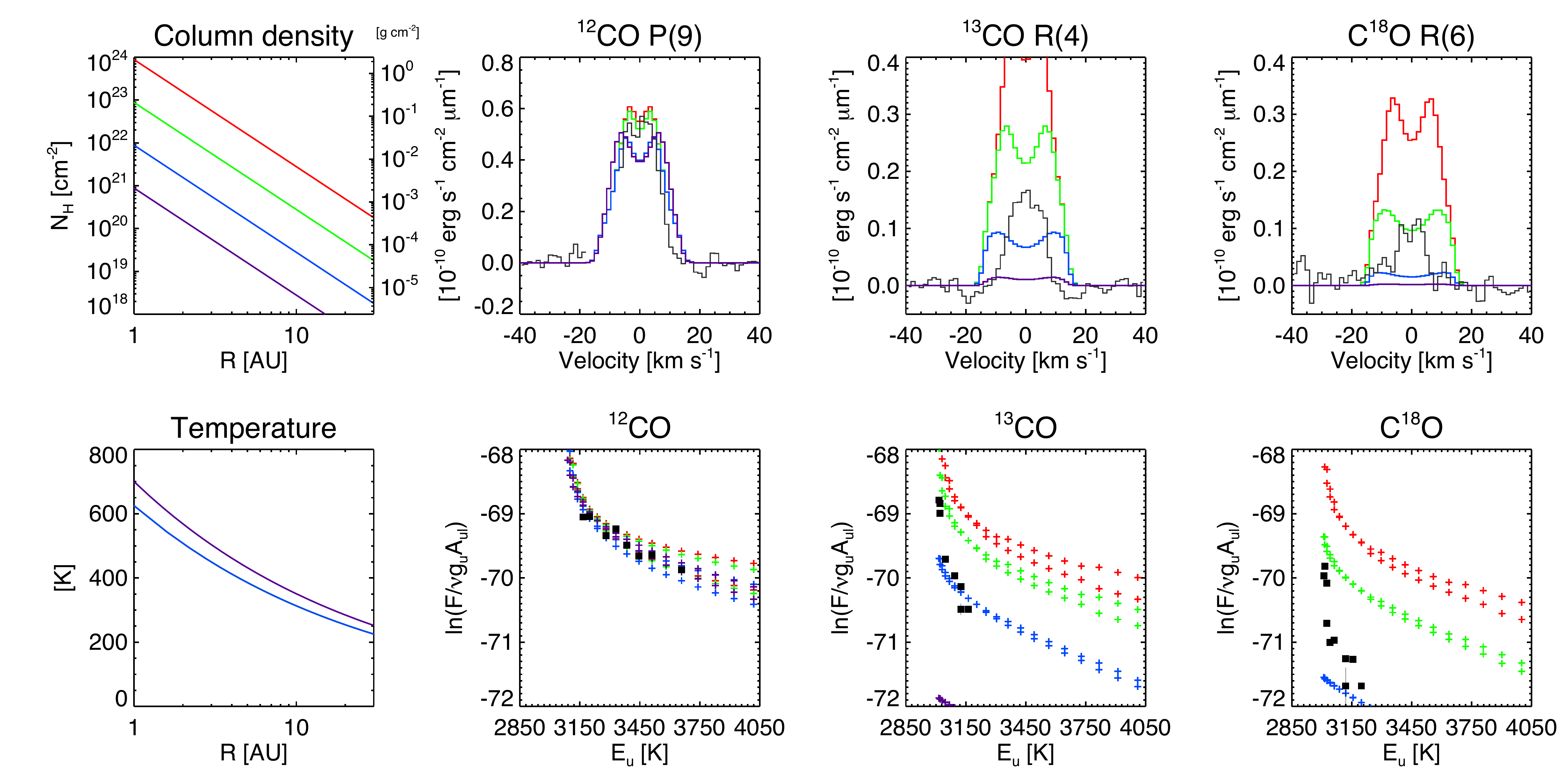}
\caption{Examples of the predicted $^{13}$CO R(4) and C$^{18}$O R(6) emission for a continuous power-law distribution 
of temperature and column density that describe the $^{12}$CO P(9) line-profile and the $^{12}$CO rotational diagram.
Colors in the spectra and rotational diagrams correspond to the different column density distributions in the first panel. 
A disk model with a single power-law surface density cannot simultaneously reproduce the $^{12}$CO, $^{13}$CO, 
and C$^{18}$O line-profiles and rotational diagrams.}
\label{Simple_power_law}
\end{center}
\end{figure*}

\subsubsection{Thermally excited or UV-pumped emission?}
The detection of $\upsilon=2\rightarrow1\,^{12}$CO emission raises the question whether the observed 
CO ro-vibrational emission is thermally excited or if it is due to UV-pumping as observed in some other Herbig Ae/Be stars \citep[e.g.,][]{Brittain2007,vanderPlas2015}.
The $\upsilon=2\rightarrow1\,^{12}$CO / $\upsilon=1\rightarrow0\,^{12}$CO average line ratio of $\sim$0.2 
is lower than in Herbig Ae/Be stars with flared disks, 
but it is at the higher end of the T Tauri sample with single-component CO ro-vibrational emission \citep[see Table 3 in ][]{BanzattiPontoppidan2015}.
The non-detection $\upsilon=3\rightarrow2\,^{12}$CO emission and the upper limit of 0.04 on the $\upsilon=3\rightarrow2\,^{12}$CO /$\upsilon= 1\rightarrow0\,^{12}$CO line
ratio indicate that the UV-pumping, if present, is not as strong as in other Herbig Ae/Be stars with CO fluorescent emission.
For example, 
in the case of HD 100546 \citet[][]{vanderPlas2015} measured an average $\upsilon=3\rightarrow2\,^{12}$CO / $\upsilon=1\rightarrow0\,^{12}$CO line ratio
of 0.3.
The A7V spectral type of HD~139614 is later than most of the Herbig Ae/Be stars studied in  \citet[][]{Brittain2007} and \citet{vanderPlas2015}.
This might in part explain the weaker effect of UV-pumping in HD~139614 with respect to other Herbig Ae/Be stars studied previously.

We explored models around the best solution of the combined fit to the 
$\upsilon=1\rightarrow0\,^{12}$CO, $^{13}$CO, and C$^{18}$O rotational diagrams
and calculated the LTE $\upsilon=2\rightarrow1\,^{12}$CO and $\upsilon=3\rightarrow2\,^{12}$CO emission.
We found that the observed $\upsilon=2\rightarrow1^{12}$CO emission and the non-detection of the $\upsilon=3\rightarrow2\,^{12}$CO emission can be well 
described by an LTE model, with $T_{\rm gas}$= 460 K and $N_H=10^{22}$ cm$^{-2}$, 
in which the rotational temperature is equal to the vibrational temperature\footnote{ 
The model with $T_{\rm gas}$= 450 K  and $N_H=10^{22}$ cm$^{-2}$ of Fig.~\ref{combined_rotational} 
generated slightly weaker $\upsilon=2\rightarrow1\,^{12}$CO emission.}.
We show this model in Fig.~\ref{2-1rotational_diagram}, where we display the rotational diagrams
of $\upsilon=1\rightarrow0\,^{12}$CO and $\upsilon=2\rightarrow1\,^{12}$CO emission of the model and the observations. 
This model generates $\upsilon=3\rightarrow2\,^{12}$CO emission lines with integrated fluxes on the order of $10^{-18}-10^{-17}$ 
erg s$^{-1}$ cm$^{-2}$, which is consistent with the non-detection of these lines in our CRIRES spectra.
{We conclude that the observed CO emission is most likely thermally excited}.

\subsection{Deriving the surface density and temperature distribution}

{The different temperatures and column densities of each CO isotopolog 
and the difference on line widths between the CO isopotologs,
suggest that the emission of each isotopolog is produced at different radial distances and/or vertical heights}.  
The narrower and colder $^{13}$CO and C$^{18}$O lines pose an interesting puzzle for the interpretation of the data.
If the CO ro-vibrational emission is modeled with a 1D power-law column density and temperature distribution 
($N_H \propto R ^{\,\alpha_{\rm N_H}}$ and $T_{\rm gas} \propto R^{\,\alpha_{T_{\rm gas}}}$) 
and both distributions have no discontinuities (i.e. no gaps nor density drops or jumps in the temperature),
and if the $^{12}$CO/$^{13}$CO  and $^{12}$CO/C$^{18}$O abundance ratios are constant,
then, a disk model able to reproduce the $^{12}$CO line-profile and $^{12}$CO rotational diagram 
would generate $^{13}$CO and C$^{18}$O lines that are too broad, too strong, and too warm to be consistent 
with the observations (see Fig.~\ref{Simple_power_law}). In the next sub-section we provide the details of the model.
Since $^{13}$CO and C$^{18}$O emission  is optically thin up to relatively high column densities 
($N_H\sim10^{22}$ cm$^{-2}$ and $N_H\sim10^{23}$  cm$^{-2}$ respectively), 
the 1D single power-law models generate $^{13}$CO and C$^{18}$O lines that are too broad, 
too strong, and too warm because the column of gas at small radii is too large.

\begin{figure*}
\begin{center}
\includegraphics[width=0.7\textwidth,angle=0]{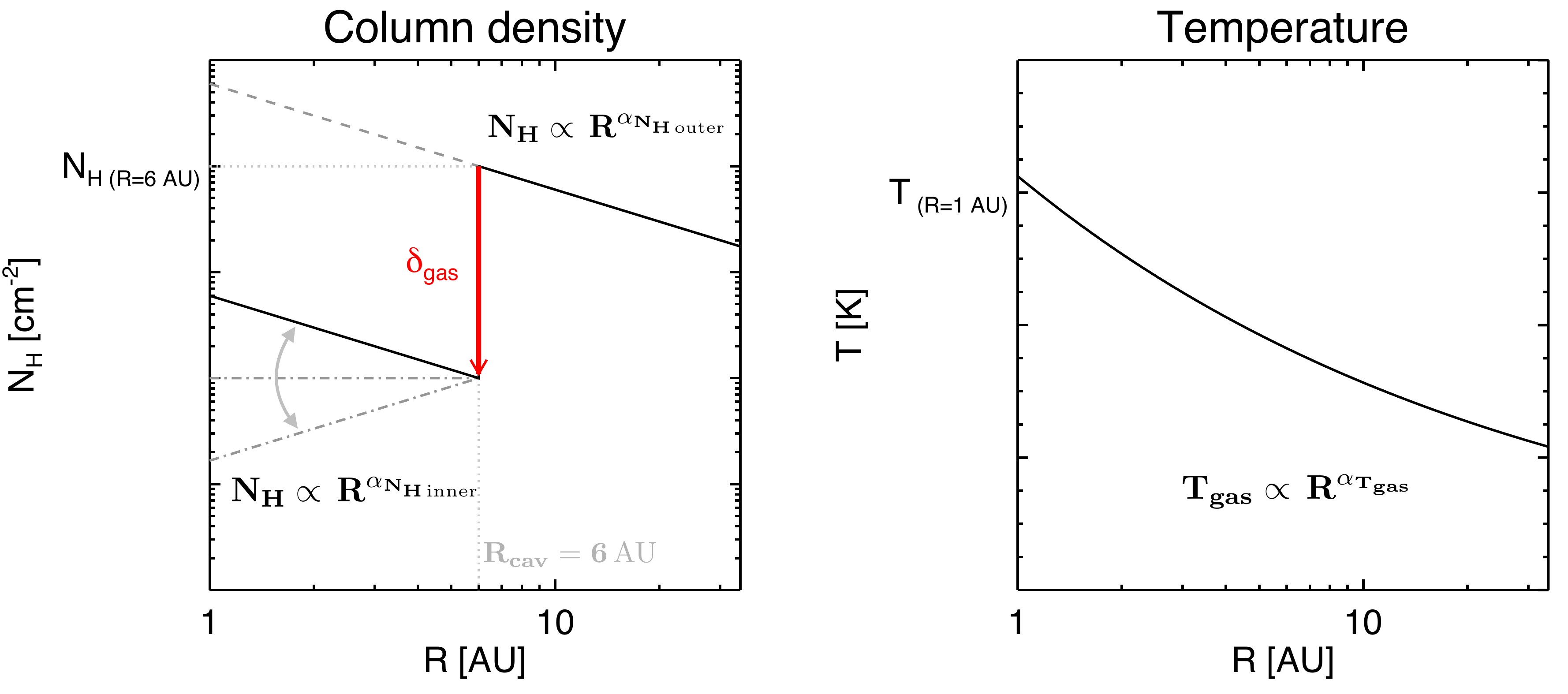}
\caption{Schematic illustration of the free parameters in the flat Keplerian disk model with a power-law column density and temperature.
The grid includes models with with decreasing ($\alpha_{inner}<0$), 
flat ($\alpha_{inner}=0$) and increasing ($\alpha_{inner}>0$) surface density distributions
in the inner 6 AU.  The anchor point for the surface density is at 6 AU and for the temperature is at 1 AU. 
In all the models the cavity radius is at 6 AU.
}
\label{cartoon}
\end{center}
\end{figure*}

\subsubsection{Power-law temperature and column density Keplerian disk model with a depleted inner region}

Near- and mid-IR interferometry and the SED \citep[][]{Matter2016}
indicate a depletion of dust mass on the order of  10$^{-4}$  in the inner 6 AU 
with respect to the extrapolated surface density at $R>$  6 AU. 
If a similar behavior were also be followed by the gas,
the narrower and colder C$^{18}$O and  $^{13}$CO  profiles could be explained as the result of
a gas density drop in the inner 6 AU of the disk.
A gas density drop can cause the C$^{18}$O and  $^{13}$CO lines to be optically thin
in the inner 6 AU and cause their emission be dominated 
by the contribution at R$>$6 AU where the density is higher and the gas colder.

To test the hypothesis of the gas density drop,
we modeled the observed CO ro-vibrational emission with a flat disk in Keplerian rotation,
in which the gas column density and temperature are described by a power-law distribution:
\begin{equation}
N_H\,(R\geq6~{\rm AU}) = ~N_{\rm H\,(R=~6\,{\rm AU})}~\left(\frac{{R}}{6~{\rm AU}}\right)^{\,\alpha_{N_{\rm H~outer}}} 
\end{equation}
\begin{equation}
T_{\rm gas}\,(R\geq1~{\rm AU}) = T_{\rm 0~(R=~1\,{\rm AU})}~\left(\frac{{R}}{1~{\rm AU}}\right)^{\,\alpha_{T_{\rm gas}}}.
\end{equation}

\noindent We allowed the models to have a reduced surface density at $R<6$ AU by a factor $\delta_{\rm gas}$ 
\begin{equation}
N_H\,(R < 6 ~{\rm AU}) = ~\delta_{\rm gas} \cdot N_{\rm H\,(R=~6\,{\rm AU)}}~\left(\frac{{R}}{6~{\rm AU}}\right)^{\,\alpha_{N_{H\,inner}}} .
\end{equation}

A reference radius of 6 AU was set for the gas column density to compare the gas distribution with that of the dust.
A reference radius of 1 AU was selected for the temperature, given that the modeling of the $^{12}$CO line with a power-law intensity
suggested that 1 AU is the radius of the maximum intensity.

We chose $\alpha_{N_{H\,inner}}$ to range from $-$2.5 up to +3 to cover a wide range of possible surface density distributions
in the inner disk\footnote{Previous models of CO ro-vibrational emission in transition disks have assumed a flat surface density \citep[][]{Pontoppidan2008,Pontoppidan2011} 
or a surface density decreasing with a $-3/2$ exponent \citep[][]{Salyk2009}. 
\citet{Carmona2014} found an increasing surface density profile for the gas with 
exponent +0.2 in the inner disk of the transition disk HD~135344B.
\citet[][]{Andrews2011}, \citet[][]{Bruderer2014}, and \citet[][]{vanderMarel2015} modeled SMA or ALMA observations of transition disks 
assuming the same surface density exponent of $-1$ for the inner and the outer disk.  
\citet[][]{Matter2016} found that the surface density of the dust in the inner 2.5 AU of HD~139614 increases 
radially with an exponent +0.6.}.
As \citet[][]{Matter2016} found a dust depletion factor of 10$^{-4}$,  
we tested models with $\delta_{\rm gas}$ ranging from 1 to 10$^{-4}$.

The choice of modeling the temperature as a power-law instead of using a full radiative transfer calculation \citep[e.g.,][]{Woitke2009,Thi2013,Carmona2014,Bruderer2013, Bruderer2014} 
enables us to describe the CO temperature in the emitting region independently of that of the 
dust with a minimum number of free parameters.
This permitted us to explore a large portion of the parameter space.

The disk was modeled with a flat geometry using a radial and azimuthal grid.
The model is analogous to the model described in Sect.~\ref{COemitting_region} and Sect.~\ref{flat_disk_model} in the Appendix,
with the difference that the intensity at each radius was calculated using the local $T_{\rm gas}$ and $N_H$ 
using the CO slab model previously described,
\begin{equation}
I(R) = I (T_{\rm gas},N_H)_{\rm slab}.
\end{equation}
The local broadening of the line is the convolution of the turbulent broadening (0.1 km s$^{-1}$),
the local thermal broadening and the spectral resolution\footnote{Including the instrument resolution in the convolution 
kernel of the thermal and turbulent broadening enable saving hundreds of convolutions per model, 10$^7$ convolutions
in the grid, and it is equivalent to convolving the final data-cube by the instrument resolution.}. 
We recall that the CO slab model assumes LTE excitation, 
which is a good approximation because CO emission is most likely thermally excited.
We set the outer radius equal to 30 AU. Observations set an upper limit of 15 AU to the emitting region,
but, we used a larger outer radius to permit some combinations of $N_H (R)$  and $T_{\rm gas}(R)$ inside the grid
to have a sufficiently large radial extent to let the intensity decrease to low levels.  
A model in which the radial calculation grid ends artificially early would generate a line-profile and a line flux
that does not correctly represent the selected $N_H (R)$  and $T_{\rm gas}(R)$.
In the flat-disk parametric intensity models that describe the CO emission in HD~139614,
we saw no significant change in the line-profiles with or without slit (lines are dominated by the contribution in the inner 15 AU).
Therefore, we did not include the slit effects in our model to enable the calculation of a large number of models.

\begin{table}[b]
\caption{Parameter space of the power-law $N_H$ and $T_{\rm gas}$ models.}
\begin{center}
{
\begin{tabular}{llll}
\hline
\hline
\\[0.1mm]
Parameter & Units & Values \\[1mm]
\hline
\\[1mm]
$N_{H\,~(R=~6\,{\rm AU})}$ &[cm$^{-2}$]  & 10$^{20}$, 10$^{21}$, 10$^{22}$, 10$^{23}$, 10$^{24}$, 10$^{25}$  \\[3mm]
$\alpha_{N_{H\,inner}~(R<~6\,{\rm AU)}}$ & & $-2.5, -2.0, -1.5, -1.0, -0.5,$ \\
&& $ ~~0.0,~0.5,~1.0,~1.5,~2.0,~2.5,~3.0$ \\[3mm]
$\alpha_{N_{H\,outer}~(R\geqslant~6\,{\rm AU})}$  & &  $-2.5,-2.0,-1.5,-1.0,-0.5$ \\[3mm]
$\delta_{gas ~(R=~6\,{\rm AU})}$  && $1,~10^{-1},~10^{-2},~10^{-3},~10^{-4}$ \\[3mm]
$T_{0~(R=~1\,{\rm AU})} $ &[K] & 550, 575, 600, 625, 650, 675, 700, \\
                                          &      & 725, 750\\[3mm]
$\alpha_{T_{\rm gas}}$ & & $-0.40,-0.35,-0.30,-0.25,-0.20$ \\[3mm]
{\bf Total} && 81000 models\\

\hline
\end{tabular}
}
\end{center}
\label{table_grid}
\end{table}

\begin{table*}
\caption{Parameters of the best-fit grid model}
\begin{center}
\begin{tabular}{lccccccccccc}
\hline
Model & $N_{\rm H\,(R=6 {\rm AU})}$ & $\alpha_{N_{\rm H\,inner}}$ & $\alpha_{N_{H\,outer}}$ & $\delta_{\rm gas}$ & $T_{(R=1\,{\rm AU})}$ & $\alpha_{T_{\rm gas}}$ \\ 
           & [cm$^{-2}$] & &&&[K] \\ 
\hline
best-fit grid model                                                          & 10$^{23}$ & +2.0 & $-2.5$ &10$^{-2}$ & 675 & $-0.35$ \\
best-fit grid model if $\alpha_{N_{\rm H\,inner}}\leq +1.0 $   & 10$^{23}$ &  ~0.0 & $-2.0$ &10$^{-2}$ &650 & $-0.35$ \\
\hline
\end{tabular}
\end{center}
\label{table_models}
\end{table*}%

\begin{figure*}
\begin{center}
\includegraphics[width=\textwidth,angle=0]{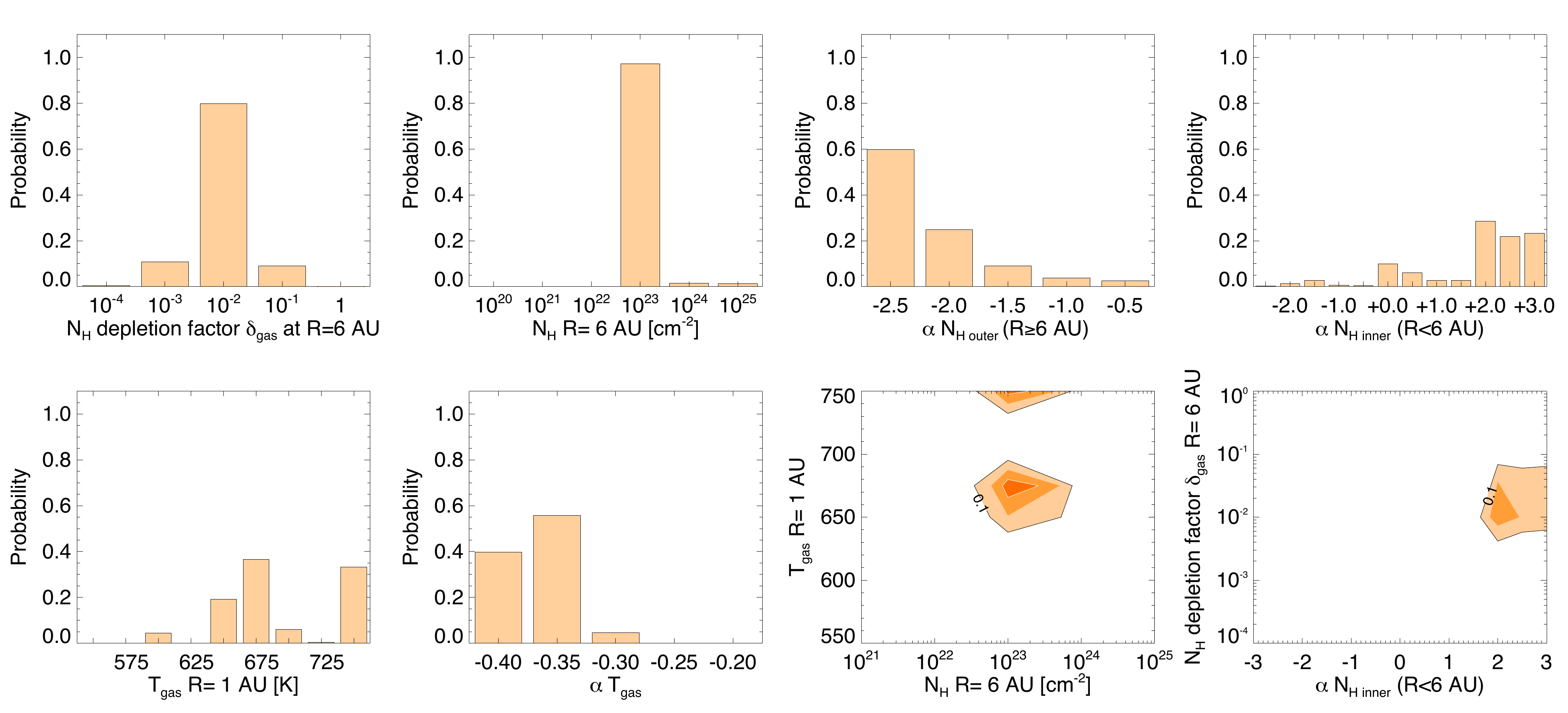}
\caption{Bayesian probability distributions of the grid of the $N_H$ and $T_{\rm gas}$ power-law models calculated (81 000 models). 
The $\chi^2$ statistic used to calculate the probability considers the rotational diagrams and line-profiles of the three isotopologs
simultaneously.
}
\label{bayesian_plots}
\end{center}
\end{figure*}

A model has six free parameters: $N_{H\,(R=~6\,{\rm AU})},  \alpha_{N_{H\,inner}}, \alpha_{N_{H\,outer}},$ 
$ \delta_{\rm gas}, T_{(R=1\,{\rm AU})}$ and $\alpha_{T_{\rm gas}}$. 
Each model produces integrated line-fluxes and rotational diagrams for the three CO isotopologs and 
synthetic line-profiles at the CRIRES resolution for the $^{12}$CO P(9) line at 4745.13 nm , 
$^{13}$CO R(4) at 4730.47 nm,
and the C$^{18}$O R(6) line at 4724.03 nm.
These three CO transitions were selected because their line-profiles have a high S/N and are less affected by telluric absorption.
The merged composite line-profile was not used 
because  the model line predictions needed to be compared with a line-profile in flux units.
In the merged composite spectra the flux information is lost.

To find the models that best describe the observations,
we ran a uniform grid of 81000 models covering the parameter space described in Table~\ref{table_grid}.
To calculate the most probable values of the parameters in the grid, 
we used a Bayesian approach \citep[see for example][and references therein]{Pinte2007}.
We provide details of the calculation of the Bayesian probabilities in Sect.~\ref{Bayesian} in the Appendix.

\subsubsection{Grid results}

Figure~\ref{bayesian_plots} displays the Bayesian probability distribution diagrams of the grid.
The first panel in the upper left displays the probability of models with and without a gas density drop.
The diagram clearly shows that
a gas column density drop in the inner 6 AU of the disk is required
to simultaneously reproduce the CO ro-vibrational
line-profiles and the rotational diagrams of the three isotopologs.
A $\delta_{\rm gas}$ =10$^{-2}$ in the column density appears as the most likely value.
To directly illustrate this, 
we show the progression from a model without a column density drop to the best-fit grid model 
which has a column density drop of 10$^{-2}$ in Fig.~\ref{progression_plot}}.
{The empirical evidence of the gas density drop emerges from both the line-profile shapes and the rotational diagrams.
Models without a gas density drop generate $^{13}$CO and C$^{18}$O lines that are too broad, strong and warm (too much warm $^{13}$CO 
and C$^{18}$O emitted at small radii) to be consistent with the observations.

\begin{figure*}
%\begin{center}
\begin{center}
{\bf \large \sc Progression of models from the grid}\\
\end{center}
{\scriptsize {\bf 1. Model with a continuous disk with a surface density with exponent $\alpha$=-1.0} : $^{13}$CO and C$^{18}$O emission is too strong,  broad, and warm.}\\
\includegraphics[width=\textwidth]{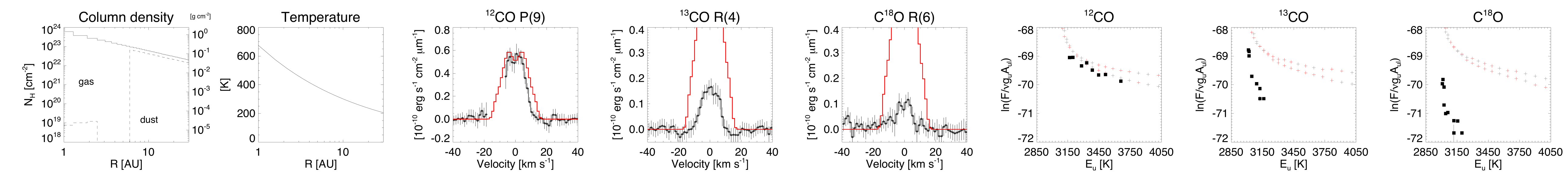}
{\scriptsize {\bf 2. Same model but with $\delta_{\rm gas}$ = 10$^{-1}$} : $^{13}$CO and C$^{18}$O emission are weaker, but still too strong} \\
\includegraphics[width=\textwidth]{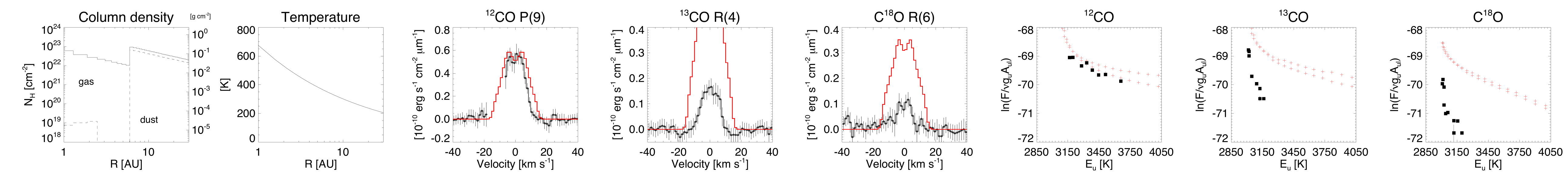}
{\scriptsize {\bf 3. Same model but with $\delta_{\rm gas}$ = 10$^{-2}$} : $^{13}$CO and C$^{18}$O lines have a flux closer to observations, but  $^{13}$CO emission is still too broad because the amount of gas at small radii.} \\
\includegraphics[width=\textwidth]{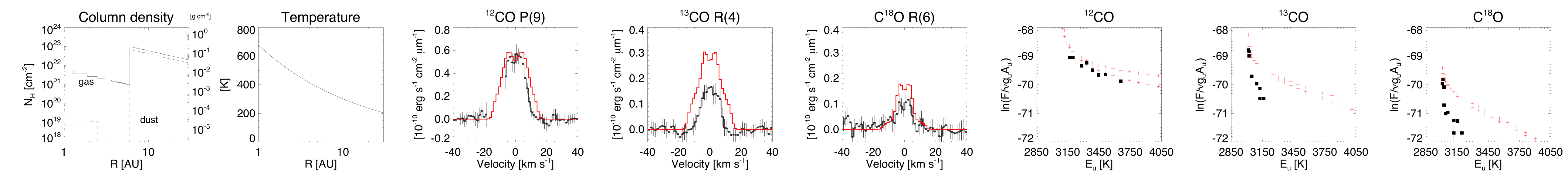}
{\scriptsize {\bf 4. Change of the $\alpha$ exponent of the inner disk from $-$1.0 to 0 (flat inner disk):} $^{13}$CO and C$^{18}$O line-profiles and rotational diagrams are more similar to the observations.}\\
\includegraphics[width=\textwidth]{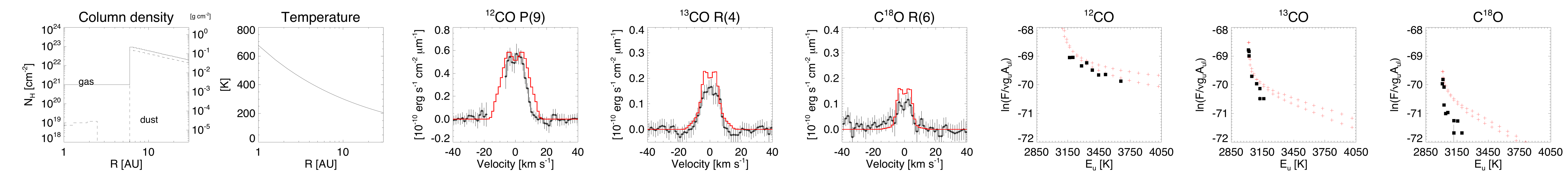}
{\scriptsize {\bf 5. Change of the $\alpha$ exponent of the inner disk to +1.0:} improvement of the fit of the $^{13}$CO and C$^{18}$O line-profiles and rotational diagrams.}\\
\includegraphics[width=\textwidth]{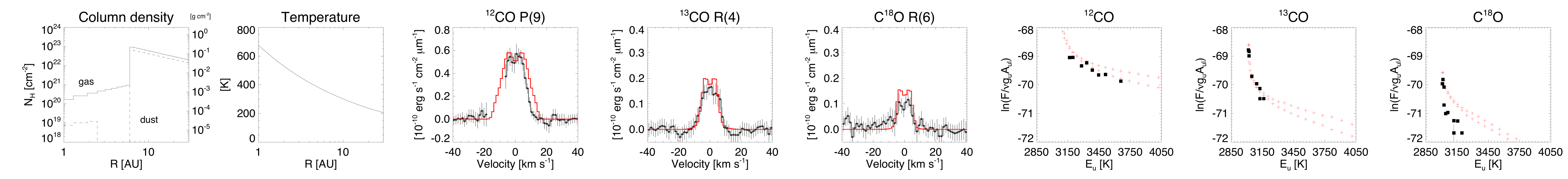}
{\scriptsize {\bf 6. Change of the $\alpha$ exponent of the inner disk to +2.0:} improvement of the fit of the wings of the $^{12}$CO line and the $^{12}$CO rotational diagram.}\\
\includegraphics[width=\textwidth]{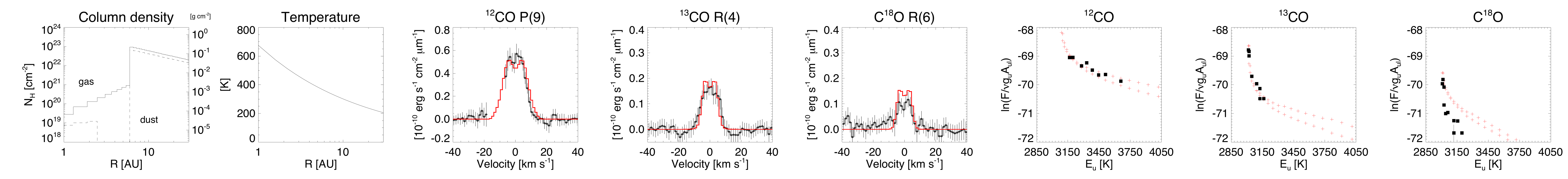}
{\scriptsize {\bf 7. Change of the $\alpha$ exponent of the outer disk from -1.0 to -2.5:} better description of the peak flux and rotational diagram of C$^{18}$O emission.}\\
\includegraphics[width=\textwidth]{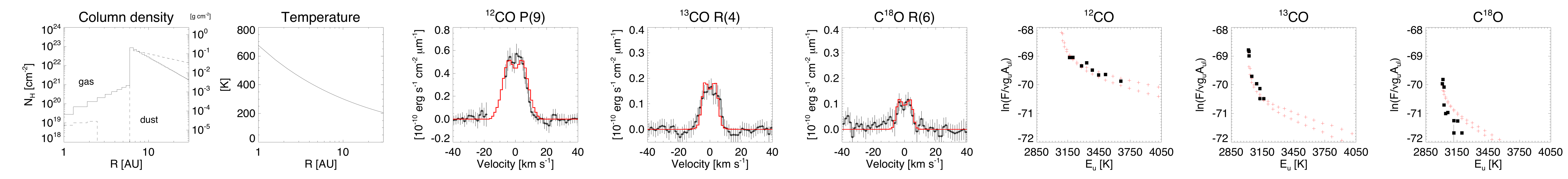}
%\end{center}
\caption{Example of a progression of  disk models taken from the grid to illustrate the effect of the change of parameters 
in the line-profiles and rotational diagrams.
The observed data are plotted in black, and the model predictions in red. 
Error bars on the line-profiles are $3\sigma$. The dashed line is the dust surface density from \citet[][]{Matter2016}.
The model starts with a continuous gas disk that follows the same power-law as the dust in the outer disk, 
and it is refined until the best description of the CO ro-vibrational observations is reached. 
The two branches seen in the rotational diagram correspond to the R and P branches of CO ro-vibrational emission.}
\label{progression_plot}
\end{figure*}

\begin{figure*}
\begin{center}
{\Large {\sc best-fit grid model:} $\alpha_{N_{\rm H\,inner}}$ =+2.0 ~~$\delta_{\rm gas}=10^{-2}$}\\[5mm]
\begin{tabular}{lll}
\includegraphics[width=0.25\textwidth]{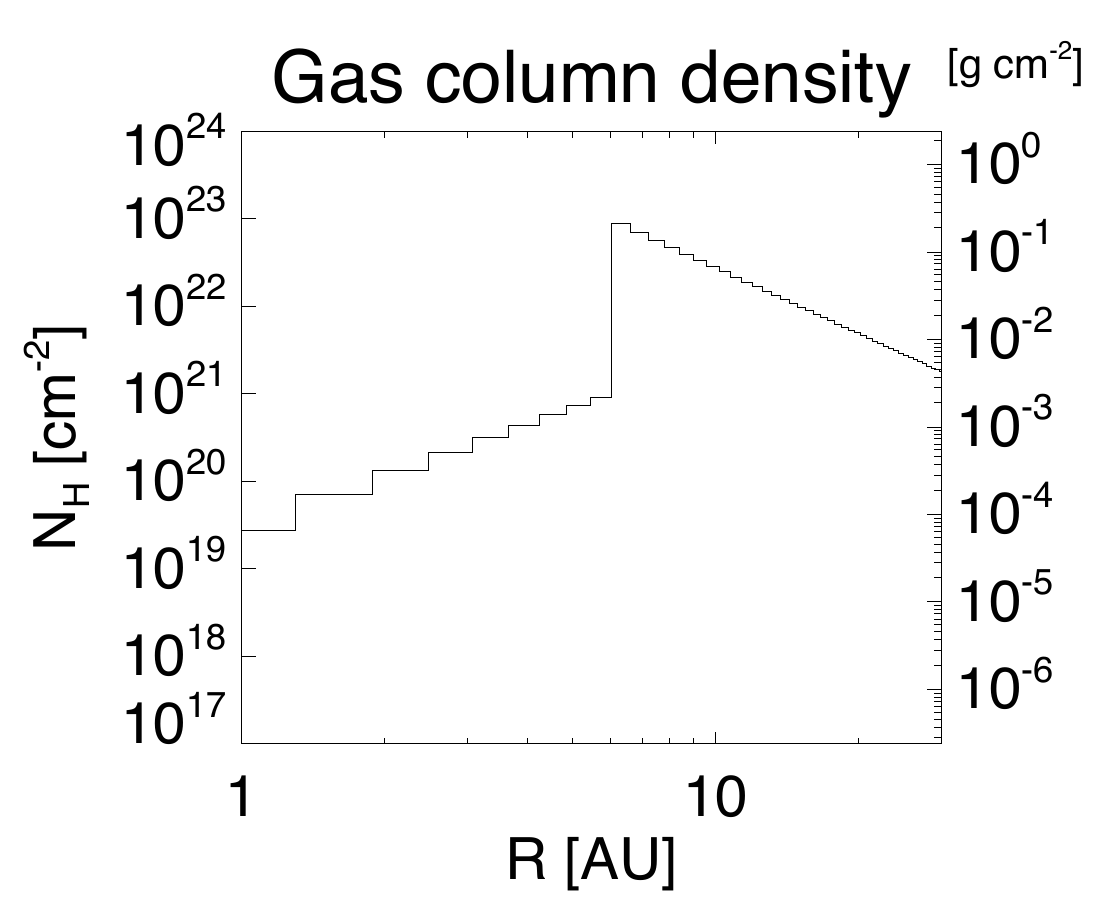}  & \includegraphics[width=0.25\textwidth]{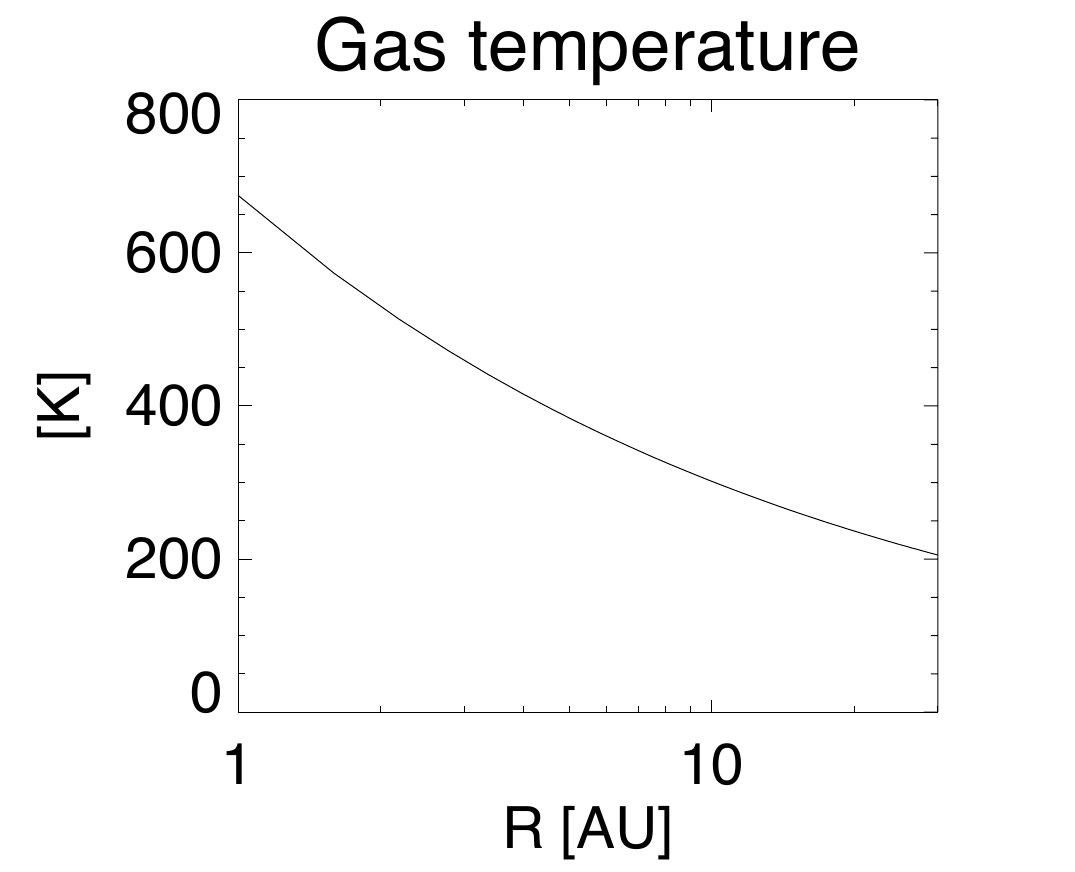} & \includegraphics[width=0.25\textwidth]{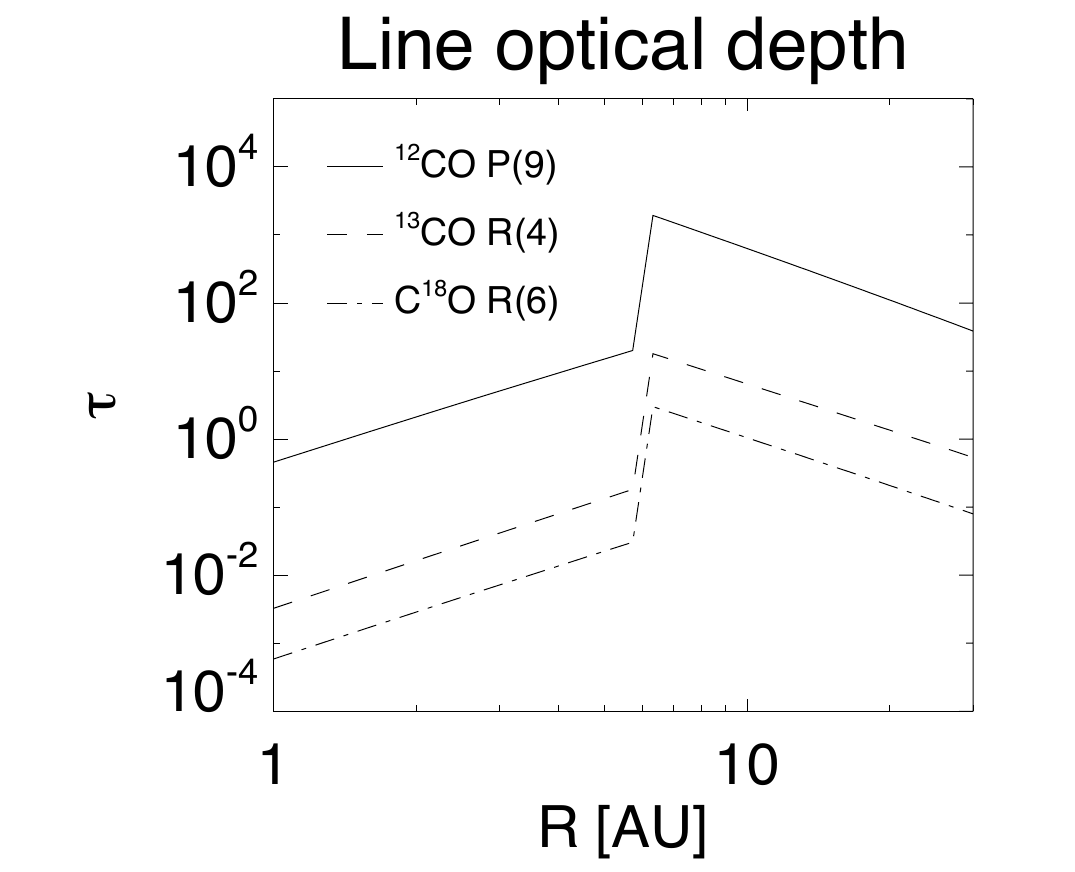} \\
\end{tabular}
\includegraphics[width=\textwidth]{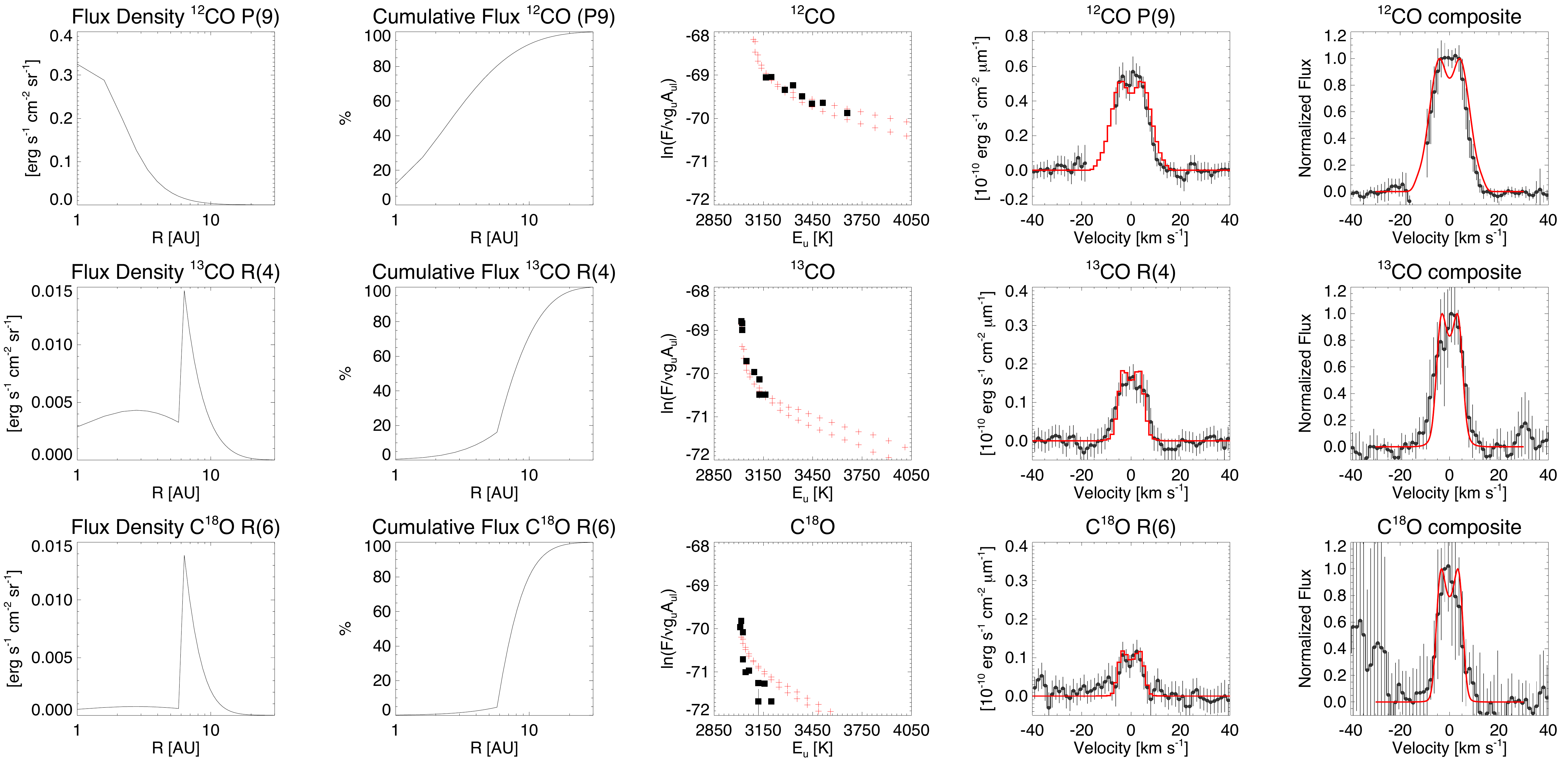} \\[3mm]
\caption{
Gas column density, temperature, CO optical depth,  flux density, cumulative line flux, rotational diagrams, and line-profiles of the $^{12}$CO P(9), $^{13}$CO R(4), and C$^{18}$O R(6)  
emissions of the best-fit grid model.
The model is shown in red, and the observations in black. 
Observed line-profiles are displayed in flux units after continuum subtraction with 3 $\sigma$ error bars. 
The two branches seen in the rotational diagram correspond to the R and P branches of CO ro-vibrational emission.
The rightmost panels compare the normalized theoretical line-profiles with the observed composite line-profile of each CO isotopolog with a 1 $\sigma$ error bar. 
}
\label{Model_plots1}
\end{center}
\end{figure*}

The column density of gas traced by CO at $R=6$ AU is well constrained to $N_H\sim$10$^{23}$ cm$^{-2}$.
This gas column density is similar to the dust column density at $R=6$ AU found by \citet[][]{Matter2016}
\footnote{We fixed the density drop location at 6 AU in the models, see the discussion section for the models 
with a varying radius of the density drop.}.
The gas column density at $R=6$ AU is higher than the column density found in the single $T_{gas}-N_H$ slab model of the three CO 
isotopologs of  Sect.~\ref{slab_section}.
This is because higher column densities describe the C$^{18}$O  and $^{13}$CO rotational diagrams better. 
{In fact, the emission of C$^{18}$O  and $^{13}$CO is optically thick in the $6-10$ AU region where 80 to 90 \%
of the line flux is emitted (see Fig.~\ref{Model_plots1})}.

CO ro-vibrational emission traces the gas in regions where the dust is optically thin or 
the disk upper layers where $T_{\rm gas} > T_{\rm dust}$.
{In Fig. \ref{tau_disk} we display the dust optical depth at 4.7 $\mu$m from the best-fit 
model of the SED and IR-interferometry data of HD~139614 from \citet[][]{Matter2016}.
At R$<$6 the dust is optically thin at 4.7 $\mu$m down to the disk mid-plane,
at R$\geq$ 6 AU the dust is optically thick at  4.7 $\mu$m (except in the disk surface layer).
As a consequence, in the inner disk at $R<$6 AU, the gas column density traced by CO ro-vibrational emission should 
be a good estimate of the total column density of gas.
Our models suggest that the column density of gas at $1<R<6$ AU ranges between 
$N_H=3\times10^{19}$ cm$^{-2}$ and $N_H=10^{21}$ cm$^{-2}$ ($7\times10^{-5} - 2.4\times10^{-3}$ g cm$^{-2}$).
In the outer disk at $R>$6 AU, the gas column density traced by CO ro-vibrational emission is  a {\it lower limit},
as we trace only the gas in the disk surface where the dust is thin and $T_{\rm gas}>T_{\rm dust}$.
If we assume a gas-to-dust mass ratio of 100 for the outer disk and use the dust column density at $R\geq6$ of \citet[][]{Matter2016},
then the total column of gas at $R\geq6$ AU can be up to a factor 100 higher than the column density traced by CO ro-vibrational emission.
Therefore, the gas density drop $\delta_{\rm gas}$ could be as large as 10$^{-4}$ depending on the total gas mass
of the outer disk.
}

The Bayesian probability distributions indicate that {the preferred values for the surface density
exponent in the inner 6 AU are flat or positive}, with  $\alpha_{N_{H\,inner}}$ ranging from 0.0 up to high values such as +3.0.
A few models with negative inner disk power-law exponents can describe the data, but, the 
largest fraction of models describing the observations
have flat or positive surface density exponents. 

The probability plots show that increasingly negative exponents of the density in the outer disk  
$\alpha_{N_{\rm H\,outer}}$ have a higher probability. 
This behavior is due to the fact that the models that best reproduce the  $^{13}$CO and C$^{18}$O 
emission are those in which a large portion of the line-flux ($\sim$60\%) is produced  between 6 and 10 AU (see Fig.~\ref{Model_plots1}). 
This suggests that the emission of $^{13}$CO and C$^{18}$O is most likely dominated by the contribution 
of the inner rim of the outer disk. This naturally explains the narrower line-profiles of these two isotopologs.

Figure~\ref{Model_plots1} displays the 
surface density profile, temperature profile, optical depth, flux density, cumulative line flux, rotational diagrams, 
and the $^{12}$CO P(9), $^{13}$CO R(4) and C$^{18}$O R(6) line-profiles
of the model in the grid that shows the best combined fit to the data.
This model has an 
$N_H$ at $R=6$\,{\rm AU} of 10$^{23}$ cm$^{-2}$,
$\alpha_{N_{H\,inner}}$ $(R< 6~{\rm AU)}=+2.0$,
$\alpha_{N_{H\,outer}}$ $(R\geqslant 6~{\rm AU})=-2.5$,
$\delta_{\rm gas}$ $(R=6~{\rm AU})=10^{-2}$,
$T_{0}$ $(R=1\,{\rm AU})=675$~K,
and
$\alpha_{T_{\rm gas}}=-0.35$.
We note that the solution is not unique,
and models with other parameters can still provide a satisfactory fit to the data. 
The Bayesian probability distribution diagrams show which values of the
model parameters are the most likely based on the observations. 

{From the pure modeling point of view, 
the higher probability of models with a flat or with an increasing surface density in the inner disk can be easily explained.
The emission of $^{13}$CO and C$^{18}$O is optically thin at R$<$6 AU (see upper-right panel in Fig.~\ref{Model_plots1}).
Thus to have a weaker flux at $v>10$ km s$^{-1}$, 
a small column of gas is needed at small radii.
A surface density with a decreasing profile in the inner 6 AU, 
even with a low column density (see Fig.~\ref{Simple_power_law}),
produces $^{13}$CO and C$^{18}$O line-profiles with line wings that are too strong.
The effect is also seen in the $^{13}$CO and C$^{18}$O rotational diagrams.
Flat and increasing surface density profiles provide colder rotational diagrams that better describe  the observations.
A visualization of this is provided in Fig.~\ref{progression_plot},
which illustrates the effect of changing the exponent of
the surface density profile in the inner disk from -1.0 to +2.0 
while keeping a constant $\delta_{\rm gas}=10^{-2}$ at $R= 6$ AU. 
As soon as $\alpha_{N_{H\,inner}}$ = 0 is reached, 
the strong high-velocity line wings of the $^{13}$CO and lines C$^{18}$O disappear.
The $^{12}$CO line-profile hardly changes in all the models with $\alpha_{N_{H\,inner}}$ -1.0 to +1.0
because it is optically thick.
However, when $\alpha_{N_{H\,inner}}$ is higher than +1.5, $^{12}$CO becomes optically thin at R$<2$ AU,
the wings of the $^{12}$CO line become weaker, and the $^{12}$CO line-profile is well fitted.
}

\begin{figure}
\begin{center}
\includegraphics[width=0.45\textwidth]{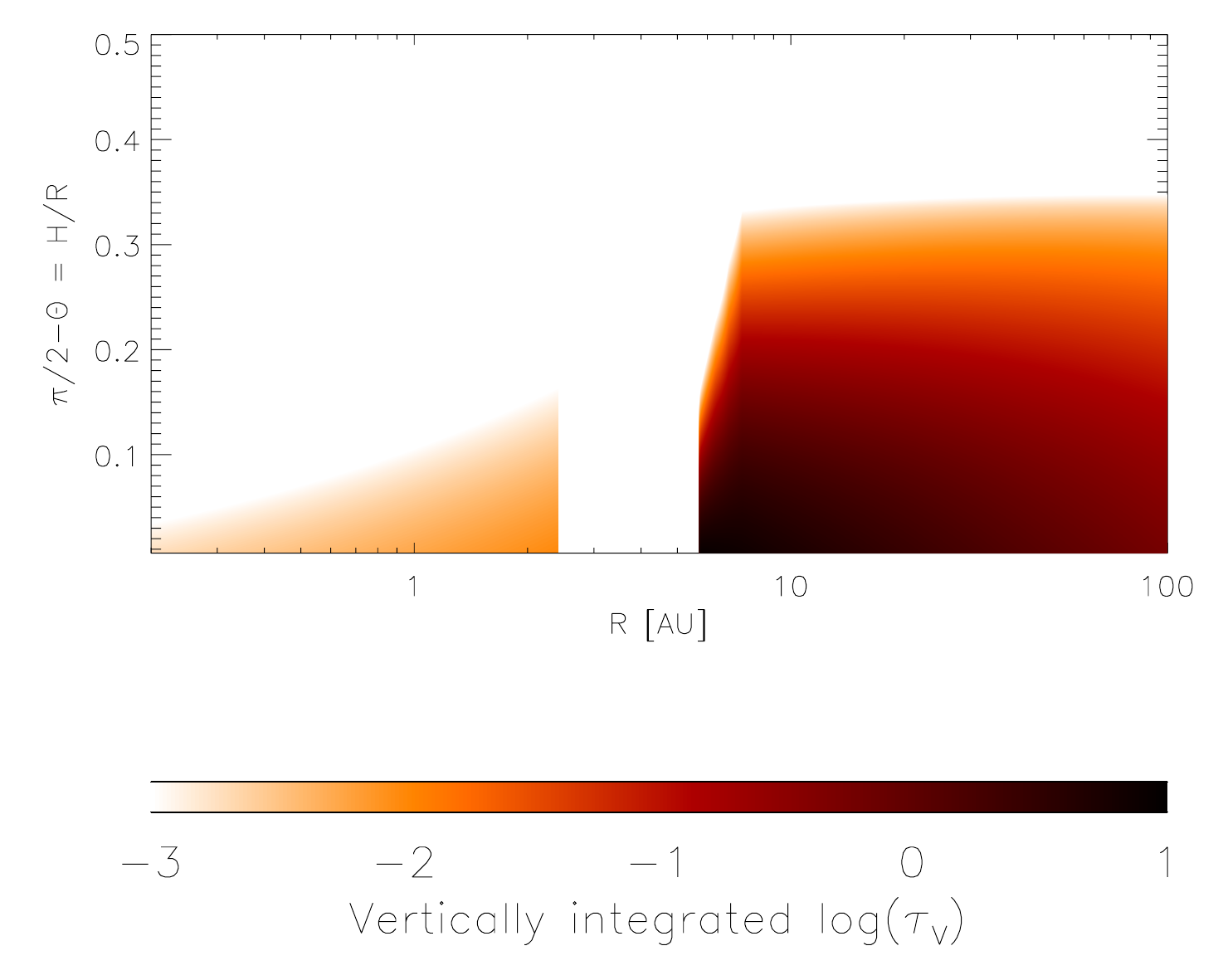}
\caption{Vertically integrated optical depth of the dust at 4.7 $\mu$m from the HD~139614 model of 
\citet[][]{Matter2016}, which best describes the SED and VLTI IR-interferometry data.
The dust is optically thin at 4.7 $\mu$m in the inner 6 AU of the disk.
At R$>$6 AU the dust is optically thick except in the uppermost layers of the disk.
}
\label{tau_disk}
\end{center}
\end{figure}

\subsection{Quantitative constraints on the width and column density depth of a gas gap}
\label{Gap_section}
{As discussed in Sect.~\ref{gap_section},
the $^{12}$CO ro-vibrational composite line-profile does not display evidence
of a gap (i.e. a zone devoid of gas) of size larger than 2 AU.
To derive quantitative limits on the gap width and the column density of the gas that could be {inside} a potential gap,
we calculated the expected $^{12}$CO P(9) line for a series of 
models around the best-fit grid model for gaps of increasing width 
(i.e., from 5 to 6 AU,  from 4 to 6 AU, from 3 to 6 AU, and from 2 to 6 AU) and 
varying  $N_H$ inside the gap (at $R= $6 AU)   
from 10$^{21}$ to 10$^{17}$ cm$^{-2}$.
The normalized theoretical $^{12}$CO P(9) spectrum was compared with the high S/N composite  $^{12}$CO spectrum.

Figure~\ref{gap_grid_plot} display the results.
The models confirm the suggestion of the simple power-law intensity model (Sect.~\ref{gap_section}).
A gap of 2 AU or smaller remains undetected.
Gaps of width larger than 2 AU and with an 
$N_H$ inside the gap lower than 10$^{18}$ cm$^{-2}$ (yellow and red, $\delta_{\rm gas}<10^{-5}$) 
would have been seen in the $^{12}$CO composite spectrum as a line-profile with shoulders  at $\pm$15 km s$^{-1}$.
}

\subsection{Upper limits to the gas column density at $R<1$ AU}
\label{NH_upper_limits_1au}
$^{12}$CO ro-vibrational emission requires relatively low column densities ($N_{\rm CO}\sim10^{15}$  cm $^{-2}$)
to be optically thick. Therefore, if the gas is sufficiently warm, CO ro-vibrational emission is relatively easy to detect.
As the gas in the inner 1 AU of a disk around a Herbig Ae star has a temperature warmer than 300 K,
the lack of strong CO ro-vibrational emission in the inner 1 AU of HD~139614 suggests a low column density 
of gas at $R\leq$ 1 AU.
To derive upper limits to the column of gas at $R\leq$ 1 AU in HD~139614, 
we calculated the expected emission from gas between 0.1 and 1.0 AU assuming 
a flat surface density profile and the temperature profile of the best-grid model.
We found that gas column densities higher than $N_H=5\times10^{19}$ cm$^{-2}$ ($1.2\times10^{-4}$ g cm$^{-2}$) 
would have produced line-profile wings ($v>$ 15 km s$^{-1}$) stronger than 5$\sigma$ 
 the noise of the $^{12}$CO P(9) line (Fig. \ref{12COupperlimits}).
As we assume standard abundances, our CRIRES observation set a 5$\sigma$ upper limit to the CO column at $R\leq$ 1 AU of 
$N_{\rm CO}=5\times10^{15}$ cm$^{-2}$.

UV photodissociation could be responsible for the destruction of CO in the inner 1 AU of the disk around HD~139614.
In the absence of dust, CO self-shields against photodissociation in the vertical and radial direction if $N_{\rm CO}>10^{15}$ cm$^{-2}$ 
\citep[][]{vanDishoeckBlack1988}.
Therefore, from the self-shielding perspective, the absence of CO ro-vibrational emission from $R<$1 AU
suggests that $N_H\leq10^{19}$ cm$^{-2}$ at R$\leq1$ AU, assuming standard abundances.
This value is consistent with the upper limit derived from the CO ro-vibrational line-profile modeling.

\begin{figure*}
\begin{center}
\includegraphics[width=0.95 \textwidth]{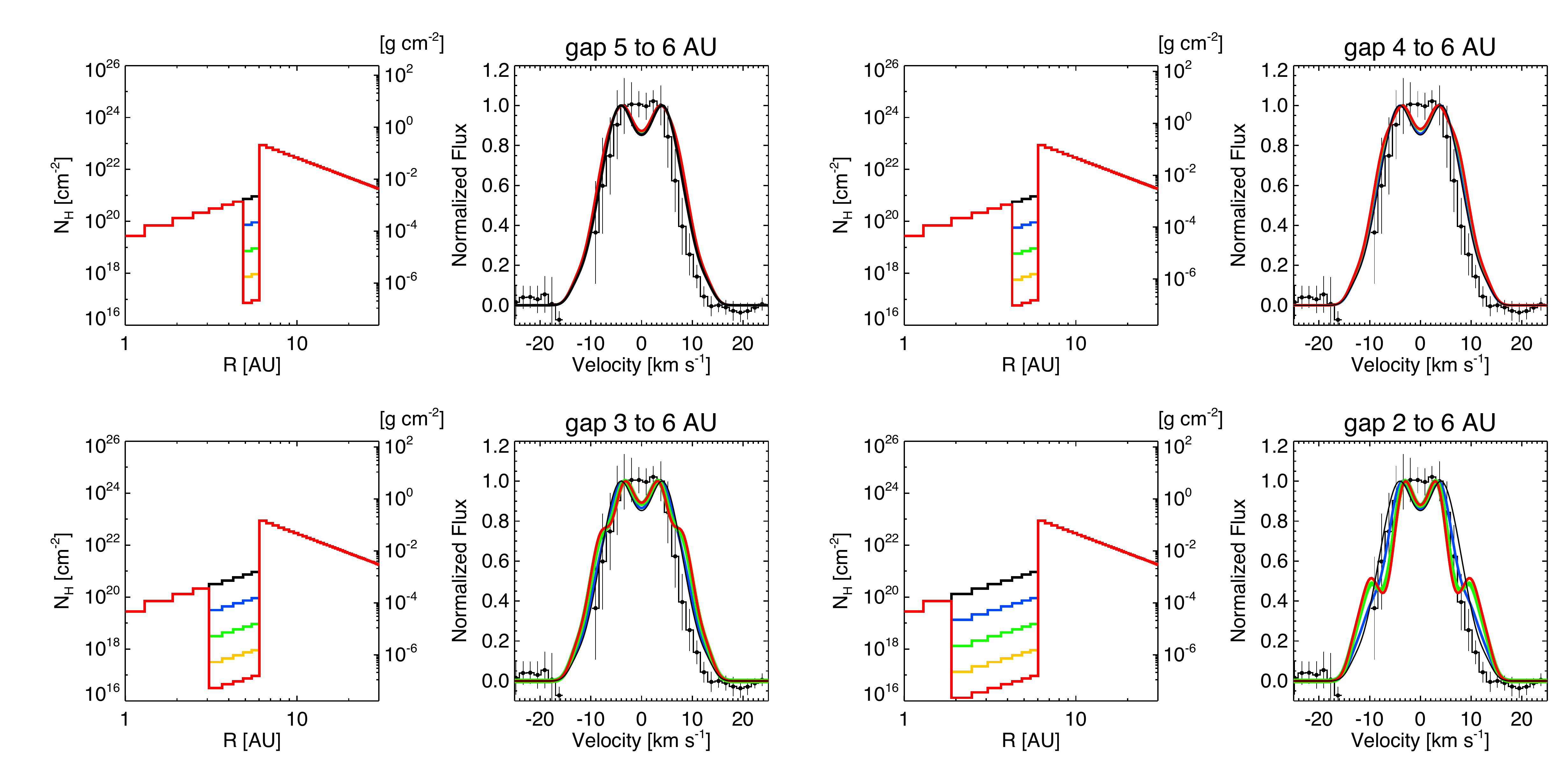}
\caption{$^{12}$CO P(9) line-profiles predicted for different gap sizes and diverse column densities inside the gap,
for the best-fit grid model (see Fig.~\ref{Model_plots1}), compared with composite $^{12}$CO line-profile. 
Observed and modeled line-profiles are normalized such that the continuum is at zero and the line peak at 1.
The color code in the surface density density panel corresponds to the color in the line-profile plot.
The solution without gap is displayed in black. The temperature profile is kept constant in all the models.
Error bars are 1$\sigma$.
}
\label{gap_grid_plot}
\end{center}
\end{figure*}

\begin{figure}
\begin{center}
\includegraphics[width=0.45\textwidth]{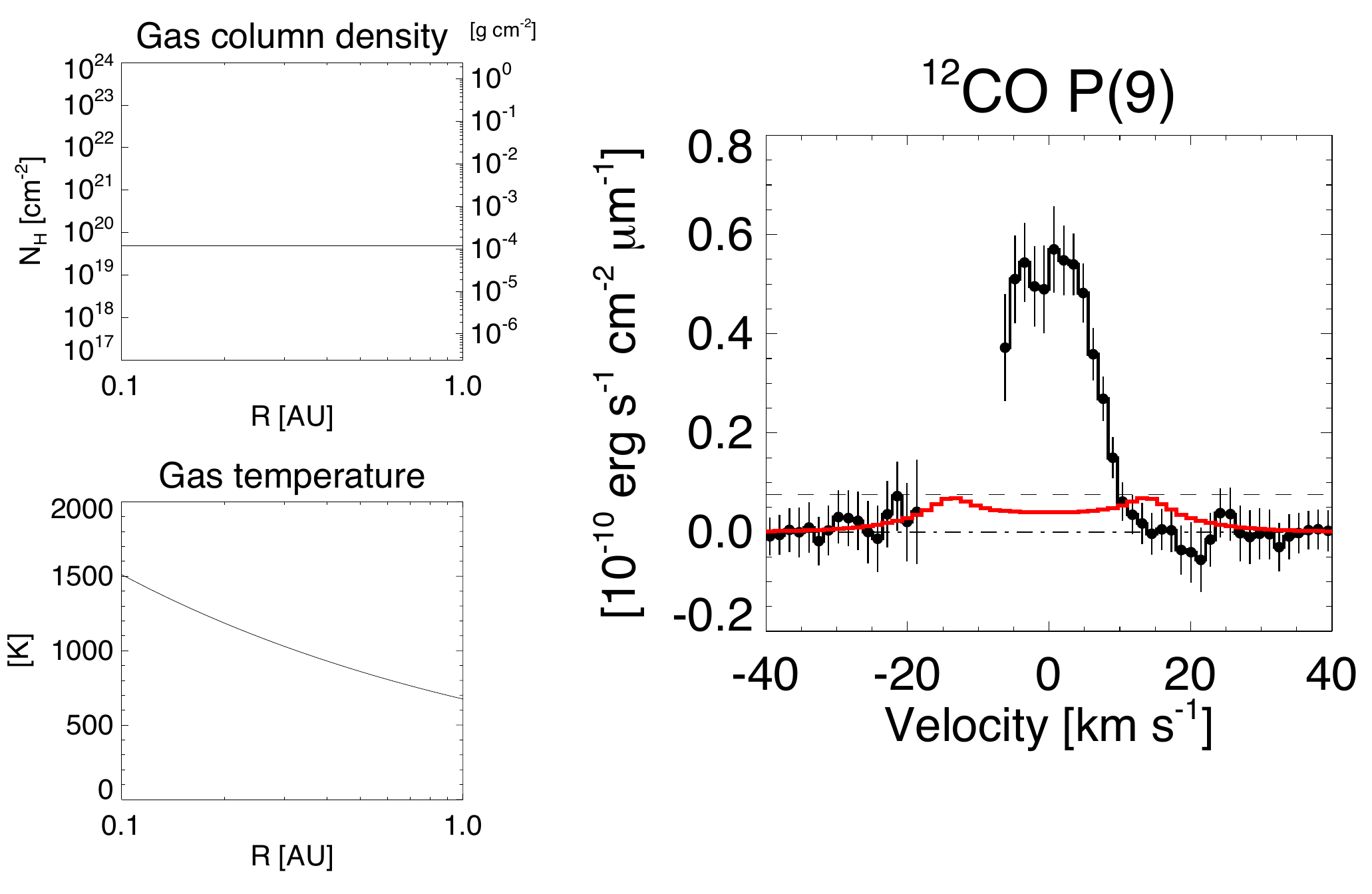}
\caption{Predicted $^{12}$CO P(9) line-profile for a disk extending from 0.1 to 1.0 AU 
with a flat surface density $N_H=5\times10^{19}$ cm$^{-2}$ and the extrapolated temperature
profile from the best-fit grid model. 
Error bars are 3$\sigma$. The horizontal dashed line is the 5$\sigma$ limit.}
\label{12COupperlimits}
\end{center}
\end{figure}

\subsection{Uncertainties, limitations and robustness tests}
Optical depth effects explain the higher probabilities of models with $\alpha_{N_{H\,inner}}$ of +2 or +3. 
{However,  such high  $\alpha_{N_{H\,inner}}$  are hard to justify physically}\footnote{X-ray photoevaporation models \citep[e.g.,][]{Owen2012,Mordasini2012} evolve to a flat surface density.
Planet-disk interaction models \citep[e.g.,][]{Crida2007} can give a positive density gradient with radius in the inner disk,
but only in the innermost radii.}. 
To test the robustness of the modeling conclusions,
we recalculated the Bayesian probability plots for a sub-grid selecting only models with $\alpha_{N_{H\,inner}}\leq +1$  (54 000 models, Fig.~\ref{grid_short_bayesian}).
We found that a gas density drop and gas surface density at $R<6$ AU that is flat are still needed to explain the observations.
In Fig. \ref{Model_plots2} and Table~\ref{table_models},
we display the characteristics of the best-fit model of the $\alpha_{N_{H\,inner}}\leq +1$ sub-grid.
The model has $\delta_{\rm gas} = 10^{-2}$ and has a flat ($\alpha_{N_{H\,inner}}=0.0$) surface density at $R<$6 AU\footnote{We note, however, that models with flat surface density profiles produce a poorer fit to the observations than models with 
positive $\alpha_{N_{H\,inner}}$ because of the 
stronger $^{12}$CO high-velocity wings (Fig.~\ref{progression_plot}).}.

C$^{18}$O emission is detected at lower S/N than $^{12}$CO and $^{13}$CO emission.
To further test the reliability of the gas density drop,
we recalculated the Bayesian probability diagrams
only taking into account the $^{12}$CO and $^{13}$CO  emission  (Fig.~\ref{bayesian_12CO_13CO}). 
We retrieved the same results:
the models with the highest probabilities are those with a gas density drop of at least a factor 100 in the inner 6 AU and 
that have a surface density at $R<6$ AU that is flat or increasing with radius.

Our models assume constant  $^{12}$CO/H$_2$, $^{12}$CO/$^{13}$CO and $^{12}$CO/C$^{18}$O
ratios with radius.
The CO ro-vibrational lines trace warm CO, therefore freeze-out onto dust surfaces and
fractionation reactions  are not a concern.
Photodissociation by UV-photons might be relevant, 
because the self-shielding of $^{13}$CO and C$^{18}$O requires higher column densities than for $^{12}$CO.
Models of disks including selective photodissociation \citep[e.g.,][]{Miotello2014} showed 
that the gas masses in the outer disk can be underestimated by up to an order of magnitude\footnote{For an application 
of selective photodissociation to the Solar Nebula see \citet[][]{LyonsYoung2005}.}.
However, an underestimation of the column density that is due to selective photodissociation by a factor of ten in the inner disk 
would not change the conclusion that a surface density drop is required to describe the CO ro-vibrational data.
Moreover, \citet[][]{vanderMarel2016} recently modeled ALMA CO sub-mm rotational emission from the gas in transition disks with large dust cavities.
They found that selective-photodissociation does not significantly
affect the CO isotopologs rotational emission from gas inside the dust cavity.

We have used a single vertical temperature for each radius, but disks have a vertical gradient of temperature.
At $R<$ 6 AU, the dust (see Fig. \ref{tau_disk}) and the $^{13}$CO and C$^{18}$O lines (see Fig.~\ref{Model_plots1})
are optically thin\footnote{An additional argument for an optically thin inner disk comes from the detailed radiative transfer
modeling of CO ro-vibrational emission in the Herbig F4V pre-transition disk HD~135344B in \citet[][]{Carmona2014}.
We found that,  to have CO ro-vibrational emission from inside the dust cavity, the dust in the inner-most disk
should be optically thin.
}.
Therefore, at $R<6$ AU, the $^{13}$CO and C$^{18}$O transitions trace the whole vertical column of gas.
Although in a disk the temperature increases from the mid-plane to the surface, 
in the inner 6 AU, the $^{13}$CO and C$^{18}$O lines are not dominated by the hottest 
gas (T$>$ 500-1000 K) located in the upper most layers near the surface 
because the amount of gas in those upper regions is very small.
The $^{13}$CO and C$^{18}$O ro-vibrational lines are emitted lower down,
in the region where the CO gas is the densest and where the gas temperature 
in the vertical direction varies by a few 10 K at most down to the mid-plane\footnote{See Fig. A.4 in \citet[][]{Carmona2014}  for the temperature distribution of the gas inside an optically thin dust cavity of a transition disk.}.
The $^{12}$CO emission is optically thick at $R<$6 AU, 
which means that on average, it is emitted higher up in the disk.
The dominant emitting regions for $^{12}$CO, $^{13}$CO and C$^{18}$O ro-vibrational emission 
have differences in vertical height but, in fact, they overlap in the vertical direction.
The temperature profile in our simple 1D models should be understood as a representative {\it average} 
vertical temperature\footnote{The best-fit grid model has a radial temperature 
that is a compromise between the $^{12}$CO that requires a higher temperature (emitted higher up)
and $^{13}$CO and C$^{18}$O emission that require lower temperatures (emitted farther down).
The best-fit model combining $^{12}$CO, $^{13}$CO, and C$^{18}$O has a gas $T_{0~(R=~1\,{\rm AU})} =675$ K.
This temperature is in between the temperature of the best-fit grid model for $^{12}$CO alone ($T_{0~(R=~1\,{\rm AU})} $ = 725 K)
and the temperature of the best-fit grid model for C$^{18}$O emission alone ($T_{0~(R=~1\,{\rm AU})} $ = 625 K).}.  
The column of C$^{18}$O (thus $N_H$) could be underestimated because the 1D models use a higher average temperature.
However, the column density in the inner 6 AU should be lower than $N_H$=10$^{21}$ cm$^{-2}$.
Higher $N_H$, even with $T_{0~(R=~1\,{\rm AU})} $  = 575 K (100 K lower than the best-fit grid model), 
would generate $^{13}$CO and C$^{18}$O lines with high-velocity wings which 
would be too strong to be compatible with the CRIRES spectrum.

We  assumed a smooth temperature profile with radius, without bumps or discontinuities. 
However,
as the dust density is much lower at $R<$ 6 AU, the temperature at the inner rim of the outer disk might be higher,
as seen in thermochemical disks models \cite[][]{Thi2013,Carmona2014,Woitke2016, RosinaPhD}.
The question in HD~139614 is which fraction of the column of the emitting CO around 
6 AU is at higher temperatures.
The smoothness and width of the $^{12}$CO line-profile, 
and the fact that the average temperatures of $^{13}$CO and C$^{18}$O lines 
(380 and 350 K respectively) are similar to that of the power-law temperature at $6<R<10$  AU of the best-fit grid model (350 - 300 K, where most of the $^{13}$CO and C$^{18}$O flux is produced)
suggest that the column of CO at temperatures much higher than 400 K in the inner rim of the outer disk AU should be small.
We conclude that a smooth-temperature power-law
describes the temperature of the largest column of gas emitting  the CO ro-vibrational lines.

We assumed that the gas density drop occurs at the same radius as the dust density drop.
But the gas density drop does not have to occur at 6 AU.
We have tested a sub-grid of models (28 800 models) in which we varied  the 
radius of the gas density drop  between 4.0 and 6.0 AU (see Fig. \ref{bayesian_gap_rmax})\footnote{That dust is seen at $R\sim6$ AU and that the
dust requires a scale height to fit the near-IR data (SED and visibilities)
implies that gas should be present at the location of the dust. For this reason larger radii than 6 AU for the gas density drop were not considered in the modeling.}.
The grid shows that the most likely value for the gas density drop is 6.0 AU.
Models with a gas density drop down to 5.0 AU can also describe the data but are less likely\footnote{The change 
in the gas-density drop radius can be compensated for by changes in $\delta_{\rm gas}$ or
the surface density or temperature exponent.}.

The model grid was calculated assuming $R_{\rm in}=1$ AU because the power-law intensity model 
indicated that the radius of the maximum intensity is close to 1 AU (Sect.~\ref{COemitting_region}).
We have tested models with gas down to 0.1 AU extending the power-law temperature and density profile. 
The fit was satisfactory for surface density exponents at $R<6$ AU ($\alpha_{N_{H\,inner}}$) between +1 and +3. 
Models with $\alpha_{N_{H\,inner}}$ smaller than +1 gave too strong line wings for the  $^{12}$CO emission.
The best-fit grid model, which has $\alpha=+2.0$ surface density exponent, 
gave a good fit when we extended it down to 0.1 AU (Fig.~\ref{Plots_model_0.1au}).
This model is consistent with the upper limits on the surface density at R$<1$ AU 
derived in Sect.~\ref{NH_upper_limits_1au}.

{In the grid of models we did not fit the $\upsilon=2\rightarrow$1~$^{12}$CO  lines.
We checked the predicted $\upsilon=2\rightarrow$1~$^{12}$CO P(3) and P(4) lines for the best-fit grid model. 
We found that the model is able to reproduce the observed {\it FWHM} of the $\upsilon=2\rightarrow$1 lines,
but the model has weaker  $\upsilon=2\rightarrow1$ line fluxes than the observations.
We explored models around the best-fit grid solution.
We found that a model with an $N_{H}$ at 6 AU three times larger ($N_{H~(R=6~{\rm AU})}$ = 3$\times$10$^{23}$ cm$^{-2}$),
the same density profile at $R<$6 AU (thus $\delta_{\rm gas}$=3.3$\times10^{-3}$), 
and the same temperature profile described the $\upsilon=2\rightarrow1$ lines well while having a good fit to the $\upsilon=1\rightarrow0$  lines.
The model indicates that the $\upsilon=2\rightarrow1$  lines are dominated by the contribution at $6<R<10$ AU.
We present the predicted emission lines of this model in Fig~\ref{CO2-1} in the Appendix.

{Our model grids were calculated before the Gaia distance release, and therefore assumed a distance of 140 pc. 
The new Gaia distance of  131 $\pm$ 5 pc translates into a location of gas (and dust) density drop 0.4 AU closer in. 
This difference is within the uncertainties of the data modeling, and therefore the conclusions we reach are not affected.
}

\section{Discussion}

{The main results of the observations and modeling (see Table~\ref{table_analysis} for a modeling overview) 
of the CO ro-vibrational emission lines in the HD~139614 disk are as follows.}
\begin{enumerate}
\item CO ro-vibrational emission extends from 1 AU to 15 AU in HD~139614, which means that the dust gap observed in IR-interferometry data contains molecular gas.
\item C$^{18}$O lines are a few km s$^{-1}$  narrower than $^{12}$CO lines.
\item $^{13}$CO and C$^{18}$O emission are 50 - 100 K colder than $^{12}$CO emission.
\item The observed CO emission is very likely thermally excited.
\item A drop of 10$^{-2}$ to 10$^{-4}$ in the gas column density at {$R<5-6$} AU is required to simultaneously reproduce
the line-profiles and rotational diagrams of the three CO isotopologs.
\item The gas surface density $N_H$ at $1< R< 6$ AU ranges between $3\times10^{19}$ and  $10^{21}$ cm$^{-2}$ and has
a distribution that most likely is flat or that increases with radius. 
\item The 5$\sigma$ upper limit on the CO column density $N_{\rm CO}$ at $R\leq1$ is $5\times10^{15}$ cm$^{-2}$, which corresponds to
a gas column density $N_H<5\times10^{19}$ cm$^{-2}$ if standard abundances are assumed.
\item Our data does not show evidence of a gap (devoid of gas) in the gas distribution. 
The width of any possible gap is constrained to be smaller than 2 AU.
\end{enumerate}

\subsection{Gas vs. dust surface density distribution}
{In Fig.~\ref{surface_density_plot} we display the dust surface density (in blue)
derived by \citet[][]{Matter2016} from ESO/VLTI near and mid-IR interferometry 
observations.
In the same figure we illustrate the gas surface density derived from the CO ro-vibrational emission.
In red we plot the best-fit grid model, in orange the best-fit grid model if $\alpha_{N_{\rm H\,inner}} \leq 1.0$.

We find that the $\alpha_{inner}=+0.6$ deduced for the dust is consistent with the range of $\alpha_{N_{\rm H\,inner}}$ 
of the gas observations.
The near-IR continuum and the CO ro-vibrational observations suggest gas-to-dust mass ratios for the inner disk at R$<$ 2.5 AU 
ranging from a few up to 100, depending on the gas surface density exponent at R$<$6 AU. 
Concerning the depletion levels of the gas and the dust ($\delta_{\rm gas}$ and $\delta_{\rm dust}$), 
if the gas-to-dust mass ratio is 100 in the outer disk, then the level of depletion for the gas and the dust would be similar.
However, 
it is likely that given the age of HD~139614 \citep[9 Myr, ][]{Alecian2013} and the weak [\ion{O}{i}] 
63 $\mu$m line flux \citep[$4.5\times10^{-17}$ W m$^{-2}$, among the weakest of the whole Herbig Ae sample of ][]{Meeus2012},
that the gas-to-dust mass ratio in the outer disk is below one hundred\footnote{In the detailed model of the transition disk
HD~135344B (which has a dust cavity of 30 AU) by \citet[][]{Carmona2014}
we found that to reproduce its weak [\ion{O}{i}] line  
\citep[$4.7\times10^{-17}$ W m$^{-2}$,][]{Meeus2012} the gas-to-dust mass ratio in the outer disk needed to be much lower than 100
with a best value below 10.}. 
In this case, the density drop in the gas would be less deep than 
the density drop in the dust. }

\begin{figure*}
\begin{center}
\includegraphics[width=0.55\textwidth]{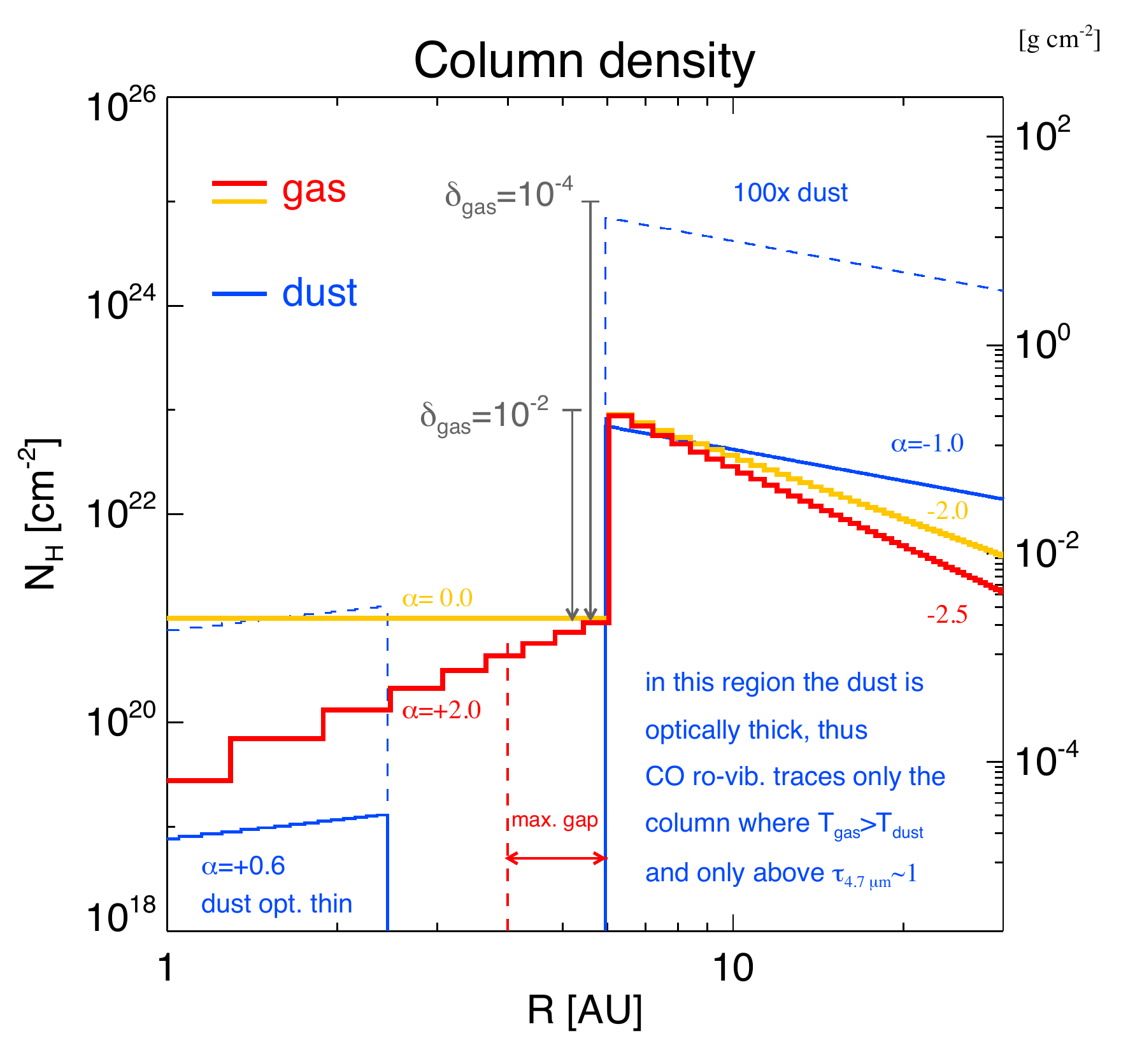}
\caption{Gas and dust surface density. 
In blue we show the dust surface density derived from modeling the SED + near and mid-IR observations,  
taken from \citet[][]{Matter2016}.
In red we plot the gas surface density of the best-fit model of the whole grid.
In orange we show the gas surface density of the best-fit grid model restricted to only models with $\alpha_{N_{\rm H\,inner}} \leq +1.0$.
Note that the value of $\delta_{\rm gas}$ depends on the gas-to-dust ratio assumed for the outer disk. 
{This plot assumes a distance of 140 pc for HD~139614. The new Gaia distance of 131 $\pm$ 5 pc implies
that the location of the gas and dust density drop is 0.4 AU closer in. However, this difference is within the uncertainties of
the modeling of the data.}
}
\label{surface_density_plot}
\end{center}
\end{figure*}

\subsection{CO ro-vibrational emission in HD~139614 and other protoplanetary disks}
{The integrated $^{12}$CO line fluxes of HD~139614 are on the same order (10$^{-14}$ erg s$^{-1}$ cm$^{-2}$) 
as previously observed primordial and transition disks \citep[e.g.,][]{Najita2003, Salyk2011,Brown2013, BanzattiPontoppidan2015}.
The widths of the $^{12}$CO line-profiles of HD~139614 (and other transition disks)
are, however, significantly narrower and lack the high-velocity wings (emission at $v>15$ km s$^{-1}$) 
that are observed in primordial disks\footnote{The $^{12}$CO ro-vibrational emission from primordial disks is two to five
 times broader than that of transition disks \citep[][]{Brown2013}.}.
This difference arises because in primordial disks
the $^{12}$CO emission is dominated by gas inside 1 AU, 
while in HD~139614 (and most transition disks) the CO ro-vibrational emission is dominated by gas at R$>$ 1 AU.

Given that  $^{12}$CO ro-vibrational emission becomes optically thick at relatively low column densities 
($N_{\rm CO}\sim10^{15}$ cm$^{-2}$), 
the lack of high-velocity wings in the $^{12}$CO line-profiles of HD~139614 (and other transition disks)
directly indicates that at R$<$1 AU the column density of gas is much lower than in primordial disks.
The 5$\sigma$ upper limit of $N_H\sim5\times10^{19}$ cm$^{-2}$ on the column of gas at $R<1$ AU (assuming standard abundances) 
derived for HD~139614 is significantly lower than the $N_H= 10^{24} -10^{26}$ cm$^{-2}$ typical of primordial disks. 

In HD~139614, the $^{13}$CO emission is narrower and colder than the $^{12}$CO emission.
Differences between CO isotopologs line widths and temperatures have been 
reported in a number of primordial and transition disks \citep[e.g.,][]{Brown2013}.
However, the $^{12}$CO and $^{13}$CO line width difference in primordial disks 
is on the order of tens of km s$^{-1}$, while in HD~139614 (and other transition disks) it is only a few km s$^{-1}$.
This further suggests that there is a different inner disk gas structure between transition disks and primordial disks 
 \citep[as already pointed out by previous authors, e.g.,][]{Brown2013,BanzattiPontoppidan2015}.
 
The $\upsilon=1\rightarrow0$ CO and $\upsilon=2\rightarrow1$ CO emission in HD~139614 can be described by thermal emission ($T_{\rm ex} = T_{\rm rot} = T_{\rm vib}$). 
The average CO temperature of 460 K (log T = 2.7) and the CO inner radius of 1 AU locate HD~139614
in the left side of the UV pumping regime in the recently proposed  T/R diagram of  \citet[][]{BanzattiPontoppidan2015}.
According to the T/R diagram, HD~139614 belongs to disk category 2: objects with partly devoid disk's gaps.
HD~139614  appears in the T/R diagram as an intermediate case between category 1 disks that are primordial, 
and category 3 that are transition disks that have $\upsilon=2\rightarrow1$ CO ro-vibrational emission dominated by UV pumping.

In summary,  the properties of CO ro-vibrational emission are clearly different between HD~139614 and primordial disks.
The data and models show that although there is molecular gas inside the inner 6 AU of HD~139614,
the gas distribution and gas mass is different from that of a primordial disk at R$<$ 6 AU.

\subsection{Comparison to other transition disks with quantified gas surface densities inside the dust cavity}
{HD~139614 adds to the growing number of transition disks with quantified gas density drops inside the dust cavity.
In Table~\ref{table_gasdrop} we provide a summary of the properties of these sources\footnote{We note that the number of transition disks with detections of gas inside the dust cavities from CO ro-vibrational and CO rotational emission is much larger
\citep[e.g.,][]{Pontoppidan2011,Salyk2011,Brown2013,BanzattiPontoppidan2015}.
We discuss here only the sources with published gas and dust surface density profiles.}.
The largest portion of transition disks shows stronger depletions in the dust than in the gas ($\delta_{\rm dust} < \delta_{\rm gas}$)
inside the dust cavity.
Some disks have a similar depletion level (e.g., HD~139614),
and only one source, J1604-2130, has a depletion level higher  in the gas than the dust.
Altogether, this suggests that the dust depletes faster than the gas in the inner disk.
This behavior might arise because it is hard to stop viscous accretion of gas through the disk unless the mass of the (outer) disk is very low.
This seems to be the case for transition disks around low-mass/solar-type stars \citep[][]{SiciliaAguilar2015}.
If trapping of mm-dust grains occurs at the inner rim of the outer disk,
then a mm-dust depletion higher than the gas depletion would be expected inside the dust cavity \citep[e.g.,][]{Pinilla2016}.

Table~\ref{table_gasdrop} shows that in a {large fraction of objects, the gas density drop radius is 
smaller than the sub-mm dust density drop radius}.
This is consistent with the detection of micron-sized particles at radii smaller than the sub-mm dust cavity radius 
\citep[e.g.,][]{Muto2012, Garufi2013, Follette2013, Pinilla2016} because small dust grains are expected to be coupled to the gas.
Furthermore, this also provides observational support for the scenario of dust trapping by pressure bumps due to the presence of (unseen) 
planets inside the dust cavity \citep[e.g.,][]{Rice2006,Paardekooper2006,Fouchet2010,Pinilla2012,Pinilla2015,Gonzalez2015}.
In the case of HD~139614, the dust and gas cavity radii are similar, but it is the only object in the sample with a dust cavity radius
smaller than 10 AU and the only object where the dust cavity radius is measured in the IR. Future sub-mm observations of 
HD~139614,
will enable us to test whether its sub-mm  dust cavity radius is larger than the gas cavity radius.
}

\subsection{Origin of the dust and gas density distribution in HD~139614}
\label{origin}

Photoevaporation has been suggested as a potential mechanism for the origin of the dust cavities 
in transition disks \citep[see recent reviews by][]{Alexander2014, Owen2015}.
The key factor in this scenario is when the mass accretion rate of the disk becomes lower than 
the photoevaporation mass-loss rate as a result of to the radiation of the central star.
The accretion rate of $10^{-8}$ M$_{\odot}$/yr in HD~139614 \citep[][]{GarciaLopez2006} is at the
high end of the mass-loss rates ($10^{-8} - 10^{-10}$ M$_{\odot}$/yr)
in which photoevaporation starts to become relevant.

The inner radius of the dust gap of $\sim$3 AU is consistent with the critical radius of the gap 
that is expected to be opened by EUV photoevaporation for a 1.7 M$_\odot$ star \citep[$R_{\rm c, EUV}\simeq1.8 \times M_\ast/M_\odot$ AU, e.g.,][]{Alexander2014}.
However,  the very fact that we detect molecular gas inside 3 AU, the lack of a gap in the gas and the accretion rate of the source  
do not favor the EUV photoevaporation scenario,
unless we are seeing the gap just beginning to form through photoevaporation.
This has a very low probability given the EUV photoevaporation timescales (10$^5$ yr). 
Although EUV photoevaporation is probably not the dominant mechanism for the formation of the
dust gap and the gas density drop,  
it cannot be completely excluded because of the age of HD~139614 ($\sim$9 Myr).

X-ray photoevaporation has been suggested to be relevant for accreting transition disks because of its high mass-loss rates \citep[e.g.,][]{Owen2011,Owen2012}.
{Herbig Ae stars are, however, generally weak X-ray emitters \citep[e.g.,][]{Stelzer2006,Stelzer2009}}.
HD~139614 has been observed with XMM-Newton in the context of a large program investigating young stars and their protoplanetary disks
(PI. M. G\"udel).
Details of the observations and data reduction will be described in a future publication of that survey.
A first analysis of the XMM data indicate an unabsorbed X-ray flux of 
5.26$\times$10$^{-14}$   erg s$^{-1}$ cm$^{-2}$, which translates into an 
X-ray luminosity of 1.2$\times$10$^{29}$  erg s$^{-1}$.
Using this X-ray luminosity and the scaling relation for X-ray photoevaporation in \citet[][]{Alexander2014},
we obtain a mass-loss rate of 5.6$\times$10$^{-10}$ M$_{\odot}$/yr.
Since this value is much lower than the accretion rate, it is likely that X-ray photoevaporation (alone) is not 
the dominant mechanism responsible for the dust gap and gas density drop in the inner disk of HD~139614.
Furthermore, in the X-ray photoevaporation model, a gap in the gas of 5$-$6 AU width 
would be expected for a gas density drop $\delta_{\rm gas}$ ranging from 10$^{-2}$ to 10$^{-4}$
\citep[e.g., see Fig. 9 in ][]{Owen2011}. This X-ray photoevaporation-induced 
gap is larger than 2 AU upper limit we derive from our CRIRES CO observations\footnote{We note, however, that the X-ray photoevaporation models described here were mostly developed for 1 M$_\odot$ stars and may not be directly applicable to A-type stars.}.
Finally, the lack of [\ion{O}{i}] 6300 \AA~emission \cite[][]{Acke2005}
as a tracer of photoevaporation winds \cite[e.g.,][]{Font2004,ErcolanoOwen2010, Gorti2011, BaldovinSaavedra2012, Rigliaco2013}
does not support photoevaporation as the main gas-depletion process taking place in HD~139614 (see also \citet[][]{SiciliaAguilar2010,SiciliaAguilar2013,SiciliaAguilar2015} 
who argued that accreting transition disks are probably not caused by photoevaporation).

The interaction of a giant planet with its parent disk causes dramatic changes in the distributions of the gas and the dust in
 the disk, which alters the planet's orbital properties (planetary migration). 
A recent review is provided in \citet[][]{Baruteau2014}. 
For this discussion, we highlight a few key aspects: 
1) a giant planet is expected to open a gap in the gas with a width of a few AU\footnote{We note that 
the gap width and depth depend on the planet mass and disk physical properties such as the temperature and turbulent viscosity near the planet orbit.};
2) the gas surface density profile inside the planet orbit could have a variety of behaviors (be lower than in the outer disk or be radially decreasing, flat, or increasing);
3) the gas at radii close to the planet orbit can have velocities that deviate significantly
from the Keplerian speed;
4) the presence of giant gap-opening planets will generate pressure bumps in the disk that will trap dust particles (for instance at  radii 
immediately outside of the planet radius).

The width of the gap in the gas (and small dust grains) opened by a planet scales with the Hill radius ($R_{\rm H}$) of the planet.
The Hill radius of a planet of mass $M_{\rm p}$  at a distance $r_{\rm p}$ from a star of mass $M_*$ 
is defined by
\begin{equation}
\label{Hill_radius}
R_{H} = r_{\rm p}\left(\frac{M_{\rm p}}{3M_*}\right)^{1/3}.
\end{equation}
The gap opened  in the gas by a planet generally does not exceed five times the Hill radius \citep[e.g.,][]{Dodson-RobinsonSalyk2011}.
The mass of HD~139614 is 1.7 M$_{\odot}$.
If we assume that the planet is located at 4.5 AU, 
according to Eq.~\ref{Hill_radius}, 
planets more massive than 3.7 M$_{\rm J}$ would be expected to open gaps larger than 2 AU in the disk.
\citet[][]{Matter2016} presented hydrodynamical simulations
adapted to HD~139614 with the objective of exploring a single giant planet 
as a cause of the dust gap observed in the IR interferometry observations. 
We refer to that paper for the details of the modeling. 
The results of the hydrodynamical simulations after 100 000 orbits (1 Myr)  are the following:
1) planets of  1.7, 3, and 6.8 M$_{\rm J}$ located at 4.5 AU produce a gas gap of width 2, 3, and 4 AU respectively;
2) the surface density profile in the inner disk (R$<$ 3 AU)  changes as a result of  a 3 M$_J$
planet from an initial $R^{-1}$ to $R^{+0.6}$ 
profile, and features a reduction factor $\sim$10 relative to the outer disk.
{In the context of the planet-disk simulations,
if a planet is responsible for the dust gap and the gas density drop it should have a mass 
lower than 2 M$_{\rm J}$\footnote{We note that the planet mass depends on assumptions such as the disk viscosity. 
Models of \citet[][]{Matter2016} assumed $\alpha=0.006$.}.}

The exponent of +0.6 observed in the inner disk in the hydro-simulations is compatible with the 
CO observations. The gas density drop in the hydro-simulations is, however, at least a factor 10 
weaker than the $\delta_{\rm gas}$ suggested from the CO-rovibrational data. 
But hydrosimulations were run for 1 Myr while HD~139614 has an age $\sim$9 Myr. 
It is indeed possible that the inner disk has lost a significant fraction of its gas mass 
due to accretion onto the star and also photoevaporation, a process which is not included in the Matter et al. 
hydro-simulations\footnote{For a study of the interplay of photoevaporation and planet formation, see, for example, \citet[][]{Rosotti2013}.}.

Moreover, a 1 $-$ 2 M$_{\rm J}$ giant planet could be responsible for the observed ~3.5 AU-wide dust gap, 
while having a gas gap smaller than 2 AU.  
A planet can generate pressure bumps in the inner and outer edge of the gap,
which could trap particles  \citep[e.g,][]{Pinilla2012}. 
The location of these dust traps are at a radius smaller (for the inner disk) and larger (for the outer disk) than the 
inner and outer edge of the gas gap opened by the planet \citep[e.g.,][]{Pinilla2016}.
{The gas-dust interaction could thus produce gaps of different width for the dust and for the gas as observed in 
HD~139614.} 

 \citet[][]{Regaly2014} have computed the CO ro-vibrational lines
profiles expected for a disk with an embedded giant planet for stars with different masses and planets at different separations.
For a 2 M$_\odot$ star harboring a 10 M$_J$ planet at 3 and 5 AU separations,
they have predicted asymmetric CO ro-vibrational lines with distortions on the order of 
10\% with respect to the symmetric line-profiles. 
The composite $^{12}$CO line-profile is symmetric and distortions are not detected at the 1$\sigma$ level. 
This indicates that the mass of the planet, if present, must be lower than 10 M$_J$, which is consistent 
with the upper limit of 2 M$_J$ from the lack of a gap in the gas distribution.

{In addition to dynamical interaction with embedded planets and photoevaporation,
various mechanisms to trap dust particles have been proposed 
to explain the gaps, lopsided shapes, and ringed structures observed recently
by ALMA in disks  \citep[e.g.,][]{Perez2014,vanderMarel2015,HLTauALMA,Andrews2016}.
Scenarios include
a magnetical origin such as 
radial pressure variations due to MHD turbulence \citep[zonal flows ][]{Johansen2009},
radial variations in the disk resistivity \citep[e.g.,][]{Flock2015,Lyra2015},
vortex formation due to instabilities at the edge of the dead-zone \citep[e.g.,][]{Regaly2013,Faure2015}.
Or a chemical origin such as the
changes in opacity that are due to migrating solids reaching the condensation fronts of volatiles 
\citep[e.g.,][]{Cuzzi2004,Brauer2008,Banzatti2015,Okuzumi2016}.
While these scenarios could in part explain the dust features discovered in disks,
it is not clear whether they are able to explain the gas density drop that the
CO ro-vibrational emission reveals in HD~139614 (and other transition disks).
The low gas column-density detected inside the dust cavities of transition disks,
combined with the low optical depth of the dust can, however, impact the ionization structure of the disk and 
have an effect on MHD phenomena.
}

\subsection{A dust trap in the inner disk?}
\label{inner_trap}
{The possibility of an increasing gas surface density profile in the inner 6 AU 
has interesting consequences from the point of view of gas and dust evolution.
If the gas surface density profile within 6~au of HD~139614
is indeed the result of a giant planet of mass between 1 and 2 M$_J$ at 4.5 AU, 
then pressure bumps would be present at the two edges of the planetary gap.
Dust can accumulate, grow, and be trapped in these pressure bumps \citep[e.g.,][]{Fouchet2010,Pinilla2012}. 
The trapping of the particles depends on how well they are coupled to the gas and therefore, it depends on their size and local gas surface density. 
Quantitatively, particles with size $a_{\rm opt}=2\Sigma_{gas}/\pi\rho_d$ 
drift the fastest toward the regions of high pressure and are the most efficient particles to trap \citep[e.g.,][]{Fouchet2010}.
Here $\rho_d$ is the internal density of the dust grains, typically $1 - 3$ g cm$^{-3}$, and $\Sigma_g$ is the gas surface density.
Using $N_H=10^{19} - 10^{21}$ cm$^{-2}$ ($2.4\times10^{-5} - 2.4\times10^{-3}$ g cm$^{-2}$)\footnote{Here the hydrogen mass fraction of gas with solar composition is assumed (0.7) to calculate the gas surface density in g cm$^{-2}$.} for the inner disk of HD~139614, 
we obtain that sub-micron and micron-sized grains are trapped at the inner gap edge. 
This can lead to an increasing dust surface density with radius in the inner disk, as the analysis of IR interferometry observations suggested \citep[][]{Matter2016}.}

\subsection{Gas surface density in the inner disk and accretion rate onto the star}
Many of the bright transition disks have stellar accretion rates ($\dot{M}_{\star}$) between 
10$^{-9}$ and 10$^{-8}$ M$_\odot$ yr$^{-1}$, very similar to accretion rates of classical T-Tauri stars 
with primordial gas-rich disks \citep[e.g.,][]{Manara2014}. However, the surface density of the gas in 
the cavities of transition disks  typically ranges from 10$^{-3}$ to 1 g cm$^{-2}$ (see Table 6), 
which is several orders of magnitude lower than at similar locations in primordial disks. 
Also quite surprisingly, transition disks with similar $\dot{M}_{\star}$ can have very 
different gas surface densities in their inner regions. This is the case of IRS 48 and 
RXJ1615, both of which have $\dot{M}_{\star} \sim {\rm a\,few} \times10^{-9}$ M$_\odot$ yr$^{-1}$, 
but estimated gas surface densities inside their cavities that differ by roughly two orders of magnitude, 
even though the cavities are of similar size in the gas \citep[][]{vanderMarel2015,vanderMarel2016}. 
These all seem to be counter intuitive facts when considering that in a steady-state model of a protoplanetary 
disk, the stellar accretion rate and the disk accretion rate should take similar values throughout the disk.

The disk accretion rate, $\dot{M}_{\rm disk}$, is related to the surface density $\Sigma$ 
and radial velocity $v_R$ of the gas at radius $R$ through $\dot{M}_{\rm disk}=-2\pi\,R\,v_R(R)\, \Sigma(R)$. 
If we now assume that $\dot{M}_{\rm disk} \sim \dot{M}_{\star}$, then at a radius of 1 AU 
and for an accretion rate of 10$^{-8}$ M$_\odot$ yr$^{-1}$, we find that $|v_R|$ should 
extend from about 70 km s$^{-1}$ down to 0.07 km s$^{-1}$ for  $\Sigma$ in the range 
$[10^{-3}-1]$ g cm$^{-2}$. At 1 AU, the Keplerian velocity ($v_{\rm K}$) is about 
30 km s$^{-1}$, which means that for the transition disks with the lowest densities inside 
the cavity, the gas radial velocity can be comparable to the Keplerian velocity. The above range 
of radial velocities is at odds with typical values expected in classical viscous disk models.
These models tell us that 
$|v_R| \sim \alpha h^2 v_{\rm K}$, were $h$ is the disk aspect ratio and $\alpha$ is  the 
dimensionless turbulent viscosity of the disk. If the disk magnetic field is able to sustain 
vigorous turbulence in the cavities of transition disks, then $|v_R|$ can reach at most  
$\sim 10^{-5} v_{\rm K}$. This is at least two orders of magnitude lower than the above 
range of radial velocities obtained when assuming $\dot{M}_{\rm disk} \sim \dot{M}_{\star}$.

Another way to visualize the problem is to think in terms of the gas column density expected 
for the accretion rates reported, and the CO ro-vibrational line-profiles that such column densities would produce.
For HD~139614,
the $\dot{M}_{\rm star} =10^{-8}$ M$_\odot$ yr$^{-1}$ would suggest an $N_H$ of $10^{25}-10^{26}$ cm$^{-2}$ 
at $R\leq$1 AU for $\alpha$ ranging from 10$^{-2}$ to 10$^{-3}$.
These gas columns are high enough for the CO to self-shield against UV photodissociation 
and produce strong CO ro-vibrational emission from $R<1$ AU (not seen in our CRIRES spectrum).
For example, if $N_H \sim 10^{25}$ cm$^{-2}$ at $R<1$ AU in a disk without dust in the cavity (best case for UV-photodissociation)
around a $T_{\rm eff}$=10000 K central star (i.e., brighter in UV than HD~139614)
$N_{\rm CO}$ would be $10^{21}$ cm$^{-2}$ at $R<$ 1 AU \citep[see Fig. 8 in][]{Bruderer2013},
a value much higher than the 5$\sigma$ limit of $N_{\rm CO}=5\times10^{15}$ cm$^{-2}$
from our CRIRES data.
According to the \citet[][]{Bruderer2013} models (for a $T_{\rm eff}$=10000 K), 
to have $N_{\rm CO} < 10^{15}$ cm$^{-2}$,
$N_H$ needs to below $10^{22}$ cm$^{-2}$ at $R<1$ AU for an inner disk {\it without} dust.
For HD~139614,
which has a T$_{\rm eff}\sim7800$ K (and lower UV, accordingly) and some dust in the inner disk\footnote{The dust at R$<$2.5 AU is optically thin at 4.7 $\mu$m, but in the UV it has higher opacity,
thus helping to shield CO.}, $N_H$ should be even lower to enable the efficient photodissociation of CO. 

\par There are probably two ways to solve these problems. 
The first is to concede that the 
inner parts of transition disks do not have to be in a steady state with $\dot{M}_{\rm disk} \sim 
\dot{M}_{\star}$.
This implies that measured accretion rates {\it should not} be used as direct proxies to 
derive the amount of gas left in the cavities of transition disks. 
The second way is to conceive that the radial flow of gas inside the cavities of transition disks is not necessarily 
driven by turbulent accretion, but by interactions with one or more massive companion(s) inside 
the cavity. HD~142527 may be a pioneering example of the latter possibility. \citet[][]{Casassus2015} 
showed that a highly inclined stellar companion to HD 142527 can generate fast radial flows inside the cavity of its transition disk.
\citet[][]{Pontoppidan2011} detected in HD~142527 a $^{12}$CO ro-vibrational spectrum and 
spectro-astrometry signature displaying asymmetries, indicating non-Keplerian contributions to the emission.
Additional resolved observations of the gas kinematics inside the cavities of transition disks will 
help assess the occurrence of this second scenario.
In our case of HD~139614, the $^{12}$CO composite line-profile is symmetric, smooth, and consistent with 
Keplerian motion, which suggests the first scenario, namely that $\dot{M}_{\rm disk}$ (1 AU) is most likely different  
from $\dot{M}_{\star}$.

\section{Summary and conclusions}
{
We have obtained VLT/CRIRES high-resolution spectra ($R\sim90\,000$) of CO ro-vibrational emission at 4.7 $\mu$m
in HD139614, an accreting (10$^{-8}$ M$_\odot$ yr$^{-1}$) Herbig Ae star 
with a (pre-) transition disk that is 
characterized by a dust gap between 2.3 and 6 AU and a dust density drop $\delta_{\rm dust}$ of 10$^{-4}$ at R$<$ 6 \citep[][]{Matter2016}.
We have detected $\upsilon=1\rightarrow$0~$^{12}$CO, $^{13}$CO, C$^{18}$O, C$^{17}$O, and  
$\upsilon=2\rightarrow$1~$^{12}$CO ro-vibrational emission. 
The lines observed are consistent with disk emission and thermal excitation. 
We find the following:

\begin{enumerate}
\item The $\upsilon=1\rightarrow$0 $^{12}$CO spectrum indicates that there is gas from 1 AU up to 15 AU,
and that there is no gap in the gas distribution. 
If a gap is present in the gas (i.e., a region devoid of gas) 
then it should have a width smaller than 2 AU.

\item The spectra of $^{13}$CO and C$^{18}$O $\upsilon=1\rightarrow$0 emission are 
on average colder and emitted farther out in the disk ($R>$6 AU) than the $^{12}$CO $\upsilon=1\rightarrow$0 emission.
Keplerian flat-disk models clearly show that a drop in the gas density $\delta_{\rm gas}$ of a factor of at least 100 at $R<5-6$ AU is needed 
to describe simultaneously the line-profiles and rotational diagrams of the three CO isotopologs.
Models without a gas density drop produce C$^{18}$O and $^{13}$CO lines that are too wide
and warm to be compatible with the data.
If the gas-to-dust mass ratio is equal to 100 in the outer disk,
the gas depletion factor $\delta_{\rm gas}$ could be as high as 10$^{-4}$. 
Moreover, we find that the gas surface density profile in the inner 6 AU of the disk is flat or increases with radius.
\end{enumerate}

The presence of molecular gas inside 6 AU and the weak X-ray luminosity
do not favor photoevaporation as the main mechanism responsible for the inner disk structure of HD~139614.
The gas density drop, a flat or increasing gas surface density profile at $R<6$ AU, 
combined with the non-detection of a gap in the gas wider than 2 AU,
suggest the presence of a single Jovian-mass planet inside the dust gap. 
If a giant planet is indeed responsible for the transition disk shape of HD~139614, 
then its location would be at around 4 AU and its mass would be lower than two Jupiter masses.
Furthermore, if a small gap in the gas (due to a planet) were to be present, 
a gas surface density profile that increases with radius in the inner disk 
might lead to a dust trap at the gap inner edge for sub-micron and micron sized grains,
which could explain that the dust surface density increases with radius at $R<2.5$ AU,
as found in IR interferometry observations.

We constrained the gas column density  between 1 and 6 AU to 
$N_H=3\times10^{19} - 10^{21}$ cm$^{-2}$ ($7\times10^{-5} - 2.4\times10^{-3}$ g cm$^{-2}$) assuming $N_{\rm CO} = 10^{-4} N_H$.
We derived a 5$\sigma$ upper limit on the CO column density at $R<$1 AU $N_{\rm CO}=5\times10^{15}$ cm$^{-2}$,
which suggests an $N_H<5\times10^{19}$ cm$^{-2}$ at $R<$1 AU .
The gas surface density in the disk of HD~139614 at $R\leq1$ AU and at $1<R<6$ AU is significantly lower than the surface density
that would be expected for  the accretion rate of HD~139614, assuming a standard viscous $\alpha$-disk model.  
Our result, and the low gas surface densities reported in the inner disks of other transition disks,
suggests that  stellar accretion rates should {\it not} be used as direct proxies to derive the amount of gas 
left inside the dust cavities of transition disks.
An investigation of the topology of the magnetic fields of young stars with transition disks
is needed to help address the question of the differences between the accretion rate and the inner disk
gas surface density.

We have discussed the ensemble of transition disks with current constraints for the gas surface density inside the dust cavity. 
The sample shows that, 
in the majority of the sources, 
the drop in the dust density is larger than the drop in the gas density ($\delta_{\rm dust}<\delta_{\rm gas}$).
This suggests that dust is depleted faster than gas in the inner disk.

The number of transition disks with a complete set of multi-wavelength observations of gas and dust (ALMA, CRIRES, {\it Herschel}, {\it Spitzer}, {VLTI}, SEDs, HiCiAO, VLT/SPHERE, and GPI) is growing. 
A homogenous  multi-wavelength  and multi-technique modeling of gas and dust observations in transition
disks would be of great help to understand the variety of gas and dust structures that these disks have, 
and to  study the possible links to planet formation.
In that respect, 
it would be of great help to have spatially resolved measurements of the dust and the rotational transitions
of CO isotopologs in the sub-mm for HD~139614 (for example with ALMA\footnote{At the time of writing, HD~139614 has not been observed  with ALMA.}).

}

\begin{acknowledgements}
A. Carmona was partly supported by the Spanish Grant AYA 2011-26202.
A. Carmona, A. K\'ospal and Zs. Reg\'aly were partly supported by the Momentum grant of the MTA CSFK Lend\"ulet Disk Research Group of the Hungarian Academy of Sciences.
A. Carmona and C. Pinte acknowledge funding from the
European Commission's 7$^\mathrm{th}$ Framework Program (EC-FP7)
(contract PERG06-GA-2009-256513) and from
Agence Nationale pour la Recherche (ANR) of France under contract
ANR-2010-JCJC-0504-01.
A. Carmona acknowledges financial support by the European Southern Observatory visitors program.
The research leading to these results has received funding from the EC-FP7 under grant agreement no 284405. 
C. Eiroa is partly supported by the Spanish Grant AYA 2014-55840-P
L.A. Cieza was supported by CONICYT-FONDECYT grant number 1140109 and the Millennium Science Initiative (Chilean Ministry of Economy), through grant ``Nucleus RC130007''.
A.C would like to thank C.P. Folsom for comments on the manuscript.
A.C and C.B would like to thank G. Lesur and S. Casassus for discussions on protoplanetary disks.

\end{acknowledgements}

\bibliographystyle{aa} % style aa.bst
\bibliography{Biblio} % your references Yourfile.bib

\Online

\begin{appendix}

\begin{table*}
\section{Measured line centers, {\it FWHM},  integrated fluxes, and upper limits.}
\caption{CO ro-vibrational line fluxes, upper limits, and average line ratios.}
\begin{center} 
{\scriptsize
\begin{tabular}{lccccccccl}
\hline
\hline
line & $\lambda_{\rm 0}$ & $\lambda_{\rm obs}~^a$ & $\Delta$ V& Integrated flux$~^b$ & {\it FWHM}$~^b$  \\
       & [nm]               & [nm]       & [km s$^{-1}$]   & [10$^{-15}$ erg s$^{-1}$ cm$^{-2}$] & [km s$^{-1}$]\\
\hline
$^{12}$CO $1\rightarrow0$ P(6) &  4717.69 &   4717.73 $\pm$   0.01 & 2.5 $\pm$ 0.6 &  13.0 $\pm$   0.5 &  13.7 $\pm$   0.4 \\
$^{12}$CO $1\rightarrow0$ P(7) & 4726.73  &   4726.73 $\pm$   0.01 & 0.4 $\pm$ 0.6 &  15.2 $\pm$   0.4 &  14.7 $\pm$   0.3 \\
$^{12}$CO $1\rightarrow0$ P(9) & 4745.13  &  4745.15 $\pm$   0.01  & 1.3 $\pm$ 0.6 &  14.3 $\pm$   0.4 &  13.5 $\pm$   0.4 \\
$^{12}$CO $1\rightarrow0$ P(10) & 4754.50 & 4754.52 $\pm$   0.01  & 1.3 $\pm$ 0.6 &  17.5 $\pm$   0.4 &  15.1 $\pm$   0.3 \\
$^{12}$CO $1\rightarrow0$ P(11) & 4763.98 & 4764.01 $\pm$   0.01  & 1.9 $\pm$ 0.6 &  14.8 $\pm$   0.4 &  13.3 $\pm$   0.4 \\
$^{12}$CO $1\rightarrow0$ P(12) & 4773.58 & 4773.64 $\pm$   0.01  & 3.8 $\pm$ 0.6 &  13.5 $\pm$   0.3 &  13.4 $\pm$   0.2 \\
$^{12}$CO $1\rightarrow0$ P(13) &  4783.29 &  4783.33 $\pm$  0.01 & 2.5 $\pm$ 0.6 &  14.7 $\pm$   0.4 &  14.0 $\pm$   0.3 \\
$^{12}$CO $1\rightarrow0$ P(15) & 4803.07  & 4803.10 $\pm$   0.01 & 1.9 $\pm$ 0.6 &  13.3 $\pm$   0.5 &  13.3 $\pm$   0.4 \\
& & & & {\bf 14.5 $\pm$ 0.5} & {\bf 13.9 $\pm$ 0.3} & {\bf average} \\\\
$^{13}$CO $1\rightarrow0$ R(6) &  4715.04 &  4715.10 $\pm$   0.01 & 3.8 $\pm$ 0.6 &   3.5 $\pm$   0.2 &  11.8 $\pm$   0.5 \\
$^{13}$CO $1\rightarrow0$ R(5) &  4722.70 &  4722.72 $\pm$   0.01 & 1.3 $\pm$ 0.6 &   4.3 $\pm$   0.3 &  13.1 $\pm$   0.6 \\
$^{13}$CO $1\rightarrow0$ R(4) &  4730.47 &  4730.49 $\pm$   0.01 & 1.3 $\pm$ 0.6 &   4.3 $\pm$   0.2 &  13.4 $\pm$   0.5 \\
$^{13}$CO $1\rightarrow0$ R(2) &  4746.31 &  4746.36 $\pm$   0.03 & 3.2 $\pm$ 1.9 &   3.3 $\pm$   4.4 &  12.6 $\pm$   2.8 \\
$^{13}$CO $1\rightarrow0$ R(0) &  4762.56 &  4762.60 $\pm$   0.01 & 2.5 $\pm$ 0.6 &   2.4 $\pm$   0.3 &  11.2 $\pm$   0.8 \\
$^{13}$CO $1\rightarrow0$ P(1) &  4779.22 &  4779.27 $\pm$   0.01 & 3.1 $\pm$ 0.6 &   2.6 $\pm$   0.2 &  12.3 $\pm$   0.9 \\
$^{13}$CO $1\rightarrow0$ P(2) &  4787.71 &  4787.71 $\pm$   0.02 & 0.0 $\pm$ 1.3 &   4.3 $\pm$   0.5 &  15.9 $\pm$   1.6 \\
$^{13}$CO $1\rightarrow0$ P(4) + &  4804.99 & - & - & 8.4 $\pm$   0.8 &  -\\
~~~$^{12}$CO $2\rightarrow1$ P(9)\\
& & & & {\bf 3.5 $\pm$ 0.6} & {\bf 12.9 $\pm$ 0.6} &{\bf average} \\\\
C$^{18}$O $1\rightarrow0$ R(7) & 4716.46 & 4716.50 $\pm$   0.02 & 2.5 $\pm$ 1.3 &   1.2 $\pm$   0.2 &  11.1 $\pm$   1.6 \\
C$^{18}$O $1\rightarrow0$ R(6) & 4724.03 & 4724.04 $\pm$   0.02 & 0.6 $\pm$ 1.3 &   1.6 $\pm$   0.2 &  10.8 $\pm$   1.2 \\
C$^{18}$O $1\rightarrow0$ R(5) & 4731.70 & 4731.75 $\pm$   0.02 & 3.2 $\pm$ 1.3 &   1.4 $\pm$   0.3 &  10.4 $\pm$   1.2 \\
C$^{18}$O $1\rightarrow0$ R(3) & 4747.34 & 4747.38 $\pm$   0.03 & 2.5 $\pm$ 1.9 &   1.2 $\pm$   0.3 &  11.4 $\pm$   2.0 \\
C$^{18}$O $1\rightarrow0$ R(2) & 4755.31 & 4755.36 $\pm$   0.02 & 3.2 $\pm$ 1.3 &   0.9 $\pm$   0.2 &   7.9 $\pm$   1.2 \\
C$^{18}$O $1\rightarrow0$ R(1) & 4763.38 & $^c$                 		&		-	    &   1.5 $\pm$   0.2 &  $^c$ \\
C$^{18}$O $1\rightarrow0$ R(0) & 4771.56 & 4771.62 $\pm$   0.03 & 3.8 $\pm$ 1.9 &   0.9 $\pm$   0.3 &  10.3 $\pm$   2.2 \\
C$^{18}$O $1\rightarrow0$ P(1) & 4788.22 & 4788.22 $\pm$   0.02 & 0.0 $\pm$ 1.3 &   0.8 $\pm$   0.2 &   5.5 $\pm$   1.1 \\
C$^{18}$O $1\rightarrow0$ P(3) & 4805.29 & $^c$ 				&		-	    &   1.1 $\pm$   0.4 &   $^c$ \\
& & & & {\bf 1.2 $\pm$ 0.1} & {\bf 9.6 $\pm$ 0.8} & {\bf average} \\\\
C$^{17}$O $1\rightarrow0$ R(0) & 4716.96&  - &  - & $<$ 0.5 & -\\
C$^{17}$O $1\rightarrow0$ P(1) & 4733.62 & - & - & $<$ 0.5 & -\\
C$^{17}$O $1\rightarrow0$ P(3) & 4750.71 & 4750.78 $\pm$   0.04 & 4.4 $\pm$ 2.5 &   0.6 $\pm$   0.2 &   9.9 $\pm$   2.3 \\
C$^{17}$O $1\rightarrow0$ P(4) & 4759.40 & - & - & $<$ 0.6 & -\\
C$^{17}$O $1\rightarrow0$ P(9) & 4804.53 &4804.56 $\pm$   0.02 & 1.9 $\pm$ 1.2 &   0.9 $\pm$   0.3 &   7.4 $\pm$   2.3 \\
& & & & {\bf 0.8$\pm$ 0.2} & {\bf 9 $\pm$ 1.6} & {\bf average} \\\\
$^{12}$CO $2\rightarrow1$ R(0) & 4715.72 & 4715.75 $\pm$   0.01 & 1.9 $\pm$ 0.6 &   $>$0.7  &  $^d$ \\
$^{12}$CO $2\rightarrow1$ P(1) & 4732.65 & 4732.70 $\pm$   0.01 & 3.2 $\pm$ 0.6 &   1.8 $\pm$   0.2 &  12.2 $\pm$   1.0 \\
$^{12}$CO $2\rightarrow1$ P(3) & 4750.01 & 4750.05 $\pm$   0.01 & 2.5 $\pm$ 0.6 &   2.7 $\pm$   0.2 &  11.4 $\pm$   0.7 \\
$^{12}$CO $2\rightarrow1$ P(4) & 4758.86 & 4758.91 $\pm$   0.01 & 3.1 $\pm$ 0.6 &   4.0 $\pm$   0.2 &  12.7 $\pm$   0.5 \\
$^{12}$CO $2\rightarrow1$ P(6) & 4776.89 & 4776.89 $\pm$   0.03 & 0.0 $\pm$ 1.9 &  $>$3.0    & $^e$\\
$^{12}$CO $2\rightarrow1$ P(7) & 4786.07 &  4786.12 $\pm$   0.01& 3.1 $\pm$ 0.6 &   5.2 $\pm$   0.3 &  14.8 $\pm$   0.8 \\
$^{12}$CO $2\rightarrow1$ P(9) &  4804.78 &\multicolumn{2}{l}{blend $^{13}$CO $1\rightarrow0$ P(4)}  &  - \\
& & & &~{\bf 3.4$\pm$ 0.7} &{\bf  12.8 $\pm$ 0.7} & {\bf average} \\\\
$^{12}$CO $3\rightarrow2$ R(7)   &  4718.52 &  - &  - & $<$ 0.5 & -\\
$^{12}$CO $3\rightarrow2$ R(6)   &  4726.31 &  - &  - & $<$ 0.6 & -\\
$^{12}$CO $3\rightarrow2$ R(5)   &  4734.20 &  - &  - &  $<$ 0.4 & -\\
$^{12}$CO $3\rightarrow2$ R(3)  &   4750.31 &  - &  - &  $<$ 0.5 & -\\
$^{12}$CO $3\rightarrow2$ R(2)  &  4758.52 &  - &   - & $<$ 0.8 & -\\
$^{12}$CO $3\rightarrow2$ R(0)  &  4775.28 &  - &  - &  $<$ 0.8 & -\\
$^{12}$CO $3\rightarrow2$ P(1)  &  4792.48 &  - &  - &  $<$ 0.7 & -\\
$^{12}$CO $3\rightarrow2$ P(2)  &  4801.24 &  - &  - &  $<$ 1.0 & -\\
& & & &~{\bf $<$ 0.6} & - & {\bf average} \\\\
& & {\bf average $\Delta$V all lines} &   {\bf 2.1$\pm$1.2} & & \\\\
\hline
&&\multicolumn{3}{c}{\it average line ratios}\\[1mm]
\hline
&&\multicolumn{2}{l}{$^{13}$CO $1\rightarrow0$ ~/~ $^{12}$CO $1\rightarrow0$ }& 0.24 $\pm$ 0.05\\ 
&&\multicolumn{2}{l}{C$^{18}$O $1\rightarrow0$ ~/~ $^{12}$CO $1\rightarrow0$ }& 0.08 $\pm$ 0.01\\
&&\multicolumn{2}{l}{C$^{17}$O $1\rightarrow0$ ~/~ $^{12}$CO $1\rightarrow0$ }& 0.05 $\pm$ 0.01\\
&&\multicolumn{2}{l}{$^{12}$CO $2\rightarrow1$  ~/~ $^{12}$CO $1\rightarrow0$ }& 0.23 $\pm$ 0.05\\
&&\multicolumn{2}{l}{$^{12}$CO $3\rightarrow2$  ~/~ $^{12}$CO $1\rightarrow0$ }& $<$0.04\\
\hline
\end{tabular}
}
\tablefoot{
Average line ratios were calculated from the average line fluxes. $^a$ Line centers are measured in the barycentric and radial velocity corrected spectra. 
Their value is the center and 3$\sigma$ error of the Gaussian fit to the line.
$^b$ Integrated flux and {\it FWHM} of the Gaussian fit and 1$\sigma$ error; upper limits of the integrated line flux are 3$\sigma$ assuming a line width of 10 km s$^{-1}$;
the error in the average is the maximum between the standard error of the mean and $\frac{\sqrt{\sum{\sigma_i^2}}}{N}$.
$^c$ Detection but Gaussian fit not possible; 
$^d$ No meaningful {\it FWHM} due to the low sigma detection ;
$^e$ Detected but severely affected by the atmospheric transmission.
}
\label{table_fluxes}
\end{center}
\end{table*}

\clearpage

\section{Flat disk model}
\label{flat_disk_model}
\begin{figure*}
\begin{center}
\begin{tabular}{ccc}
\includegraphics[width=0.32\textwidth]{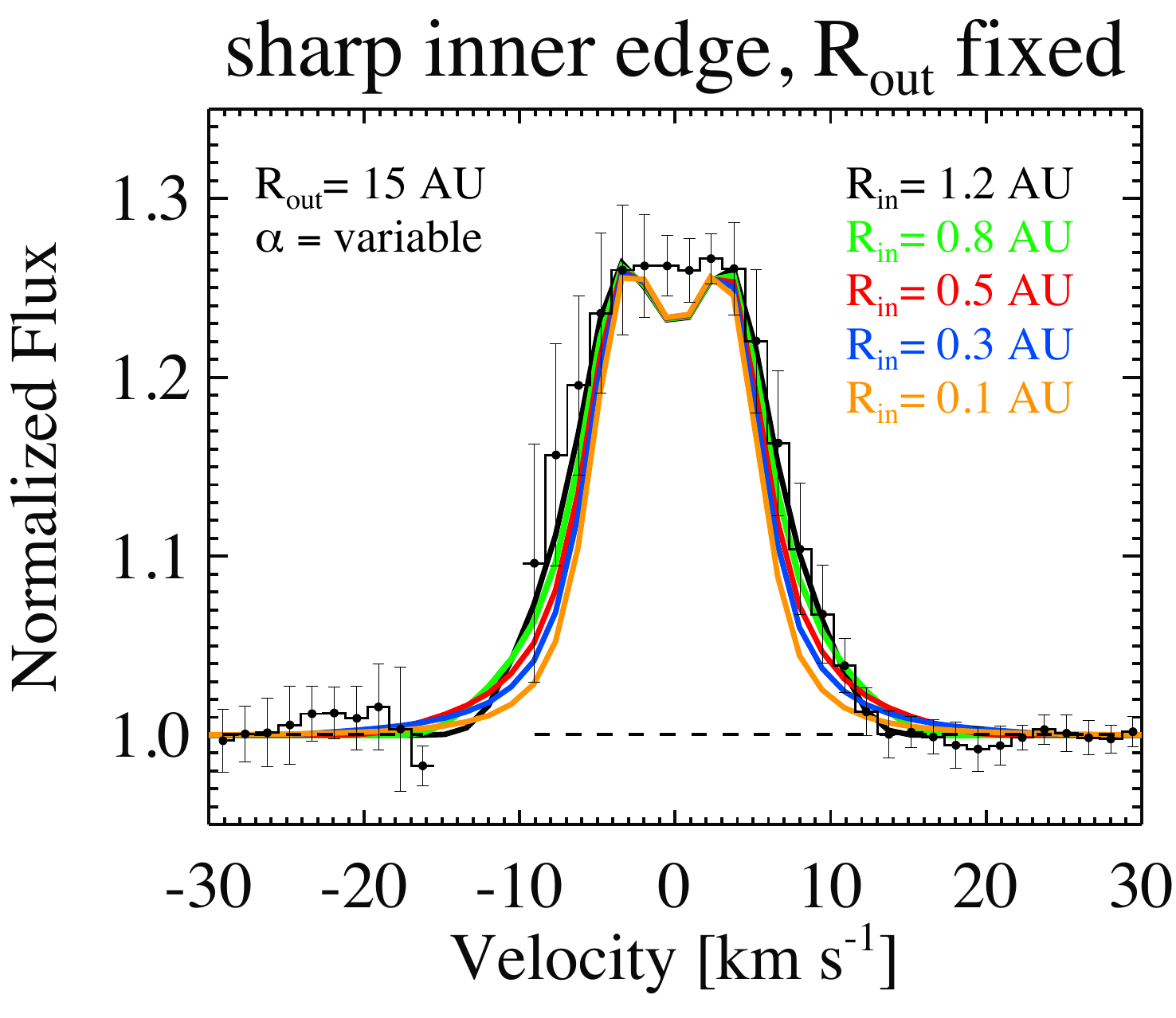} &
\includegraphics[width=0.32\textwidth]{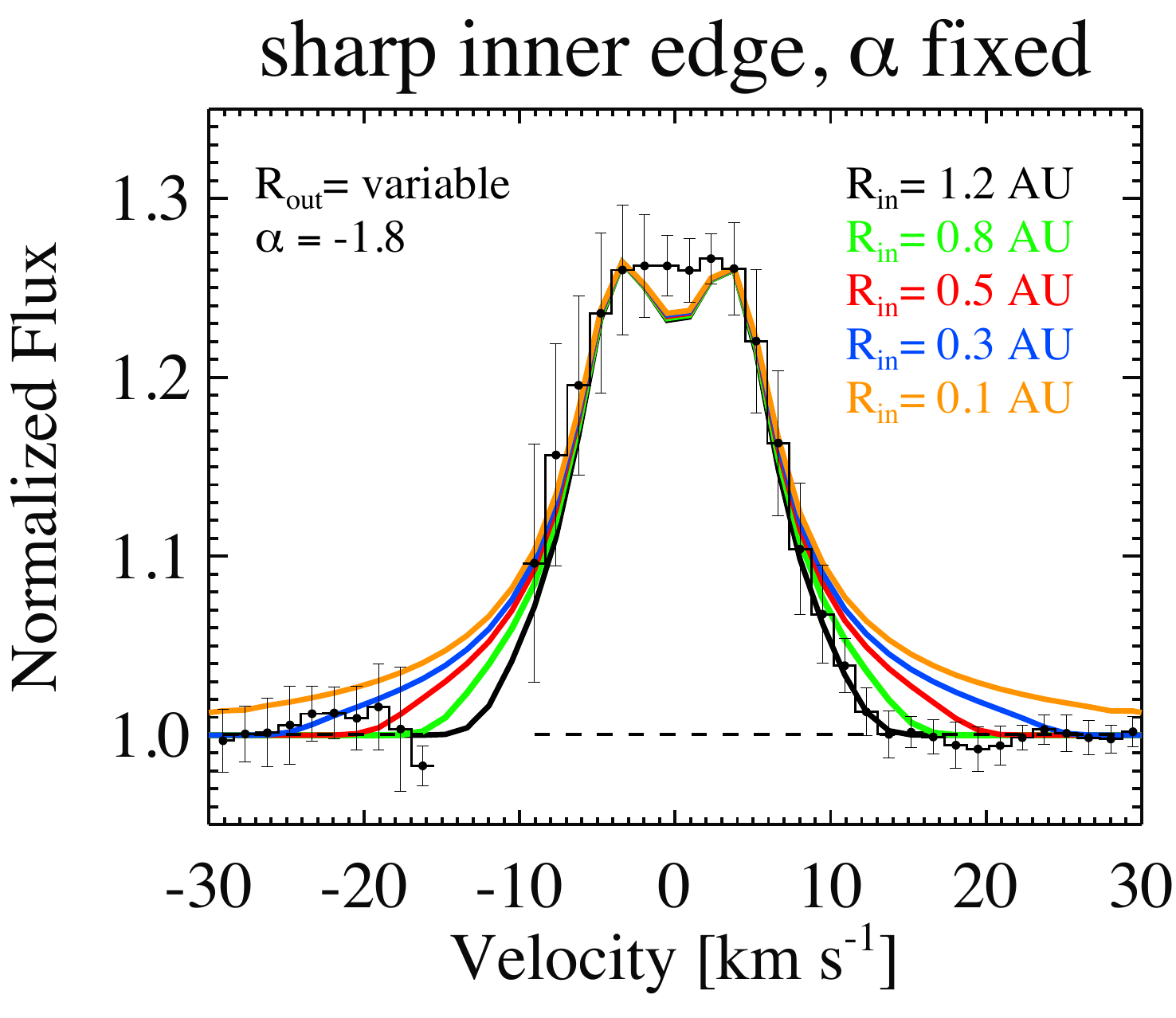} &
\includegraphics[width=0.32\textwidth]{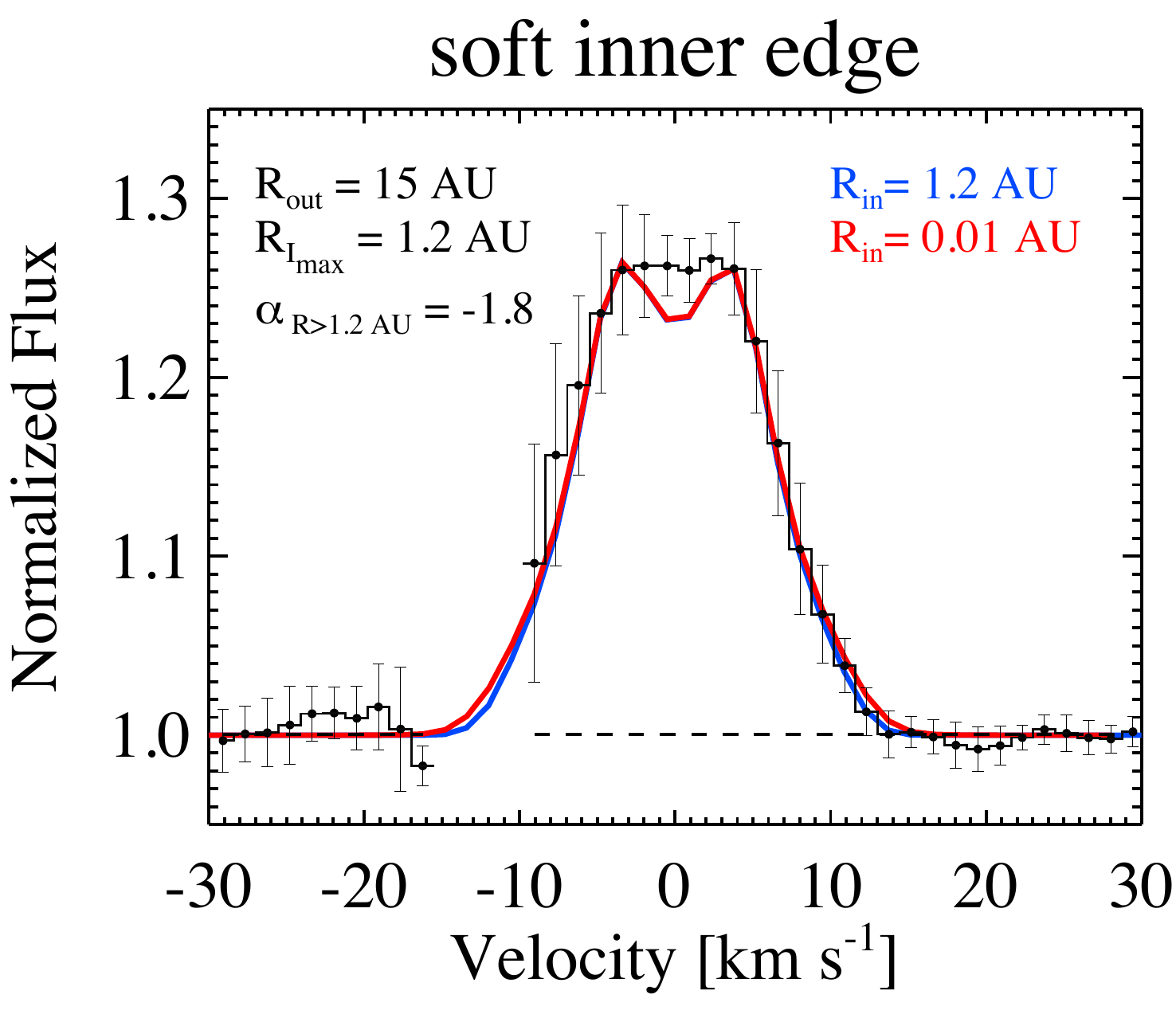} \\
{\bf\large~~~~~~a)} & {\bf\large~~~~~~b)} & {\bf\large~~~~~~c)}\\[3mm]
\end{tabular}
\caption{Line-profiles predicted for models around the best solution of the power-law intensity model: 
{\bf a)} effect of varying the inner radius with the outer radius fixed;
{\bf b)} effect of varying the inner radius keeping $\alpha$ fixed.
In the models, 
$\alpha$ or $R_{\rm out}$ are adjusted such that $I(R_{\rm out}) = 0.01\times I(R_{\rm in})$;
{\bf c)} line-profile for an intensity distribution with a sharp increase at 1.2 AU (in blue) and
the line-profile of an intensity distribution that grows as a power-law from 0.01 AU to 1.2 AU (in red).
In both models the intensity decreases with $\alpha=-1.8$ at $R>$ 1.2 AU.
Error bars on the composite spectrum are 1$\sigma$.
}
\label{inner_radius}
\end{center}
\end{figure*}

The computation of a flat disk model starts by calculating the expected integrated line flux 
of an annulus at radius $R$. This is done by multiplying the intensity
by the solid angle of the annulus projected by the inclination $i$:
\begin{equation}
F(R) = I(R)\,\Omega(R)_{\rm projected}.
\end{equation}
Here $R$ is the mid-point of two grid points in the radial grid
\begin{equation}
R=0.5\,(R_j + R_{j+1}),
\end{equation} 
and 
\begin{equation}
\Omega(R)_{\rm projected} = \pi~{\rm cos}(i)~(R_{j+1}^2 - R_j^2)/ D^2 ,
\end{equation}
where D is the distance to the source.
The integrated line flux for each cell in the azimuthal direction is then determined by dividing the total  integrated flux of the annulus 
by the number of points of the azimuthal grid ($N_{\rm \theta}$):
\begin{equation}
F(R,\theta) = F(R)/N_{\rm \theta}.
\end{equation}
We used 50 to 100 points in the radial direction and 1000 points in the azimuthal direction.
The local line-profile $\phi$ of a grid point in $R$ and $\theta$ is then obtained by convolving the integrated flux of the cell with 
a normalized Gaussian kernel, with a {\it FWHM} equal to the spectral resolution convolved with the turbulent and thermal broadening:
\begin{equation}
\phi (R,\theta,\nu) = F(R,\theta)*\phi_{\rm Gauss}(\nu)
\end{equation}
The line-profile of each cell in the azimuthal direction $\theta$ is then velocity shifted to the expected local Keplerian velocity shift 
\begin{equation}
\Delta V = {\rm cos}(\theta)~{\rm sin}(i)\sqrt{\frac{GM_\star}{R}} ,
\end{equation}
thus obtaining a $\phi (R,\theta,\nu)_{\rm shifted}$ for each cell.

When no slit effects are taken into account, the 1D spectrum of the whole disk 
is obtained by summing the contributions of each azimuthal cell in each annulus:
\begin{equation}
\phi (R,\nu) = \sum^{N_{\rm \theta}}\phi (R,\theta,\nu)_{\rm shifted},
\end{equation}
and summing the spectra of all the annuli in the radial direction
\begin{equation}
\phi (\nu) = \sum^{N_{\rm R}}\phi (R,\nu).
\end{equation}

To generate 3D channel maps the $\phi (R,\theta,\nu)_{\rm shifted}$ of each cell is 
sampled in a Cartesian data cube with coordinates $X,Y,\nu$ where
\begin{equation}
X=R~{\rm cos}(\theta),
\end{equation}
\begin{equation}
Y=R~{\rm sin}(\theta)~{\rm cos}(i).
\end{equation}
When the effect of the slit is taken into account,
the 1D spectrum is extracted from the 3D $(X,Y,\nu)$ channel map data cube 
generated by the model.
First, the image in each velocity channel is convolved with a Gaussian beam of {\it FWHM} 206 mas to model 
the spatial resolution.
Then,  to simulate the effect of the slit,
the 3D data-cube is rotated to account for the position angle of the disk on the sky 
and the slit orientation.
Then a 2D spectrum is obtained from the 3D data by summing the pixels inside a 0.2" vertical aperture.
This 2D disk model spectrum is scaled such that
in the extracted 1D disk spectrum, 
the peak of the line is equal to the peak of the flux in the normalized observed composite 1D spectrum. 

To calculate the spectroastrometric signature,
a synthetic star+disk 2D spectrum is created 
by adding to the 2D disk spectrum a 2D star spectrum broadened by the {\it PSF-FWHM}. 
The 2D star spectrum is constructed such that the continuum in the extracted 1D spectrum is equal to 1.
The model 2D star + disk spectrum is finally re-binned in the spatial and spectral directions such that its spatial and
spectral pixel scales are the same as the CRIRES data.
With this synthetic 2D star + disk spectrum, 
the theoretical spectroastrometric signature was measured
using the formalism of \citet[][]{Pontoppidan2011}.

\section{Calculation of the Bayesian probability}
\label{Bayesian}
In each model, 
a $\chi^2$ was calculated for each observational dataset (i.e., one $\chi^2$  for the $^{12}$CO P(9) line-profile, 
one $\chi^2$ for the $^{12}$CO rotational diagram, etc...) using
\begin{equation}
\chi^2=\frac{1}{N-1}\sum\limits_{i}^{}(Y_{{\rm model}~i} - Y_{{\rm obs}~i})^2/\sigma^2_{Y_{{\rm obs}~i}},
\end{equation}
where, $Y$ corresponds to the line flux per velocity channel, 
in the case of the spectrum, 
and the $Y$ axis of the plot, in the case of the rotational diagram.
$N$ is the number of channels in the spectrum or the number of data points in the rotational diagram. 
For the line-profiles we used the velocity channels from -15 to 15 km s$^{-1}$.
This enabled us to cover the wings of the line and  a small part of the continuum. 

As there are six observational datasets, 
six $\chi^2$ values were calculated per model.
Because the numerical value of $\chi^2$ can be very different for the rotational diagram and the line-profile,
before calculating the combined $\chi^2$, the $\chi^2$ of each observational dataset (i.e. line-profile, rotational diagram)
was  normalized by the minimum values of $\chi^2$ of that observational dataset in the entire grid.
The sum of the six normalized $\chi^2$  gave the final $\chi^2$  for a model.
The Bayesian probability 
\begin{equation}
p = {\rm exp}(-\chi^2/2)
\end{equation}
was calculated for each model,
and finally $p$  was divided by the sum of all $p$. In this way, a normalized Bayesian probability $p$ was obtained
for each model.
The 1D probability for each free parameter
was calculated by summing the normalized $p$ of all the models containing a particular value of the parameter in question.
Similarly, 2D probability distributions were constructed by summing the normalized $p$ of all the models
containing the pair of values for the free parameters in the plot.

\begin{figure}
\begin{center}
\includegraphics[width=0.5\textwidth]{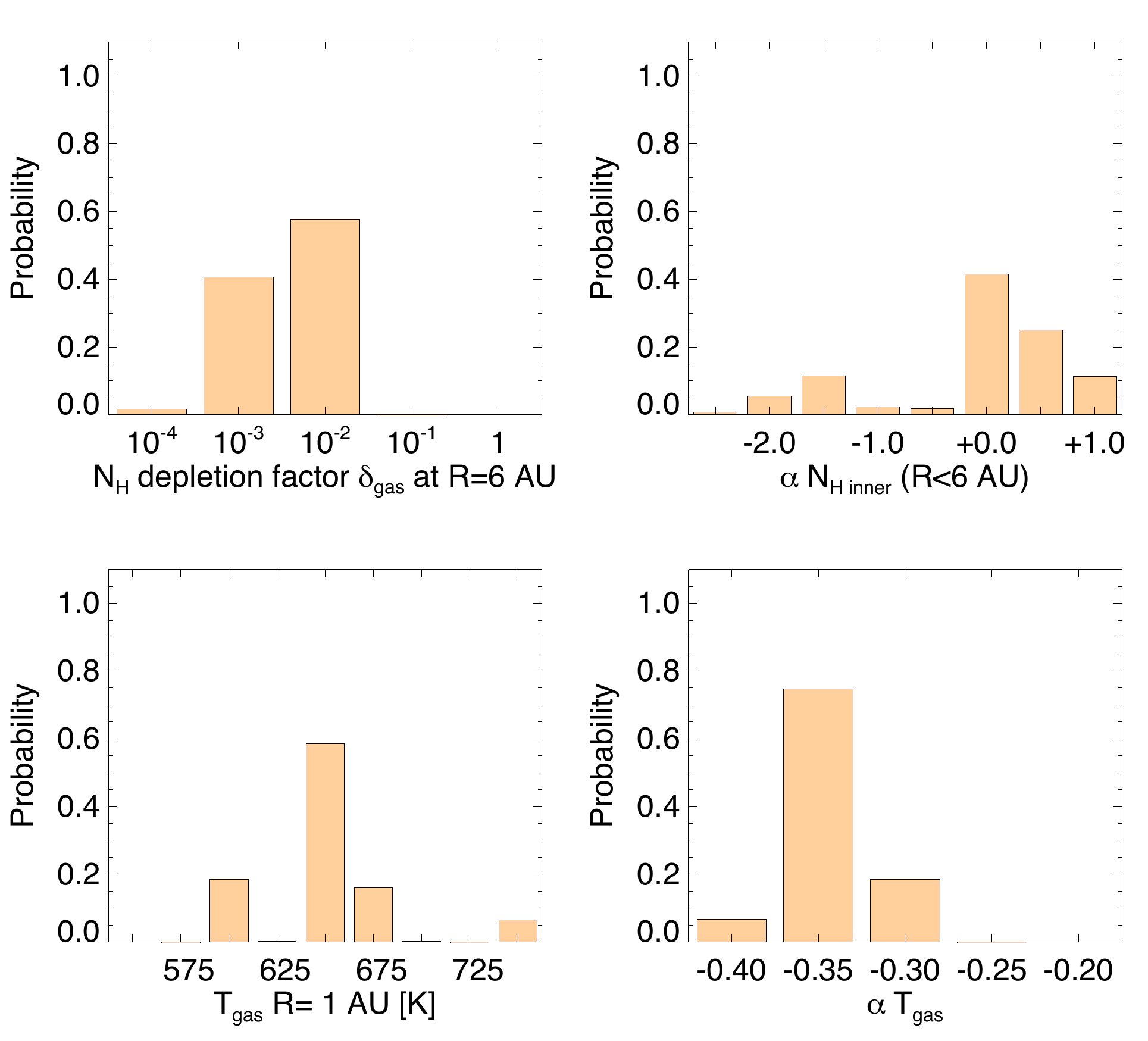}
\caption{Bayesian probability plots for a sub-sample of the grid in which only the models with $\alpha_{N_{H\,inner}}\leq +1$ are considered
(54~000 models).}
\label{grid_short_bayesian}
\end{center}
\end{figure}

\begin{figure*}
\begin{center}
{\Large {\sc best-fit grid model if} $\alpha_{N_{\rm H\,inner}}\leq +1.0$~:~~~~~$\alpha_{N_{H\,inner}}$ =0.0 ~~$\delta_{\rm gas}=10^{-2}$}\\[5mm]
\begin{tabular}{lll}
\includegraphics[width=0.25\textwidth]{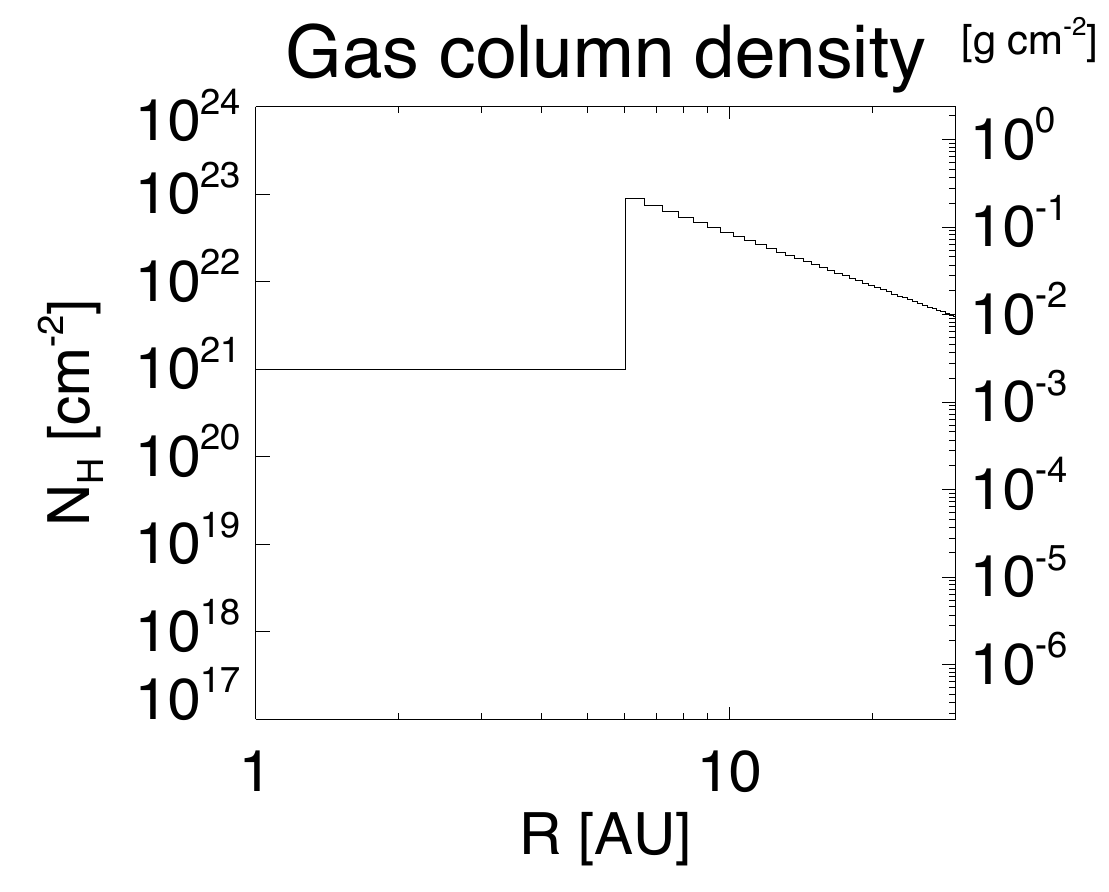}  & \includegraphics[width=0.25\textwidth]{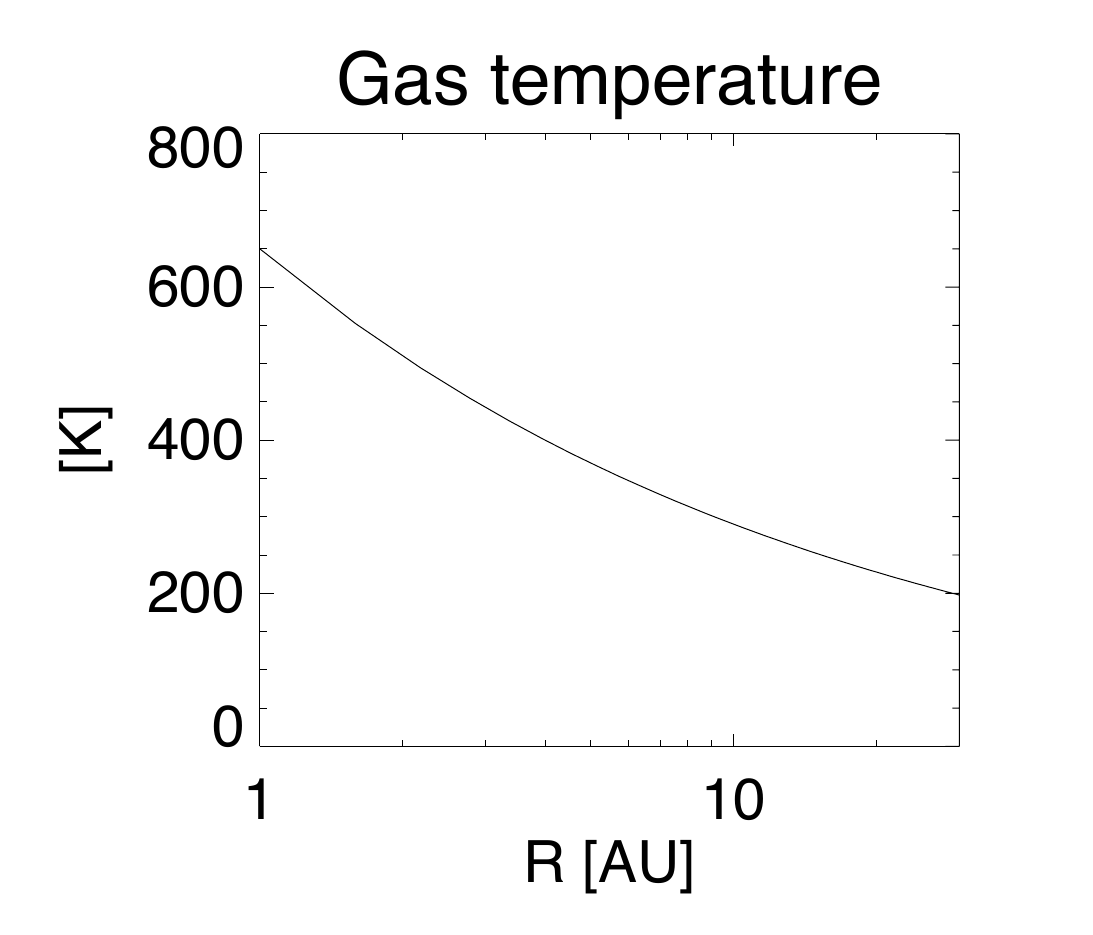} & \includegraphics[width=0.25\textwidth]{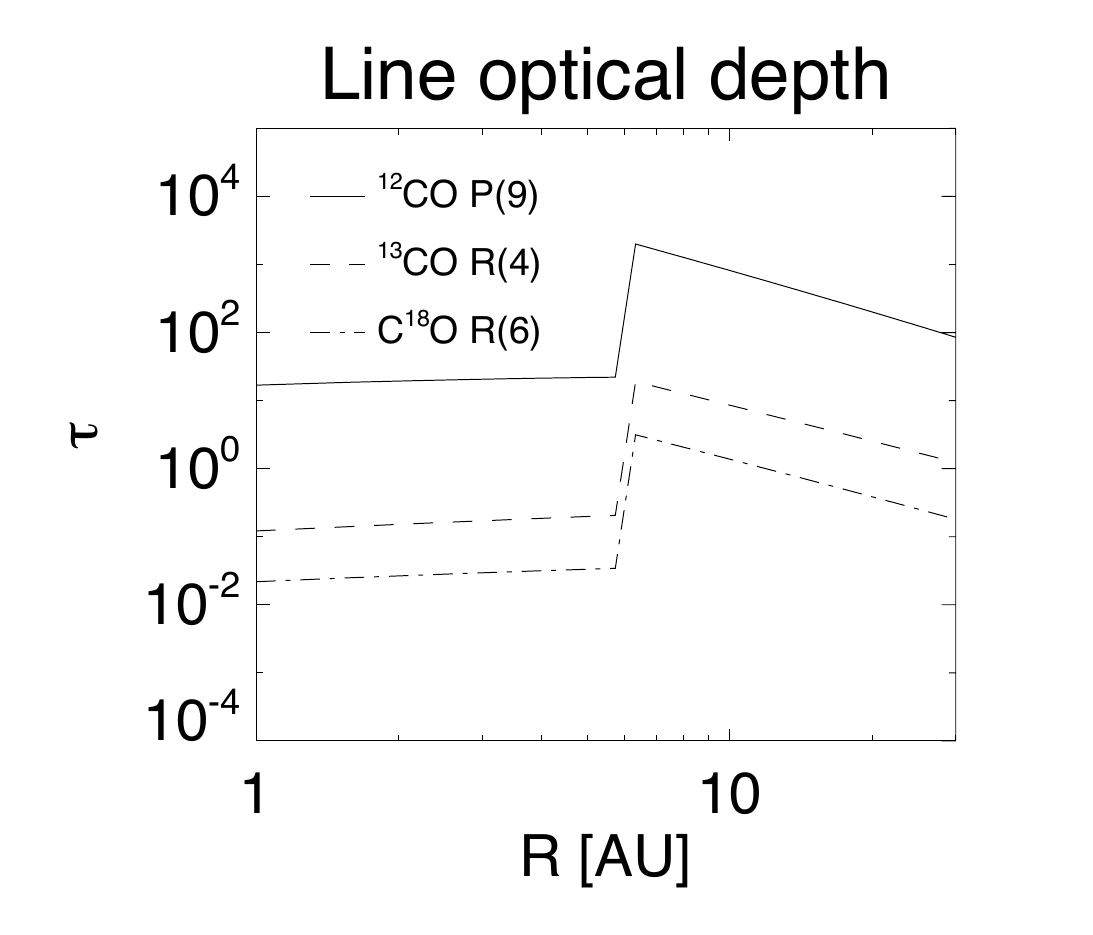} \\
\end{tabular}
\includegraphics[width=\textwidth]{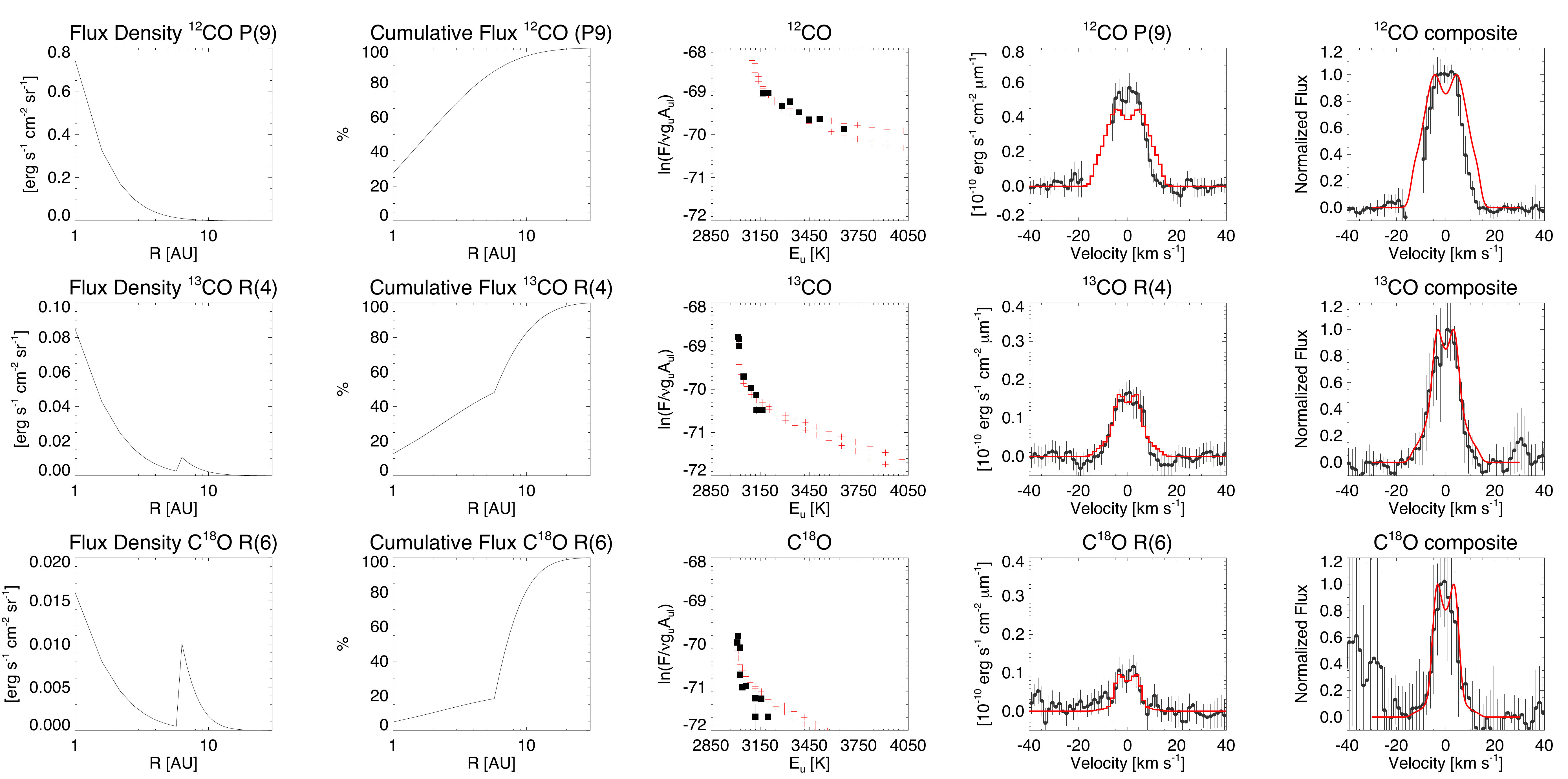} \\
\caption{
Surface density, temperature, CO optical depth,  flux density, cumulative line flux, rotational diagrams, line-profiles of the $^{12}$CO P(9), $^{13}$CO R(4), and C$^{18}$O R(6)  
emission for best-fit grid model when $\alpha_{N_{H\,inner}}\leq +1.0$.
The model is plotted in red and the observations in black. 
Observed line-profiles are displayed in flux units after continuum subtraction with 3 $\sigma$ error bars. 
The two branches seen in the rotational diagram correspond to the R and P branches of CO ro-vibrational emission.
The rightmost panels compare the normalized theoretical line-profiles with the observed composite line-profile of each CO isotopolog with a 1 $\sigma$ error bar. 
}
\label{Model_plots2}
\end{center}
\end{figure*}

\begin{figure}
\includegraphics[width=0.5\textwidth]{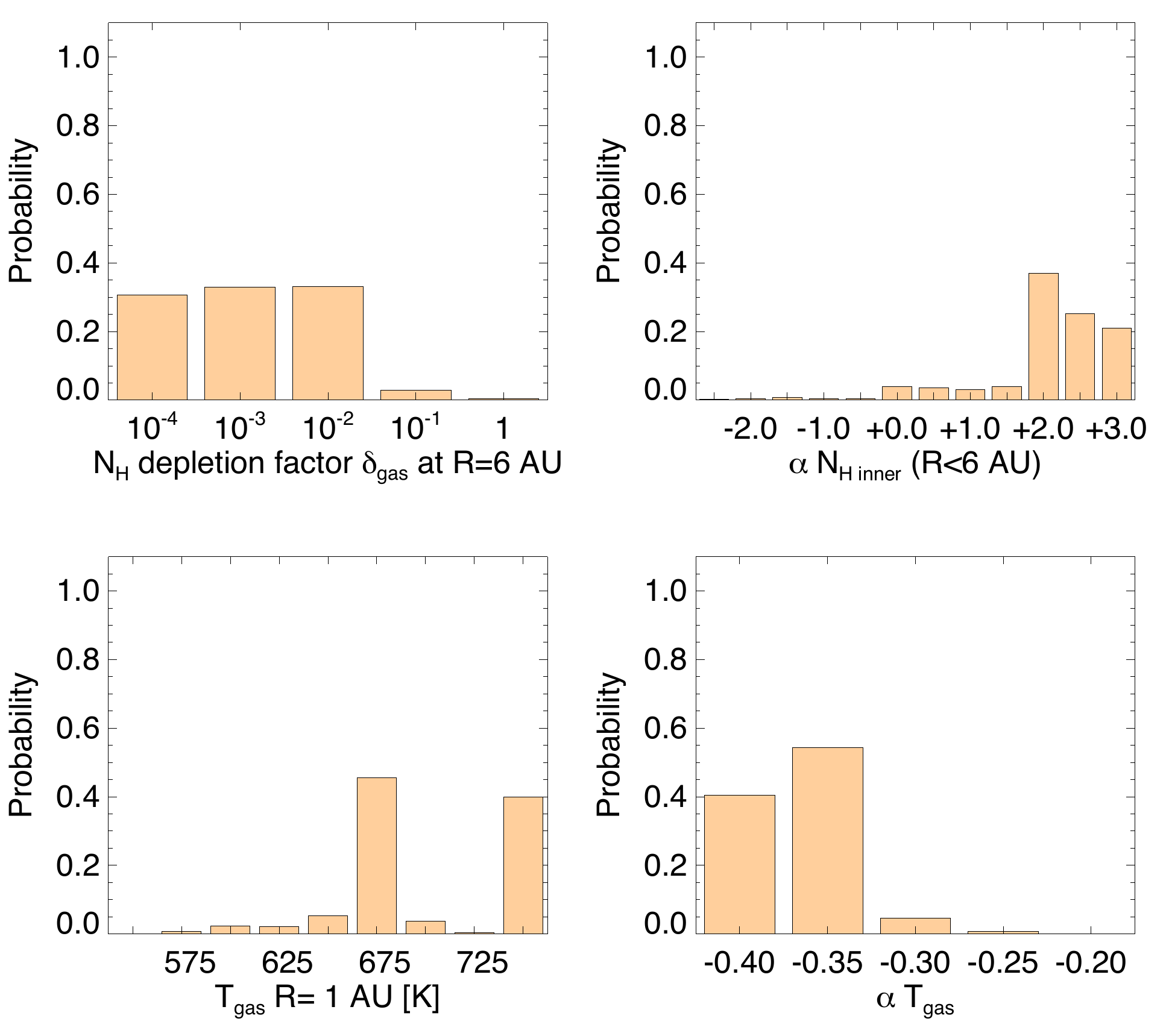}
\caption{Bayesian probability plots for the grid using only the $^{12}$CO and $^{13}$CO data (i.e. no C$^{18}$O data).
The models suggest a surface density drop of at least a factor 100 in the inner 6 AU,
and an increasing surface density profile with radius (i.e. a power-law with a positive exponent).
}
\label{bayesian_12CO_13CO}
\end{figure}

\begin{figure*}
\includegraphics[width=\textwidth]{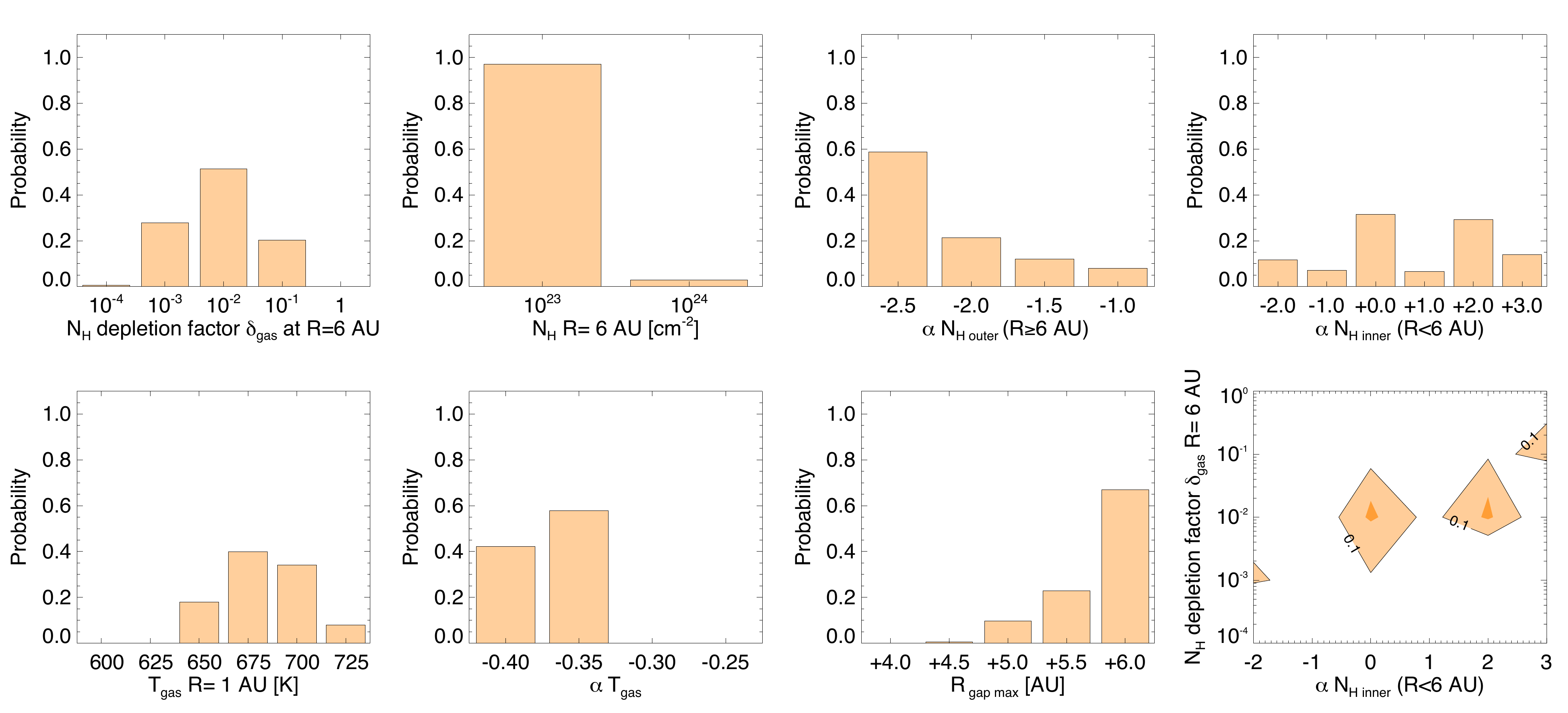}
\caption{Bayesian probability plots for a sub-grid of models (28800), in which we varied the radius of the gas density drop (R$_{\rm gap~rmax}$) between 4.0 and 6.0 AU. R$_{\rm gap~rmax}$ down to 5 AU are compatible with the data. The most likely value for the gas density drop is 6 AU, a radius similar to the dust density drop.
}
\label{bayesian_gap_rmax}
\end{figure*}

\begin{figure*}
\includegraphics[width=\textwidth]{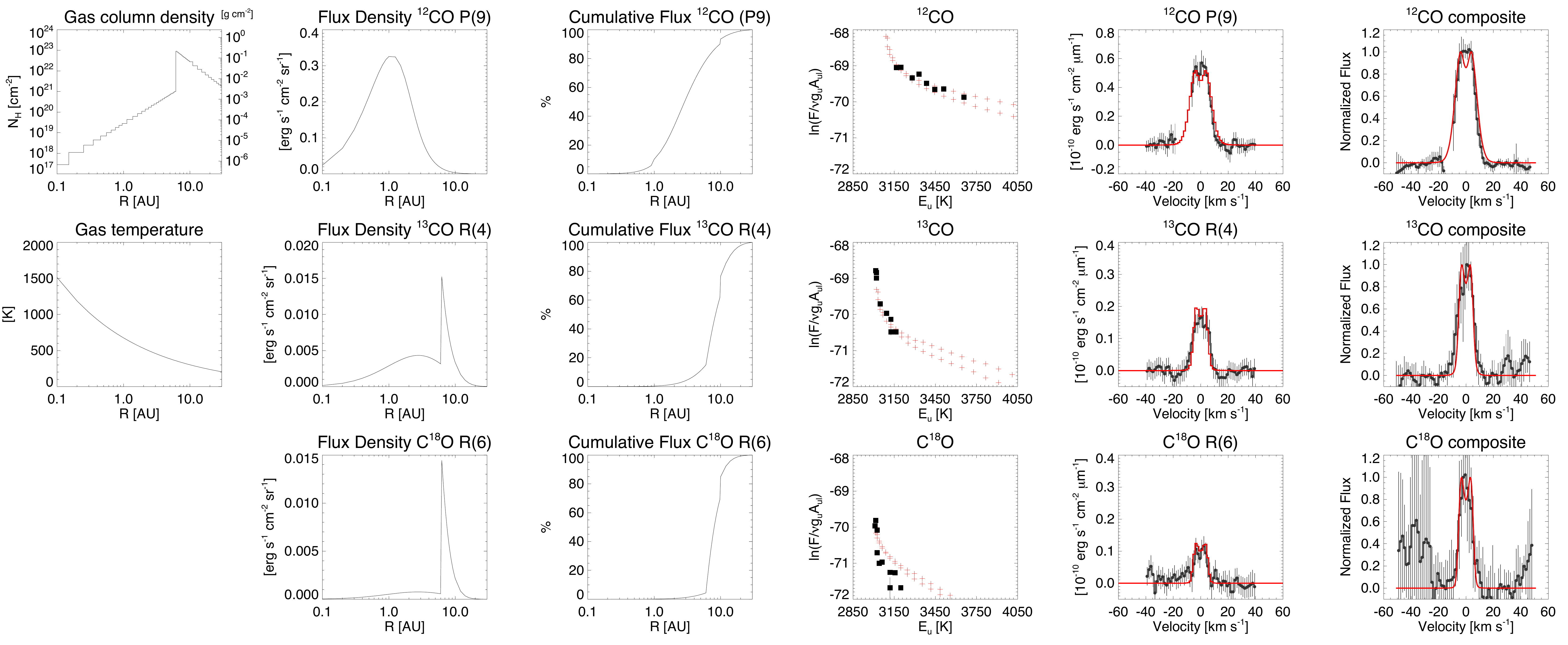}
\caption{Flux density, cumulative line flux, rotational diagrams, and the line-profiles of the $^{12}$CO P(9), $^{13}$CO R(4), and C$^{18}$O R(6)  
emission, for the model with the best combined fit to the rotational diagrams and line-profiles, extrapolating the surface density and
temperature profile down to 0.1 AU.
The model is shown in red and the observations in black.
The observed line-profiles are displayed in flux units after continuum subtraction with 3 $\sigma$ error bars. 
The two branches seen in the rotational diagram correspond to the R and P branches of CO ro-vibrational emission.
The rightmost panels compare the normalized theoretical line-profiles with the observed composite line-profile of each CO isotopolog with a 1 $\sigma$ error bar. 
}
\label{Plots_model_0.1au}
\end{figure*}

\begin{figure*}
\begin{center}
\includegraphics[width=\textwidth]{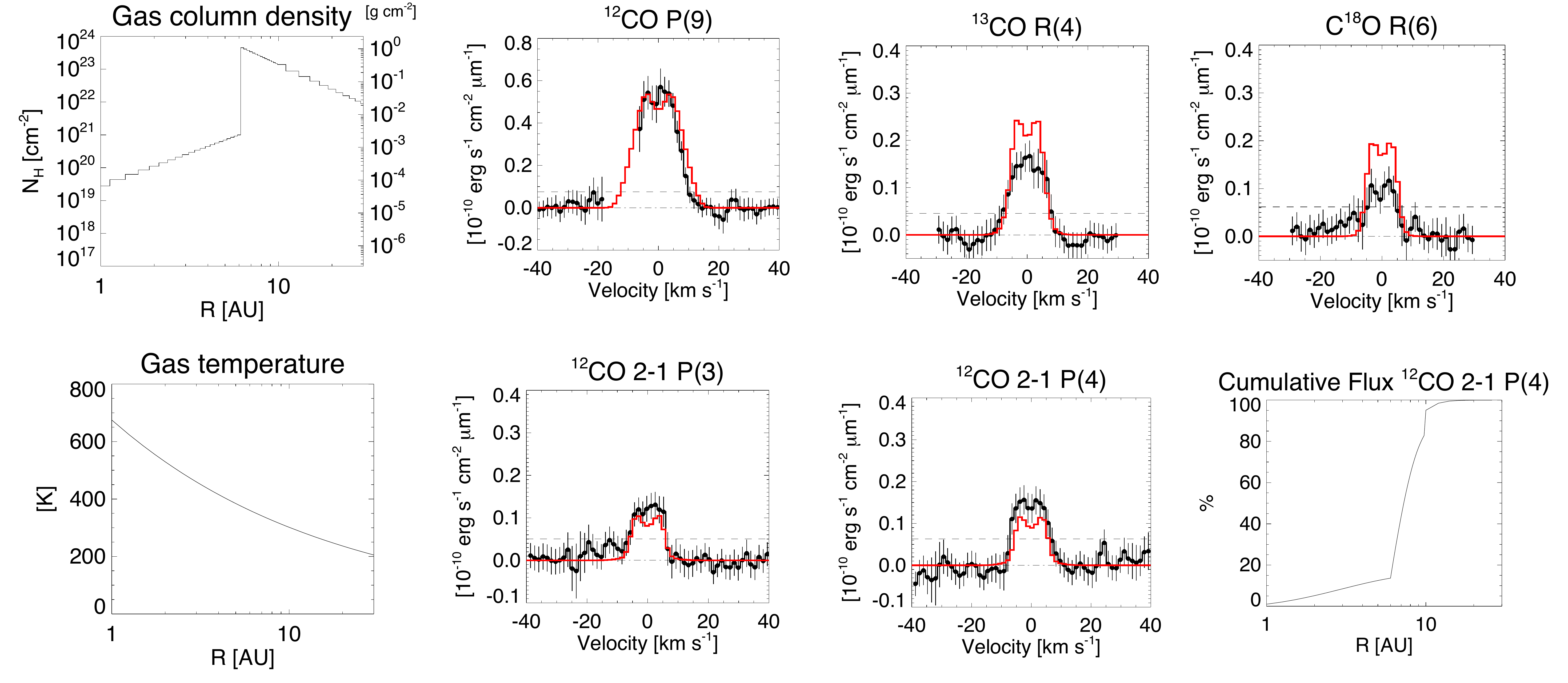}
\caption{Surface density and temperature profile, and predicted $\upsilon=1\rightarrow0$  $^{12}$CO P(9),  $^{13}$CO R(4),
C$^{18}$O R(6), and  $\upsilon=2\rightarrow1$  $^{12}$CO P(3) and $^{12}$CO P(4) line-profiles for a model with 
$N_H$ at $R=6$ AU three times larger than the best-fit grid model ($N_{H~(R=6~{\rm AU})}$ = 3$\times$10$^{23}$ cm$^{-2}$).
The higher $N_H$ enables to describe the $\upsilon=2\rightarrow1$  $^{12}$CO P(3) and $^{12}$CO P(4) line-profiles while
keeping a good  fit to the $\upsilon=1\rightarrow0$ lines. 
The model has the same temperature structure and same surface density at $R<6$ as the best-fit grid model (thus $\delta_{\rm gas}$=3.3$\times10^{-3}$).
The cumulative flux plot  shows that the $\upsilon=2\rightarrow1$  lines are dominated by the contribution at $6<R<10$ AU.
Errors in the plot are 3 $\sigma$, and the dashed horizontal line is the 5$\sigma$ limit. 
}
\label{CO2-1}
\end{center}
\end{figure*}

%\clearpage

\begin{landscape}
\begin{table}
\caption{Transition disks with a quantified gas density drop inside the dust cavity.}
\label{table_gasdrop}             % title of Table
% is used to refer this table in the text
                          % used for centering table
\centering
\begin{tabular}{l l r l l c r c l l l l l}        % centered columns (4 columns)
\hline\hline                 % inserts double horizontal lines
Star  		  &  $R_{\rm cav~dust}$  & $\delta_{\rm dust}$  & $\lambda$  & Ref. &\multicolumn{1}{l}{$R_{\rm cav~gas}$} & $\delta_{\rm gas}$ & $\Sigma_{\rm gas~inner}$ &  g/d $_{outer}$ & gas tracers 		& Ref. & $\dot{M}$ 		& Ref. \\
       			  &  [au]    	    			&  		      		& 		     & 		& [au] 			& 				& [g cm$^{-2}$] 	       &			&		 	&	   	& [$M_\odot$ yr$^{-1}$]\\
\hline                        % inserts single horizontal line
\\
 HD 139614        & 6 	 	  & $10^{-4} $ 		  & near-IR, mid-IR  & 1   & 5$-$6  	& $10^{-2}-10^{-4}$  		& $7\times10^{-5} - 2.4\times10^{-3}$	& $1-100$	& $^{12}$CO,$^{13}$CO, C$^{18}$O ro-vib.	& this work	& $1\times10^{-8}$  		& 8 	\\
                           &               &                 		  & sub-mm     	& 		&		&			&  							&		&									&			&					& 	\\[2mm]
 RXJ1615-3255  & 20 	  & $10^{-6} $ 		  & sub-mm   	& 2    	& 20 & $>10^{-4}$		& $5\times10^{0} - 1\times10^{-1}$	& 100  	& $^{12}$CO sub-mm				   	& 2 			& $3.2\times10^{-9}$ 		&  9 	\\[2mm]
 SR21 	    	  & 25 	  & $10^{-6} $ 		  & sub-mm   	& 3    	& 25 & $5\times10^{-2}$	& $1\times10^{1} - 1\times10^{0}$	& 100    	& $^{12}$CO,$^{13}$CO sub-mm 			& 3                    & $1.3\times10^{-8}$		&  9 	\\
 			  &              &                   		  &                  	&       	& 7   	& $\leq10^{-5}$    	& $1\times10^{-1} - 1\times10^{-2}$ 	&		&  	                                              			&  			&  					&	\\[2mm]
 DoAr 44             & 32         &  $10^{-2} $ 		  &  sub-mm   	& 3    	& 32	& $10^{-2}$  		& $5\times10^{-1} - 1\times10^{-1}$ 	& 100	& $^{12}$CO,$^{13}$CO sub-mm 			& 3 			& $6.3\times10^{-9}$ 		& 9	\\
			  &          	 &                  		  &                    	&       	& 16 & $\leq10^{-4}$	        &  $1\times10^{-2} -5\times 10^{-4}$  &  	         &  	                                              			& 			&					& 	\\[3mm]
 HD135344 B	  &  30, 45	 & $10^{-2}$  		  & near-IR, sub-mm	&4,5,6& 45 & $10^{-1}$             & $1\times10^{-2} -2\times10^{-2}$    & 4 		& $^{12}$CO ro-vib. \& sub-mm 			& 6$^{a}$ 		& $2\times10^{-8}$		& 10	\\
 			  &	    	&				  & 			&		&	&				&		 					&		& [OI] 63 $\mu$m						& 			&\\[2mm]
 HD135344 B  	 & 40		&  $2\times10^{-4} $   & sub-mm       	& 5,3		& 30 	&  $2\times10^{-4}$  & $7\times10^{-1}  - 2\times10^{-2}$	& 80 		&  $^{12}$CO,$^{13}$CO sub-mm			& 3$^{a}$ 		&  $2\times10^{-8}$		& 10\\[3mm]
LkCa 15 	     	 & 45		& $10^{-5}$     		 & sub-mm        & 2		& 45	& $10^{-1}$ 		& $1\times10^{3} - 1\times10^{1}$	& 100 	&  $^{12}$CO sub-mm 					& 2			& $4\times10^{-9}$ 		&  9 \\[2mm]
IRS 48		 & 60   	& $10^{-3}$   		 & sub-mm       	& 3		& 25	& $\leq10^{-3}$		& $1\times10^{-2} - 5\times10^{-4}$ 	& 12 		&  $^{12}$CO,$^{13}$CO sub-mm			& 3 			& $3.2\times10^{-9}$ 		&  11	\\[2mm]	
J1604-2130 	& 70   	& $10^{-1}$          	& sub-mm 	& 2		& 30	& $10^{-5}$  		& $1\times10^{-2} -1\times 10^{-4}$ 	& 100 	& $^{12}$CO sub-mm					& 2			& $<10^{-11}$			& 12	\\[2mm]
HD 142527$^b$ & 130 	& $10^{-5}$ 	 	&  sub-mm	& 7		& 90 &  $2\times10^{-2}$	& $1\times10^{2} - 4\times10^{0}$	& 100 	& $^{12}$CO,$^{13}$CO, C$^{18}$O sub-mm 	& 7			& $2\pm1\times10^{-7}$		& 13	\\
\hline                                   %inserts single line
\end{tabular}
\tablefoot{$R_{\rm cav~dust}$ is the dust cavity radius resolved by near-IR or sub-mm observations. When two values appear, the first corresponds to the small dust particles in the near-IR and the second to the large dust grains in the sub-mm.
  $R_{\rm cav~gas}$ is the reference radius for the drop in the gas surface density. An additional  $R_{\rm cav~gas}$ is given if a second drop in the gas surface density is suggested in the literature. 
  $\Sigma_{\rm gas~inner}$ is generally described as a power-law, the interval of values given correspond to $\Sigma_{\rm gas~inner}$ at the inner disk's radius $R_{\rm in}$, and at $R_{\rm cav~gas}$.
  If two drops for the surface density are suggested in the literature, two intervals for the surface density in the inner disk are given.
  $^{a}$ Two models are available in the literature for HD~135344B. Although they have dust mass estimations similar within a factor 2 
  ($1- 2\times10^{-4}$ M$_\odot$) the gas mass and gas surface density in the outer disk differ.  \citet[][]{Carmona2014} find a gas-to-dust mass ratio of 4 and a gas mass of 8$\times10^{-4}$ M$_\odot$ for the outer disk,  \citet[][]{vanderMarel2016} 
  find a gas-to-dust mass ratio of 80 and a gas mass of 1.5 $\times10^{-2}$ M$_\odot$. Both models also used a different disk outer radius, 200 AU in \citet[][]{Carmona2014} and 125 AU in \citet[][]{vanderMarel2016};
  $^{b}$  $\delta_{\rm dust}$ was calculated from the surface density profile provided by \citet[][]{Perez2015HD142527}, using a gas-to-dust mass of 100 at R$>$130 AU, such that the dust mass between 0.2 and 6 AU is 10$^{-9}$ M$_{\odot}$ as reported by the same authors.
 {\it References}: 
(1) \citet[][]{Matter2016}, 
(2)  \citet[][]{vanderMarel2015},
(3) \citet[][]{vanderMarel2016},
(4) \citet[][]{Andrews2011},
(5) \citet[][]{Muto2012},
(6) \citet[][]{Carmona2014},
(7) \citet[][]{Perez2015HD142527},
(8)  \citet[][]{GarciaLopez2006},
(9) \citet[][]{Manara2014},
(10) \citet[][]{Sitko2012},
(11) \citet[][]{Follete2015},
(12) \citet[][]{Mathews2012},
(13) \citet[][]{Medigutia2014}.
}
\end{table}
\end{landscape}

\end{appendix}
\end{document}